\RequirePackage{fix-cm}
\documentclass[twocolumn,natbib]{svjour3}          

\smartqed  

\usepackage[utf8]{inputenc}
\usepackage{graphicx}
\usepackage{tabularx}
\usepackage{booktabs}
\usepackage{amsmath,amssymb,amsfonts}
\usepackage{subfigure}
\usepackage{tikz}

\usepackage{hyperref}
\hypersetup{
    bookmarksdepth=2,
    breaklinks=true,
    colorlinks=true,
    citecolor=blue,
    linkcolor=blue,
    filecolor=blue,      
    urlcolor=blue,
}

\newcommand*{\affaddr}[1]{#1} 
\newcommand*{\affmark}[1][*]{\textsuperscript{#1}}

\begin{document}

\title{A New Robust Multivariate Mode Estimator for Eye-tracking Calibration}


\author{Adrien Brilhault\affmark[1]  \and
         Sergio Neuenschwander\affmark[2]   \and
         Ricardo Araujo Rios\affmark[1] }

\authorrunning{Adrien Brilhault et al.} 

\institute{
  \affaddr[1] Department of Computer Science,\\
  Federal University of Bahia, Salvador, Brazil\\ 
  \affaddr[2] Brain Institute,\\
  Federal University of Rio Grande do Norte, Natal, Brazil
}

\date{Received: date / Accepted: date}

\maketitle

\begin{abstract}
We propose in this work a new method for estimating the main mode of multivariate distributions, with application to eye-tracking calibrations. When performing eye-tracking experiments with poorly cooperative subjects, such as infants or monkeys, the calibration data generally suffer from high contamination. Outliers are typically organized in clusters, corresponding to the time intervals when subjects were not looking at the calibration points. In this type of multimodal distributions, most central tendency measures fail at estimating the principal fixation coordinates (the first mode), resulting in errors and inaccuracies when mapping the gaze to the screen coordinates. Here, we developed a new algorithm to identify the first mode of multivariate distributions, named \href{https://adrienbrilhault.github.io/BRIL/}{BRIL}, which rely on recursive depth-based filtering. This novel approach was tested on artificial mixtures of Gaussian and Uniform distributions, and compared to existing methods (conventional depth medians, robust estimators of location and scatter, and clustering-based approaches). We obtained outstanding performances, even for distributions containing very high proportions of outliers, both grouped in clusters and randomly distributed. Finally, we demonstrate the strength of our method in a real-world scenario using experimental data from eye-tracking calibrations with Capuchin monkeys, especially for distributions where other algorithms typically lack accuracy.

\end{abstract}


\section{Introduction}\label{sec:intro}

\sloppy
Eye-tracking systems were developed to provide estimates of a subject's gaze, by processing data collected from a variety of devices, which can be classified into three main categories: i) electrooculography (EOG); ii) scleral search coil; and iii) camera-based systems \citep{duchowski_eye_2007}. EOG was the most commonly used device in the early years of eye-tracking and relies on a set of electrodes placed around subjects' eyes to measure changes of skin electric potentials corresponding to eye movements. Although this technique offers high-frequency measurements, it usually suffers from poor spatial precision. On the other hand, scleral coils provide both high temporal (up to $1000$Hz) and spatial resolution (with errors lower than $0.08^o$). Search coils consist of a metallic wire, either embedded in a contact lens or surgically implanted on the sclera. Eye positions are estimated from the voltage signals induced by a set of orthogonal Helmholtz coils placed around the head \citep{robinson_method_1963}. Although this method is usually considered a gold standard, it is often not suitable due to its invasive nature and the complexity of the experimental set-up. Camera-based eye-trackers, in contrast, are non-invasive. The estimation of gaze direction is obtained from the pupil shape and position, derived from video recordings of the eye. With advances in hardware and image processing techniques, today, high-end camera-based eye-trackers can reach temporal and spatial precision comparable to search coils, thus becoming the most popular technique \citep{schmitt_comparing_2007}.

Most eye-tracking applications require a calibration procedure to estimate the model parameters and map the eye positions into the application space (typically a screen in front of the user). This step is critical to ensure accurate estimates and require the adjustment of individual parameters. Calibration methods usually rely on the presentation of fixation targets at different known screen positions, while the subject is required to maintain fixation at each of them. A minimum of four calibration points is generally required to compute the mapping function \citep{hansen_eye_2005}, but due to non-uniform variations across the visual field, a higher number of positions is often preferable to improve precision. Given the time and attentional demands, calibration procedures generally consist of 5 to 15 positions, as a trade-off between precision and subject effort.

\begin{figure}
    \centering
    \begin{tikzpicture}
    \draw (0, 0) node[inner sep=0] {\includegraphics[width=.48\linewidth]{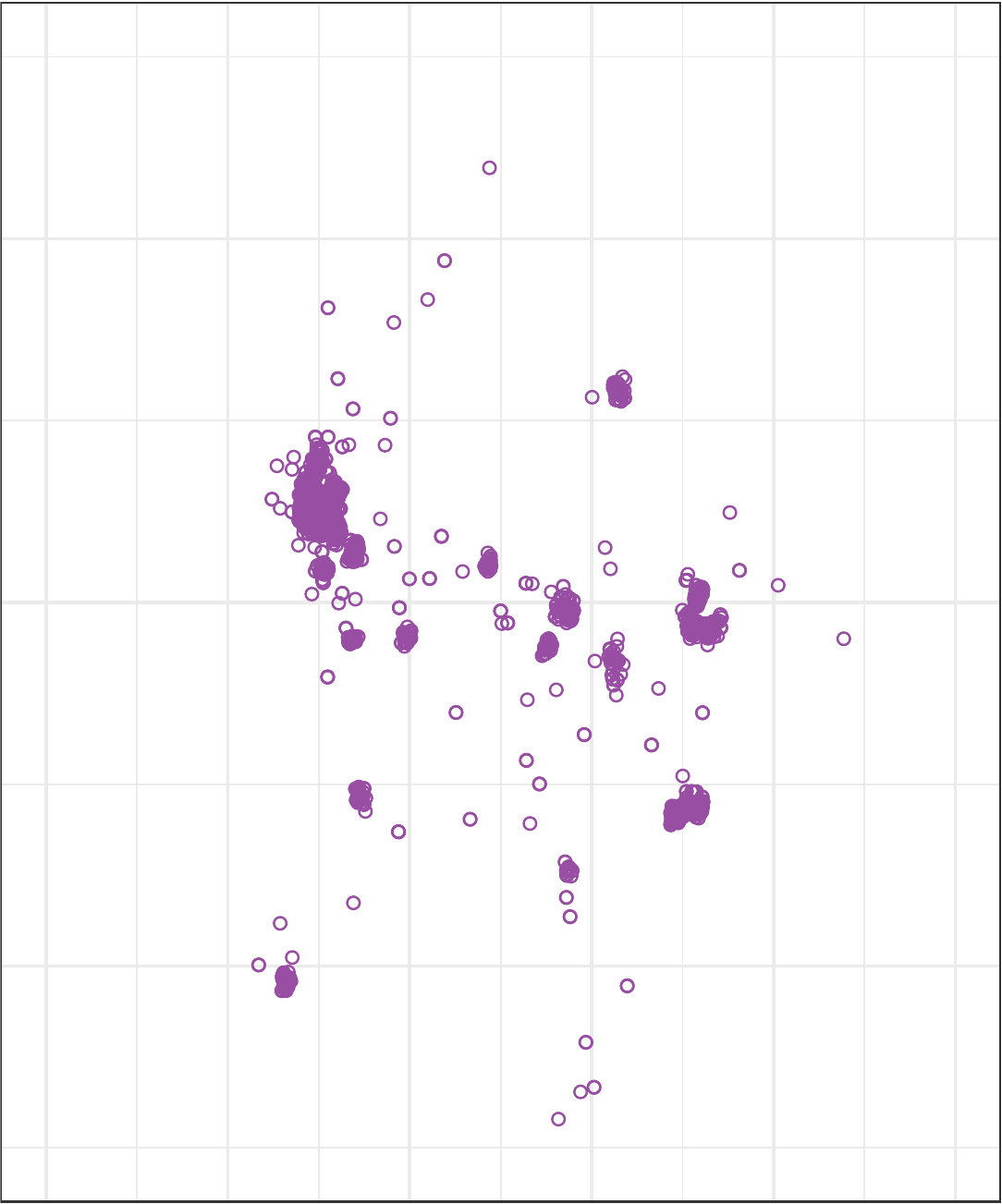}
      \includegraphics[width=.48\linewidth]{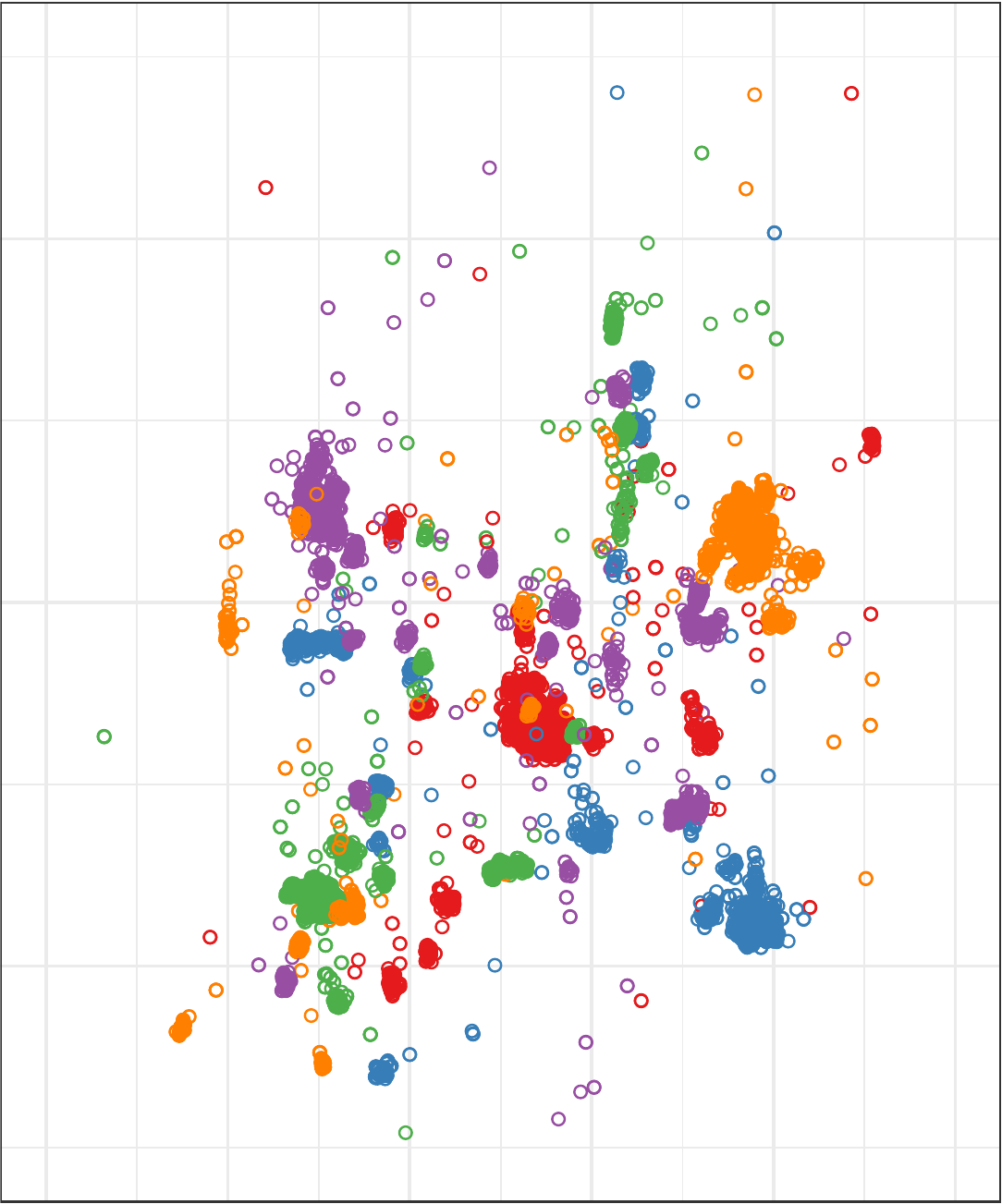}};
    \draw (-3.7,2) node {A};
    \draw (0.5,2) node {B};
    \end{tikzpicture}

    \caption{Gaze positions recorded during a calibration task performed with a Capuchin monkey. The calibration consisted of 50 trials of 1700ms, with one of 5 targets presented in each, in a randomized manner. In Figure A, the gaze positions correspond to all the presentations of the top-left target. In Figure B, we show the eye-tracking recordings for all 5 targets of the calibration procedure (colors corresponding to the target displayed at the time).}
    \label{fig:exampleCalib}
\end{figure}

For most studies with humans, calibration procedures are straightforward. Instructions can be given directly to the subjects, who will maintain fixation at the different calibration points appearing successively on the screen and confirm each target by pressing a key. For poorly or non-cooperative subjects (such as infants, people with a mental disorder, or animals), calibration procedures can be challenging. Subjects may not understand or follow the instructions, leading to errors in the parameters adjustment and gaze estimation. To illustrate these issues, we show in Figure~\ref{fig:exampleCalib} the gaze positions for a calibration task performed by a monkey. Figure~\ref{fig:exampleCalib}A corresponds to the coordinates recorded during repeated presentations of a single calibration point. The presence of different clusters indicates that the monkey looked at several positions rather than maintaining fixation on the target (which corresponds, in this example, to the largest cluster located on the top-left). A similar behavior was observed for the whole calibration procedure with all five targets (Figure~\ref{fig:exampleCalib}B). Notice that, due to the high number of outliers, computing an average of eye positions, as usually done with cooperative subjects, would yield poor estimates.

A simple solution when dealing with non-cooperative subjects is the input from an operator during the experiment, who will manually assign gaze positions to calibration points while looking at the eye-tracker's live feed. The precision of this approach, however, has been shown to strongly relies on the experience and attention of the assistant \citep{nystrom_influence_2013}.

In principle, calibration-free procedures might overcome such issues. Different methods have been proposed, which essentially rely on the computation of a 3D model of the pupil. These techniques are generally based on multiple cameras \citep{kohlbecher_calibration-free_2008,smith_viewpointer:_2005}, or on a single camera and several light sources \citep{hammoud_passive_2008,morimoto_free_2002}. An inherent problem of calibration-free methods is the complexity of the experimental setup, requiring multiple cameras and/or light sources, and high computational power. Another downside is that they usually provide a poor spatial precision, with errors above 2 visual degrees \citep{model_calibration_2011}.

Here, we propose an alternative calibration procedure, that does not require manual intervention from either the subject or the experimenter. Our method relies on the assumption that subjects will fixate at the targets displayed for a sufficient part of the calibration procedure, as a result of a natural tendency to look at salient stimuli (in our experiments, colored targets shown on a black background).

A similar user-calibration-free approach was proposed by \citet{model_probabilistic_2012}. Calibration parameters were estimated by extracting the dominant bin from the histograms in both vertical and horizontal dimensions, taking into account all the gaze positions recorded during the presentation of each calibration target. The study, conducted in infants, was able to provide accurate results despite subjects attending the targets for a small fraction of the time ($47\%$). The assumption was that incorrect fixations would follow a uniform distribution across the visual field, while the actual fixations to the target would be grouped into a single cluster. In these conditions, the highest bin in every dimension would indeed coincide with the center of the cluster, allowing for the identification of the main fixation coordinates.

In our experiments with monkeys, however, we noticed that the eye positions recorded presented two types of incorrect coordinates. First, randomly distributed errors resulting from sampling artifacts or transient failures in pupil tracking due to occlusions, blinks, and light reflections. Secondly, errors that occurred when the subject was not looking at the target. We noticed, for instance, that subjects tend to look systematically at positions shown in previous trials, probably reflecting an expectation created from the repeated presentations (every target was shown multiple times in a randomized order). These incorrect gaze positions, rather than being uniformly distributed, as assumed by Model and Eizenman, were grouped in clusters, corresponding to each of the incorrect fixations.

\begin{figure}[!htbp]
    \centering
    \includegraphics[width=0.95\linewidth]{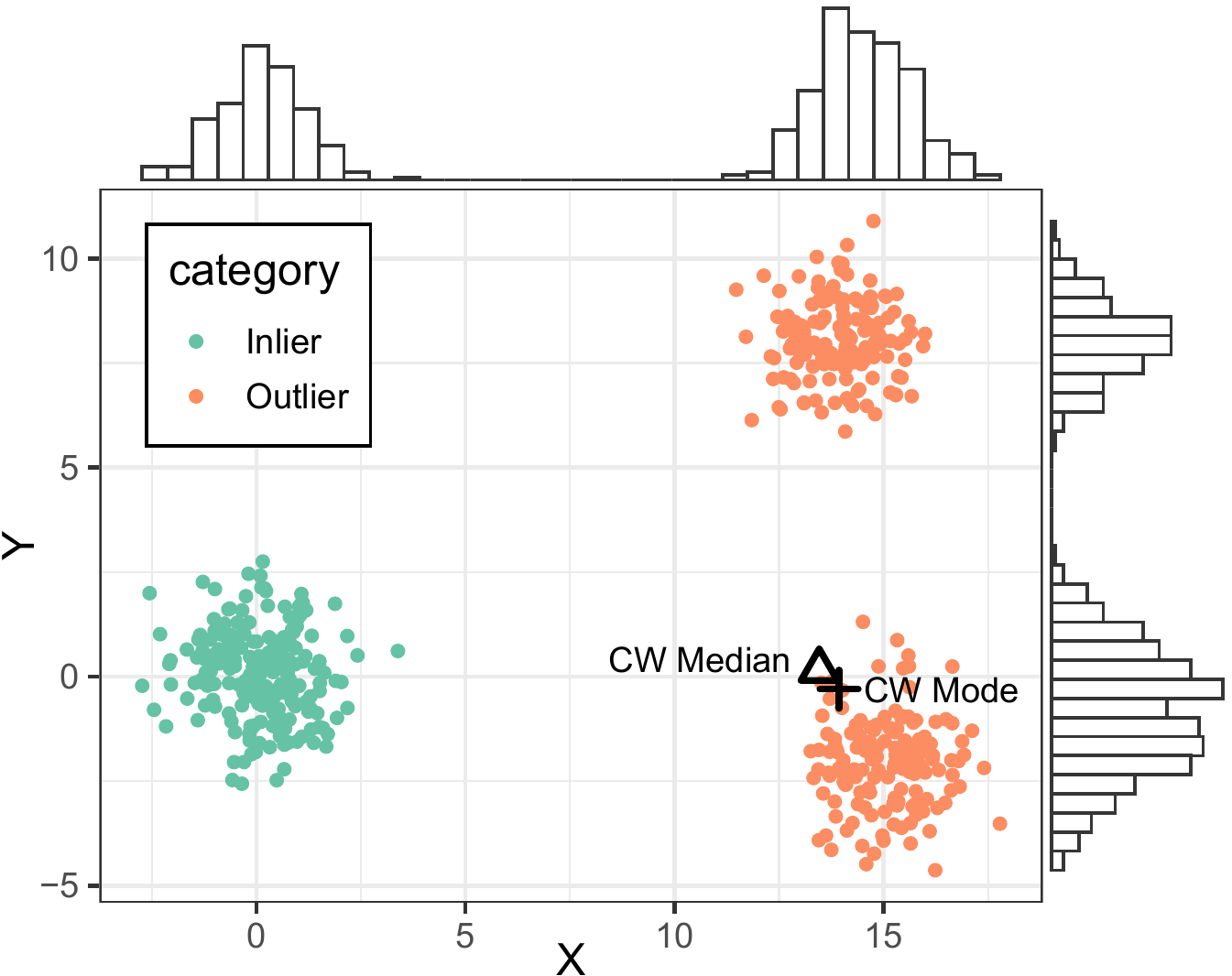}
    \includegraphics[width=0.95\linewidth]{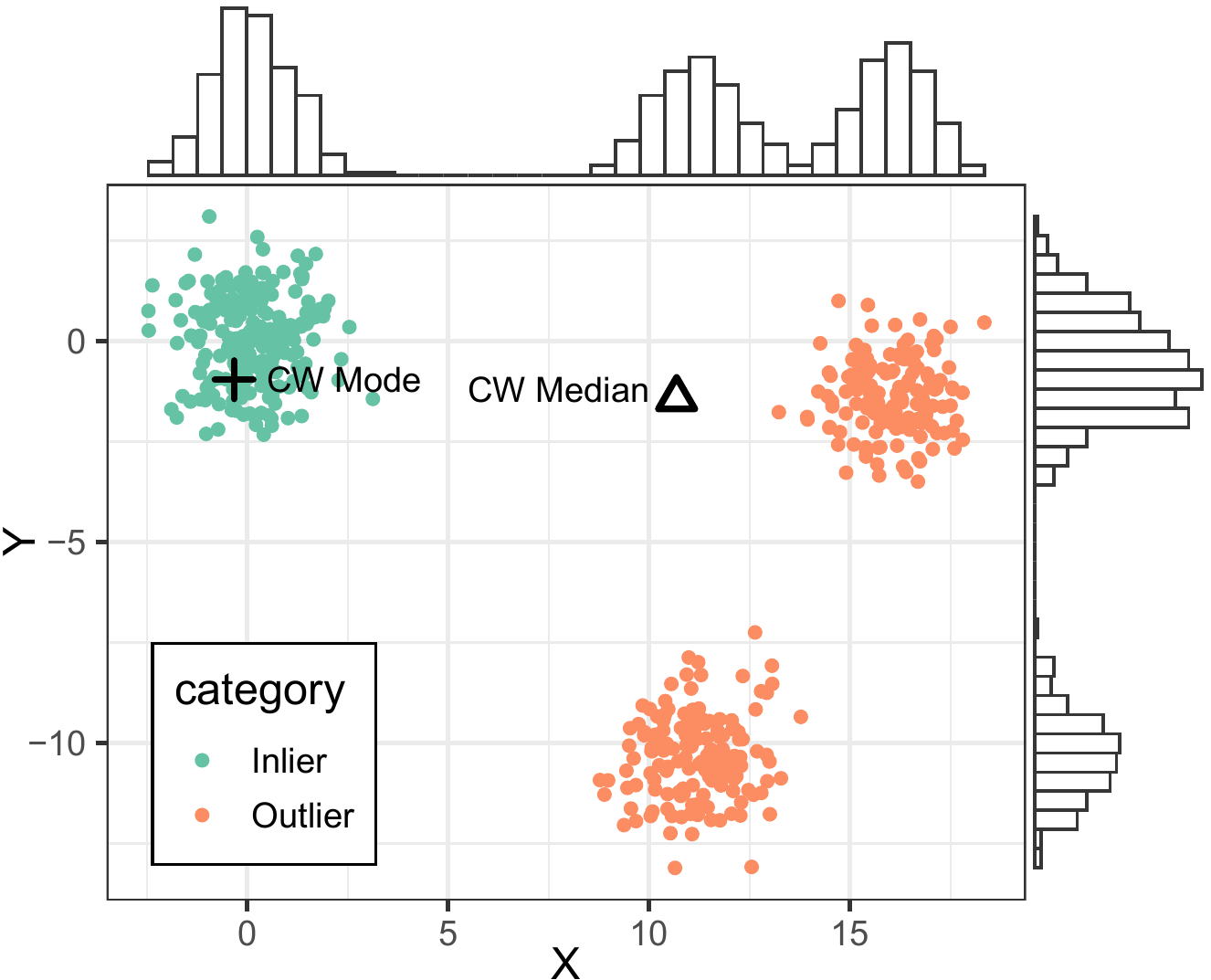}
    \caption{Coordinate-wise median and mode in multimodal distributions. Both figures present the same data points, rotated around the origin by an angle of $35\deg$. The distribution is composed of 500 samples grouped in three clusters (40\% in the main cluster, in green, and 30\% in each of the two outliers clusters, in red). These central measure tendencies are not affine equivariant and often fail to locate the main cluster when several groups share overlapping positions in one of the two dimensions.}
    \label{fig:CW-Median}
\end{figure}

An essential limitation of \citet{model_probabilistic_2012} approach is its poor performance on multimodal distributions with high contamination, such as those observed in our monkey recordings. This problem is particularly clear in a scenario where all the incorrect gaze coordinates are grouped in clusters (as in the illustrative example in Figure~\ref{fig:CW-Median}). If several clusters of outliers share an overlapping position in one of the two dimensions, extracting the highest bin for each component (which can be considered a coordinate-wise mode), or the coordinate-wise median, will often result in coordinates outside of the cluster with the highest cardinality. Both approaches fail to take into account the relation between the two dimensions, and are also sensitive to affine transformations. Considering incorrect positions exclusively distributed in clusters, Model and Eizenman's method, as well as the coordinate-wise median, would offer a tolerance to outliers only up to $50\%$.

To overcome these issues, we developed a new approach to estimate the main mode of multivariate distributions. Our method, named BRIL\footnote{https://adrienbrilhault.github.io/BRIL/} (Bootstrap and Refine Iterative Locator), can withstand high levels of contamination, and consists of the following steps:

\sloppy
\begin{enumerate}
  \item BOOTSRAP: An initial location is computed through the recursive trimming of low depth samples.
  \item REFINE: The central location from Step 1 is refined through two successive filtering processes. First, we sequentially remove the furthest outliers, ordered by Euclidean distances to the first estimate, until reaching an unimodal distribution. The center of this temporary subset is then re-estimated as in Step 1, and the ordering of the remaining samples adjusted considering the new location and a robust estimate of scatter. The second filtering step consists in recursively trimming this subset until it fails to reject the normality hypothesis. The central location for this group is finally computed by averaging the samples of the last subset.
  \item ITERATE: Steps 1 and 2 are re-applied after removing the samples selected in Step 2 from the global distribution. This procedure is repeated until all samples have been assigned to a group or until the distribution considered in Step 1 already appears unimodal before any filtering. In this case, a last execution of Steps 1 and 2 is performed, while not leading to new iterations if samples were to remain unassigned. The final estimate of the algorithm is obtained by selecting the center associated to the iteration with the group of highest cardinality.
\end{enumerate}

The paper is organized as follows. In the next section, we detail the characteristics of the artificial and experimental datasets used to evaluate our approach, along with the testing procedure and metrics considered. In Section~\ref{sec:robustcentral}, we introduce depth measures and study their performance in the context of multimodal distributions. Section~\ref{sec:BRIL} describes the different components of our approach relying on these depth functions. We then show, in the fifth section, how convex body minimizers, such as the Minimum Volume Ellipsoid (MVE) and the Minimum Covariance Determinant (MCD) locators, can be used within our framework in place of depth measures. Finally, in Section ~\ref{sec:results}, we present our results on both artificial and experimental data, showing that our approach is able to reliably estimate the center of the main cluster in multimodal distributions, even when facing a high quantity of outliers, as with eye-tracking calibrations performed by non-cooperative subjects.

\section{Methods}\label{sec:methods}

\subsection{Synthetic dataset} \label{sec:artificialdataset}

The tests on synthetic data were performed using a Monte-Carlo design, creating mixtures of bivariate normals and uniform distribution, with $500$ repetitions for each combination of the following parameters: i) the number of samples, set to $500$ (with exception of the simulations in Section~\ref{sec:size-effect}, for which the sample size varied from $250$ to $2000$); ii) the uniform noise ratio, varying from $0$ to $50\%$; iii) the number of clusters, chosen from the interval $K = [2,5]$; and iv) the proportion of inliers (i.e. the percentage of samples belonging to the main cluster over all the clustered data points, disconsidering uniform noise), set within $\delta = [100/K, 100]$ percent.

The clusters centers were randomly drawn at every Monte-Carlo repetition in the $\mathbb{R}^2$ space of coordinates $([0-50],[0-50])$, with a spatial constraint to avoid overlapping clusters, by setting a minimum distance between any pair of cluster centers of three times the sum of their standard deviations. A bivariate normal distribution was then created for every cluster, centered on each of these coordinates, with a $\sigma = 1$ for the main cluster, and a random value between $0.5$ and $1.5$ for each of the secondary clusters. The cardinality of the main cluster was defined by $|k_p| = N \times (1 - \varepsilon) \times \delta$, in which $N$ is the total sample size, $\varepsilon$ the noise ratio, and $\delta$ the inliers ratio. On the other side, the cardinalities of the other clusters were defined by $|k_i| = N \times (1 - \varepsilon) \times (1 - \delta) / (K - 1)$, with $K$ corresponding to the number of clusters.

\subsection{Real-world dataset}\label{sec:real-dataset}

 To assess the performance of our method in a real-word scenario, we analyzed eye-tracking calibrations procedures made in a series of experiments with monkeys. Data was obtained in the context of visual studies in capuchins, carried out at the Brain Institute of the Federal University of Rio Grande de Norte, Brazil.

 Eye positions were recorded using the eye-tracking setup developed by \citet{matsuda_keiji_advanced_2013,matsuda_widely_2017}. The system relies on a PointGrey Grasshopper3 infrared camera (GS3-U3-41C6NIR-C). Image acquisition was set at a frequency of $235$Hz. The eye coordinate signals generated by the system were recorded through an acquisition board (National Instruments E-Series board), at a sampling frequency of $1000$ samples/s (oversampling was due to requirements of the experimental setup). The monkeys had the head fixed during the recordings.

\begin{figure*}[!tbhp] 
    \centering
    \includegraphics[height=6cm,keepaspectratio]{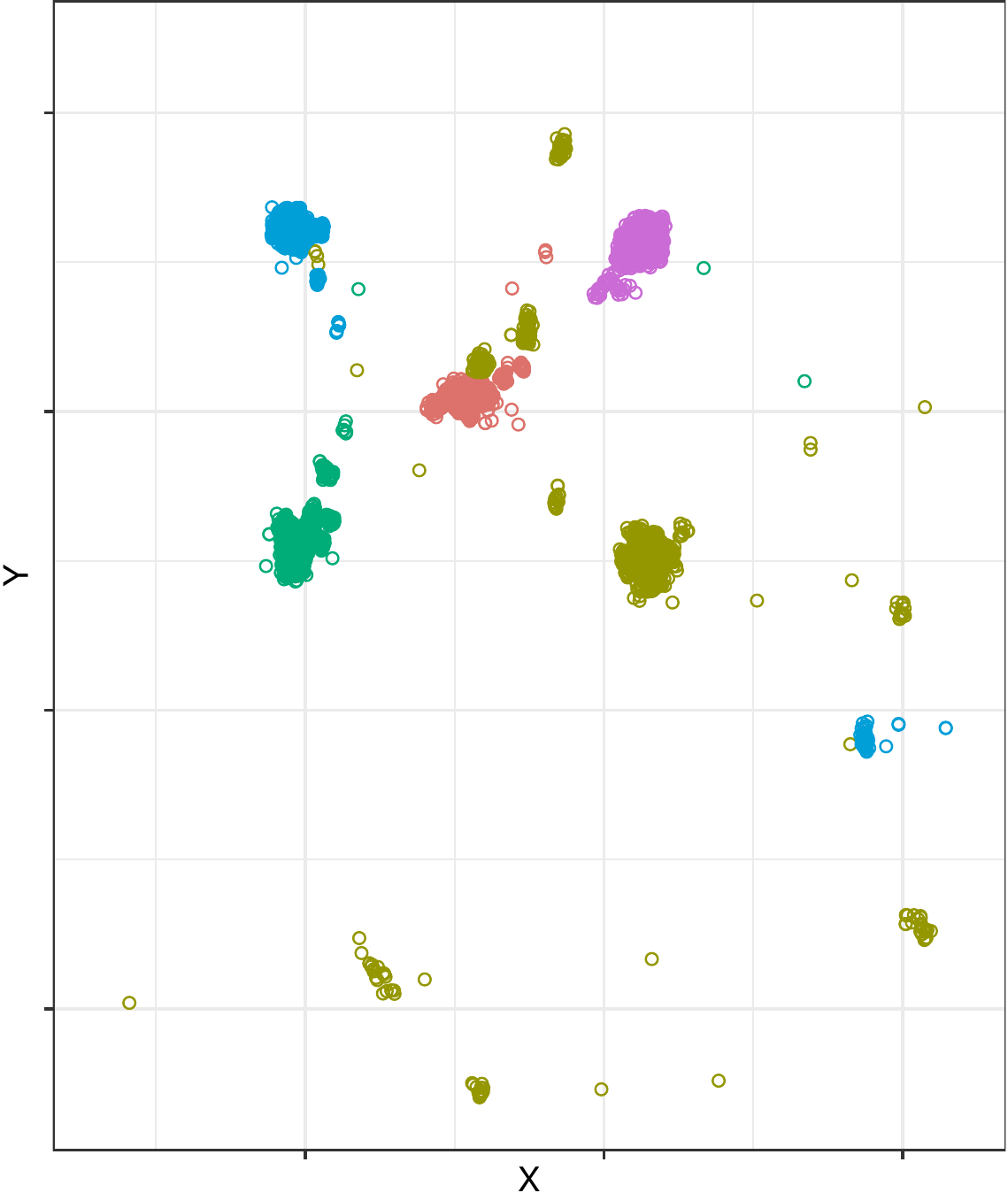} \hfill
    \includegraphics[height=6cm,keepaspectratio]{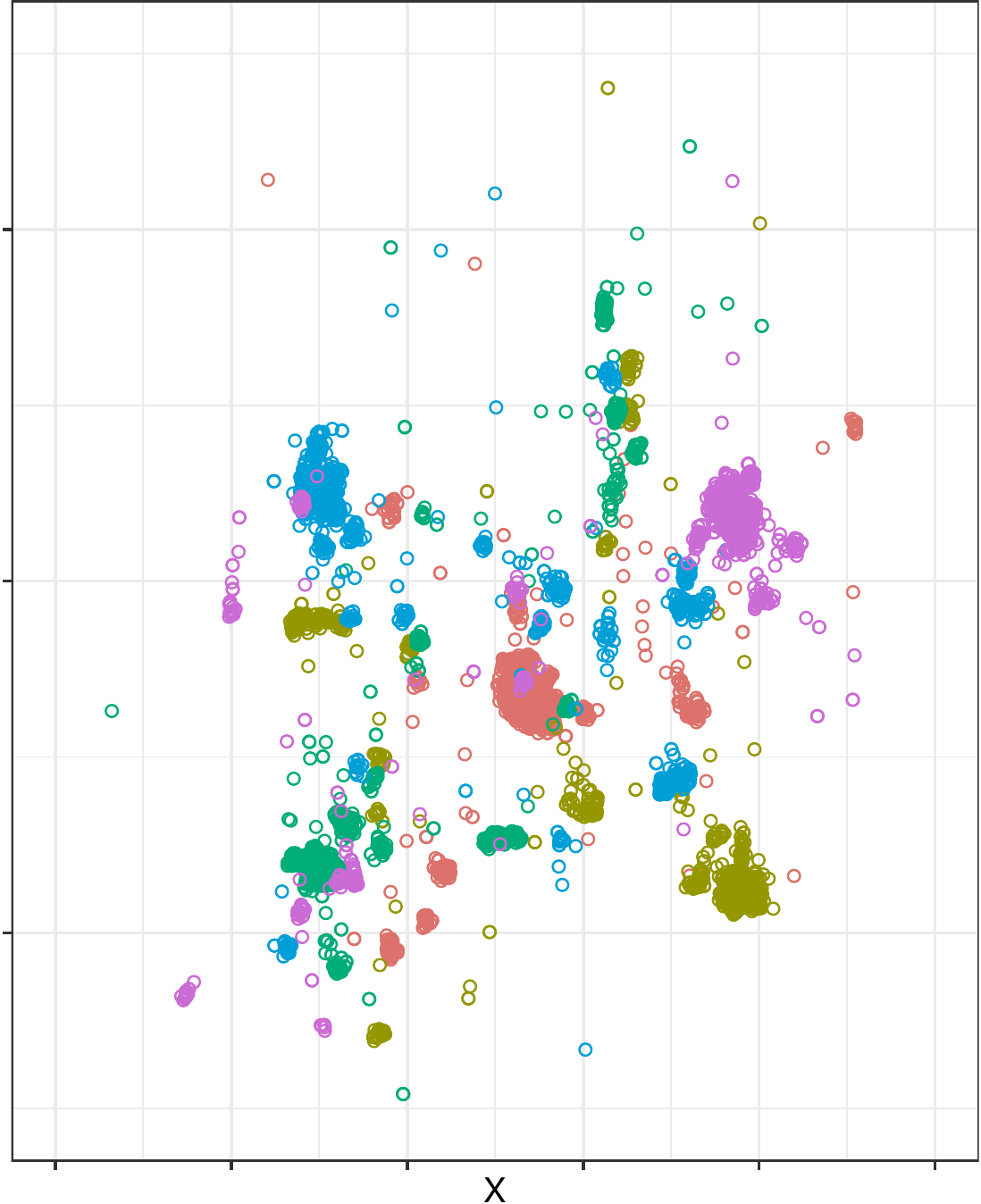} \hfill
    \includegraphics[height=6cm,keepaspectratio]{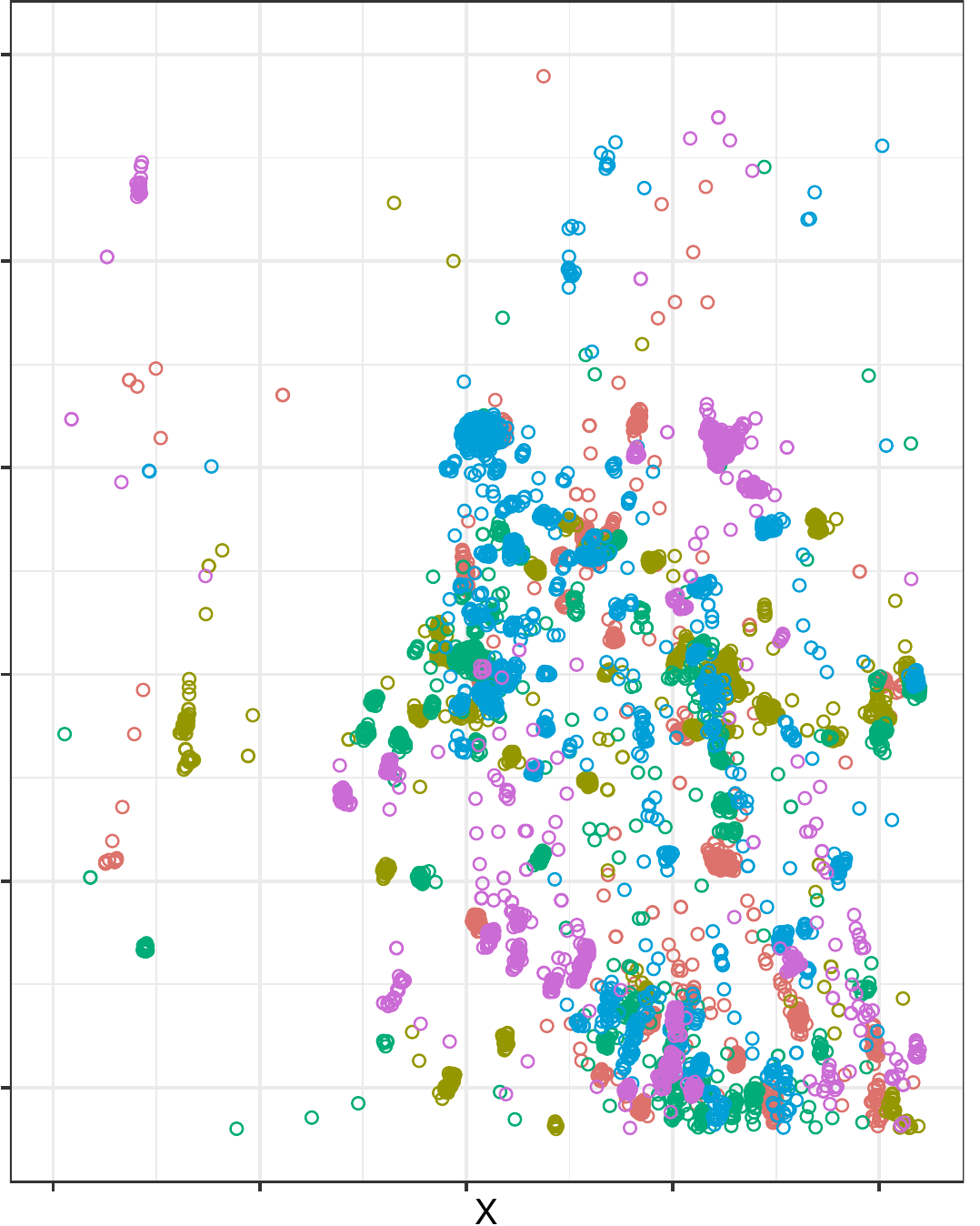}
    \includegraphics[height=6cm,keepaspectratio]{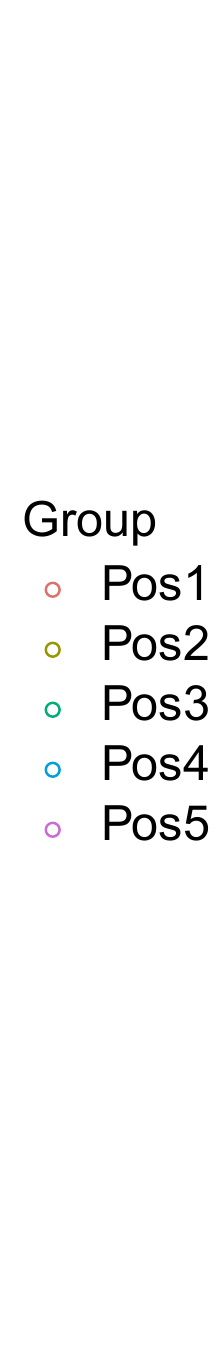}

    \caption{These three real-world datasets, showing different amounts of noise, consist of the gaze coordinates recorded during eye-tracker calibrations performed with Capuchin monkeys. Colors indicate which of the five calibration targets was displayed at the time of the recording. The data contamination can reflect both a lack of attention from the subject, attending other parts of the screen, and technical artefacts.}
    \label{fig:realDatasets}
\end{figure*}

 We analyzed three datasets of increasing contamination\footnote{Available at \href{https://github.com/adrienbrilhault/BRIL/}{github.com/adrienbrilhault/BRIL/}}, shown in Figure~\ref{fig:realDatasets}, which reflected different levels of attention and cooperation. In every trial, a target was displayed for a duration ranging from $1200$ to $3000$ms, at one of 5 possible positions \{~[0,0];[-100,100]; [100,100];[100,-100];[-100,-100]~\}. The calibration sessions consisted of 50 trials, with every target presented multiple times, in a randomized manner. The Table~\ref{table:tableDatasets} summarize the characteristics of each dataset, showing, in particular, the total number of samples collected for every calibration target, and the number of valid eye position, after removing the incorrect values due to eye blinks, artifacts, or saturated values from the analog acquisition board.

\begin{table}[htbp]
  \centering
  \caption{Summary of the real-world datasets. Set1, Set2, and Set3 correspond to calibrations with low, medium, and high contamination, respectively. For each of these sessions, the table provide the duration of trials (in ms), the number of trials in the session, its number of calibration targets, as well as the number of repetitions of each target, with the total number of samples, and the number of valid ones (once technical artefacts and out-of-range values were filtered).}
  
  \resizebox{1\linewidth}{!}{
    \begin{tabular}{cccccccc}
    \toprule
    \textbf{Session} & \multicolumn{1}{l}{\textbf{Dur}} & \multicolumn{1}{l}{\textbf{nTrials}} & \multicolumn{1}{l}{\textbf{nTargets}} & \multicolumn{1}{l}{\textbf{Target}} & \multicolumn{1}{l}{\textbf{targetTrials}} & \multicolumn{1}{l}{\textbf{Samples}} & \multicolumn{1}{l}{\textbf{Valid}} \\
   
    \midrule
    Set1  & 1700  & 50    & 5     & 1     & 7     & 11900 & 11900 \\
    Set1  & 1700  & 50    & 5     & 2     & 14    & 23800 & 23175 \\
    Set1  & 1700  & 50    & 5     & 3     & 12    & 20400 & 20400 \\
    Set1  & 1700  & 50    & 5     & 4     & 6     & 10200 & 8951 \\
    Set1  & 1700  & 50    & 5     & 5     & 11    & 18700 & 18700 \\
    Set2  & 1200  & 50    & 5     & 1     & 12    & 14400 & 11122 \\
    Set2  & 1200  & 50    & 5     & 2     & 10    & 12000 & 11016 \\
    Set2  & 1200  & 50    & 5     & 3     & 7     & 8400  & 7637 \\
    Set2  & 1200  & 50    & 5     & 4     & 11    & 13200 & 12127 \\
    Set2  & 1200  & 50    & 5     & 5     & 10    & 12000 & 11022 \\
    Set3  & 3000  & 50    & 5     & 1     & 6     & 18000 & 15607 \\
    Set3  & 3000  & 50    & 5     & 2     & 8     & 24000 & 21318 \\
    Set3  & 3000  & 50    & 5     & 3     & 10    & 30000 & 26602 \\
    Set3  & 3000  & 50    & 5     & 4     & 19    & 57000 & 46419 \\
    Set3  & 3000  & 50    & 5     & 5     & 7     & 21000 & 18536 \\
    \end{tabular}
    }
  \label{table:tableDatasets}
\end{table}

\subsection{Clustering algorithms}\label{sec:cluster}

Clustering algorithms are an intuitive choice to consider when dealing with mixtures of ellipsoidal distributions. In \citep{brilhault_eyetracking_2019}, we showed that clustering techniques are indeed highly effective in solving calibration issues with uncooperative subjects. The identification of the main fixations to a calibration target was obtained by partitioning the gaze samples using a clustering algorithm, and selecting the centroid of the cluster with the highest cardinality. The performance of various clustering methods were assessed on artificial and experimental datasets similar to those in the present work. Five algorithms were evaluated, selected from the main clustering techniques categories: 
\renewcommand{\labelenumi}{\alph{enumi})}
\begin{enumerate}
\item Partitional: \emph{PAM} (Partition Around Medoid, \citet{rousseeuw_clustering_1987}) and \emph{TClust} (a robust adaptation of the \emph{K-means} algorithm, \citet{garcia-escudero_general_2008}).
\item Hierarchical clustering: \emph{HClust}~\citep{murtagh_survey_1983}.
\item Density-based: \emph{DBSCAN} (Density-Based Spatial Clustering of Applications with Noise, \citet{ester_density-based_1996}).
\item Model-based: \emph{MClust} (Gaussian mixture model fitted via the Expectation-Maximization algorithm, \citet{scrucca_mclust_2016}).
\end{enumerate}

Among all these methods, Model-based clustering showed the best performances
on artificial data, as expected, but poor results on experimental data. \emph{PAM} and \emph{DBSCAN}, on the other hand, presented satisfying results both in real and synthetic environments. To compare \emph{PAM}, \emph{DBSCAN} and \emph{MClust} with our new approach, we applied in this work the same methodology as in \citep{brilhault_eyetracking_2019}. The clustering algorithms where applied to our datasets (both experimental and synthetic), and the main mode estimate was computed by averaging the coordinates of the samples belonging to the group with the highest count. \emph{PAM} and \emph{DBSCAN} parameters were selected through a grid search minimizing the average silhouette, while \emph{MClust} used an initialization through agglomerative hierarchical clustering and an EM fitting of Gaussian mixtures based on the Bayesian Information Criterion (BIC).

\subsection{Evaluation}

To quantify the performance of different central tendency measures and partitioning algorithms in locating the main mode of a compound distribution, we considered the following evaluation metrics:
i) $Mean Error$: average Euclidean distance between the true center of the inliers group (or the ground truth coordinates for real-word data) and the location estimate (in the case of partitioning technique, the center of the largest group provided by the partitioning algorithm); ii) $SD$: the standard deviation of measure i) across all monte-carlo repetitions; iii) $SSE$:  the sum of squared error, from the Euclidean distances used in measure i);  iv) $Hit Rate$: the proportion of simulations correctly identifying the largest cluster; v) $Miss Rate$: the opposite measure of iv), equal to $(1-HitRate)$. Regarding measures iv) and v), considering that the main cluster is drawn from a multivariate normal distribution with a standard deviation $\sigma$ equal to 1, and therefore that 99.7\% of its samples are expected to fall within $3\sigma$, we labeled as correct identifications (\emph{hits}) all the simulations with a location estimate showing an error to the true center inferior to $3$.

\section{Robust estimates of central tendency}\label{sec:robustcentral}

In this section we review different measures of central tendency which could be applied to contaminated multivariate distributions as those encountered in our context of application, and present their results on similar artificial mixtures.

Measures of central tendency and robust statistics have been extensively studied for univariate distributions. The mean, median, and mode are certainly the most common central estimates, and, when facing contaminated distributions, the mode and median are usually preferred due to their tolerance to outliers. The median can in fact sustain up to $50\%$ contamination in regard to the breakdown property defined by \citet{donoho_breakdown_1992}. Univariate definitions of central tendency, location, scale, or outlyingness are however not trivial to generalize to multivariate data. For instance, the coordinate-wise median and mode, unlike the mean, are not affine equivariant and fail to consider the dependency between the different dimensions \citep{niinimaa_multivariate_1999}. The extension of these concepts to higher dimensions has been the focus of numerous researches in the statistical community \citep{oja_descriptive_1983}. John Tukey, for instance, introduced in 1975 the notion of data depth, as a new estimate of central tendency in multivariate distributions \citep{tukey_mathematics_1975,zuo_general_2000}. Associated with a given distribution $F$ on $\mathbb{R}^p$, depth functions are designed to provide a $F$-based center-outward ordering (and thus a ranking) of points in $\mathbb{R}^p$. High and low depths corresponding to ``centrality'' and ``outlyingness'', respectively. The Multivariate Median (MM) associated to a depth measure is defined as the location that globally maximize depth \citep{serfling_depth_2006}. 

Desirable properties of depth functions include monotonicity, uniqueness, maximality at center, affine equivariance, and robustness to outliers (generally measured by the breakdown point or influence function) \citep{liu_notion_1990,zuo_performance_2000}. Several data depth functions and their corresponding MM have been proposed over the last decades, offering diverse definitions of centrality, and presenting differences in term of theoretical properties, robustness, or computational complexity. Among the most popular, we can cite:

\begin{itemize}
\item The Halfspace depth defines the depth of a point $\theta \in \mathbb{R}^p$ relative to a dataset $X = {x_1, . . . , x_n}$ as the minimal number of observations in any closed halfspace (i.e. half-planes when $p=2$) that contains $\theta$ \citep{tukey_mathematics_1975}. The Halfspace median (also called Tukey median) is the location $\theta$ with maximal depth \citep{donoho_breakdown_1992}.

\item Oja's depth, or Simplicial Volume depth: given a set $X$ of $n$ points in $\mathbb{R}^p$, the Oja depth of a point $\theta$ is the sum of the volumes of the $p-$variate simplices formed by $\theta$ and all subsets of $p+1$ elements from $X$ (in bidimensional data, it corresponds to the sum of the areas of all the triangles having $\theta$ as vertex). The point in $\mathbb{R}^p$ with the maximal depth is the Oja median \citep{oja_descriptive_1983,oja_multivariate_2013,fischer_computing_2016}.

\item Simplicial depth corresponds to the probability of a point $\theta \in \mathbb{R}^p$ to be contained inside a closed random simplex of $p+1$ observations from the distribution \citep{liu_notion_1990,liu_notion_1988}. The simplicial median, or Liu median, is the point included in the highest number of unique simplices formed by $p+1$ samples.

\item The spatial median, also known as median center or geometric median, corresponds to the point minimizing the sum of the euclidian distances to all observations \citep{mottonen_asymptotic_2010,small_survey_1990,lopuhaa_breakdown_1991,chaudhuri_geometric_1996}. Different outlyingness measures can be derived from this median, such as the $L2$-Depth from \citep{mosler_depth_2012}, based on the $L2$ distance to the spatial median (or its affine invariant version using the covariance matrix of the samples, the one used in this work), or the spatial depth introduced in \citep{vardi_multivariate_2000,serfling_depth_2002}, which relies on spatial quantiles computed from the distances of a point to each sample from the distribution
\citep{serfling_depth_2006,serfling_quantile_2002,zuo_structural_2000}. Note that the terminology used in the literature can often be ambiguous, the spacial multivariate median, despite relying on $L_2$ distances, is sometimes referred to as $L_1$-MM (and associated data depth functions called $L1$-Depth by extension), as it minimize the $L_1$ norm of the vector of euclidean distances, see \citep{zuo_performance_2000,dodge_multivariate_1999} for discussion.

\item The Mahalanobis depth of a point $\theta $ with respect to a distribution $F$ in $\mathbb{R}^p$ is defined as $MD(\theta|F)=[1+(x - \mu)^t S^{-1}(x-\mu)]^{-1}$, where $\mu$ and $S$ correspond to the the mean and covariance matrix of $F$, respectively \citep{masse_asymptotics_2004,zuo_structural_2000}.

\item Projection depth: depth is determined as the minimum outlyingness of a point in relation to the univariate median of any one-dimensional projection of the distribution  \citep{zuo_projection-based_2003,zuo_general_2000}. 

\item Convex hull trimming/peeling methods \citep{donoho_breakdown_1992,eddy_convex_1982,green_peeling_2006}, such as the zonoid depth \citep{dyckerhoff_zonoid_1996,koshevoy_zonoid_1997}: these methods usually construct nested convex hulls recursively, starting with the smallest convex hull which encloses all samples, then stripping away outlying data (the points on the hull boundary) and reiterating the process. These successive hulls delimit depth regions, the more central the higher depth. The corresponding median would be the final point remaining, or if it is not unique the centroid of the last convex hull.

\end{itemize}

\begin{figure*}[!hbt] 
    \centering
    \begin{minipage}[c]{\textwidth} 
    \hspace{1cm}
    \includegraphics[width=0.2\textwidth]{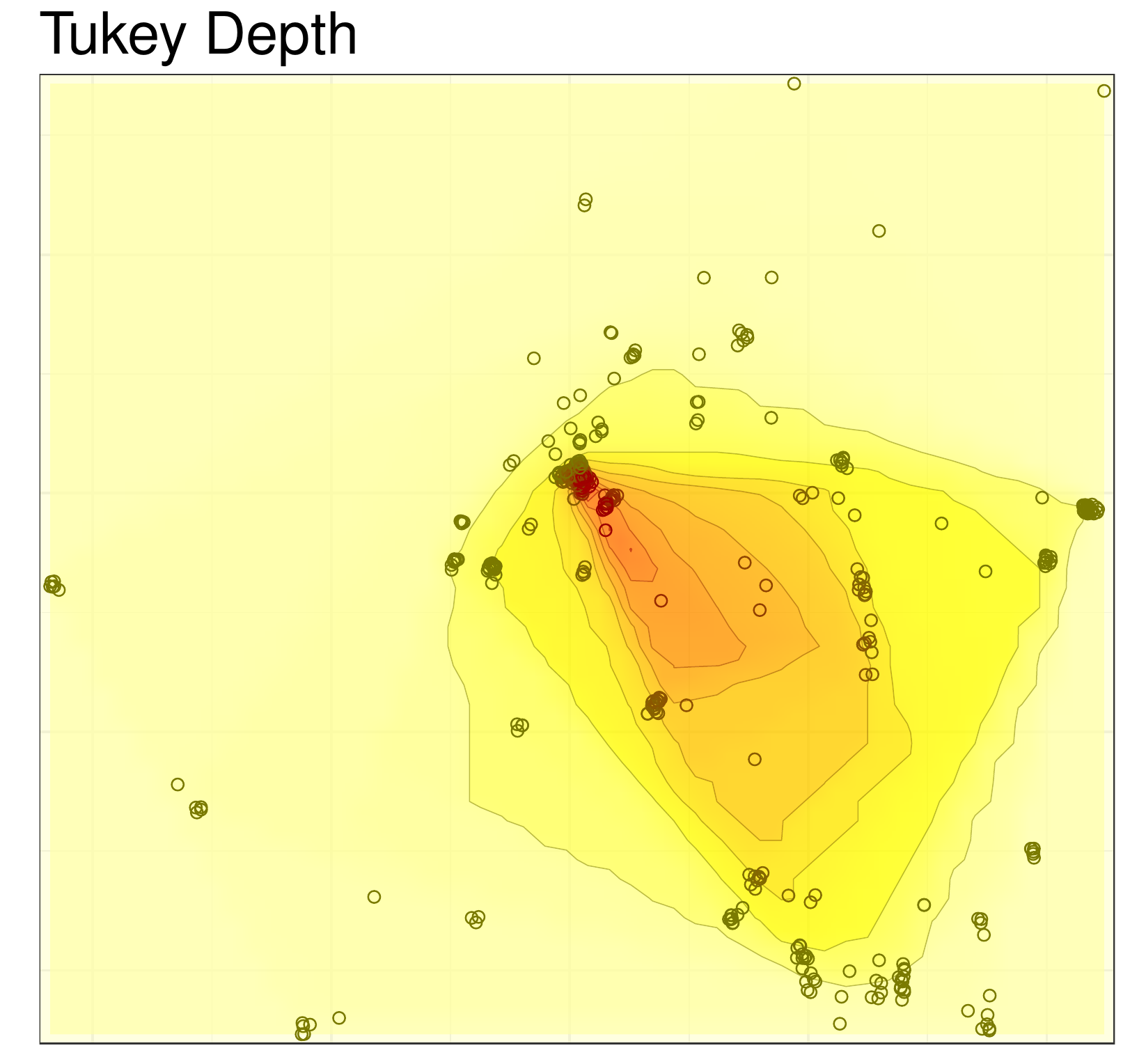} \hfill
    \includegraphics[width=0.2\textwidth]{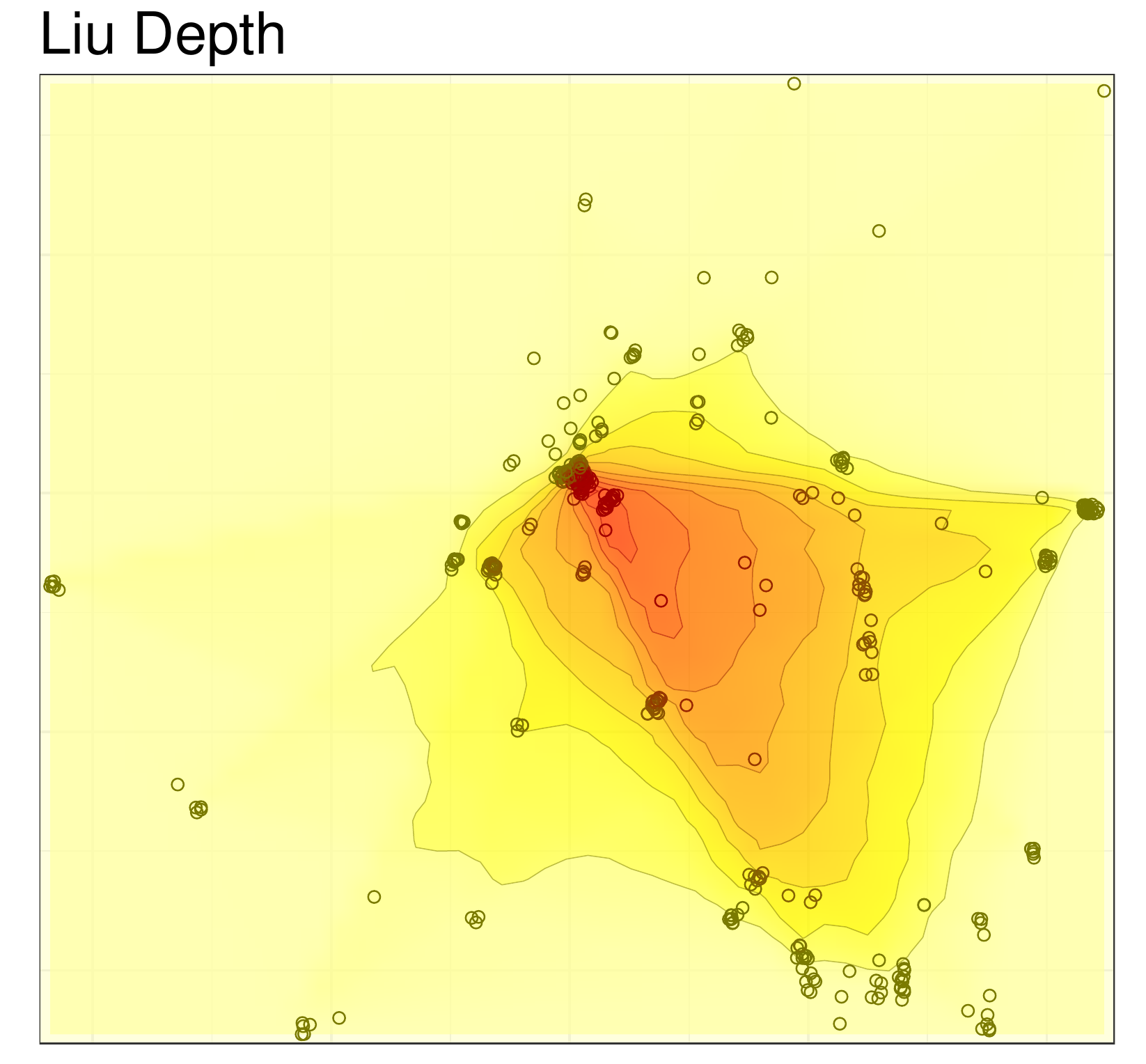}\hfill
    \includegraphics[width=0.2\textwidth]{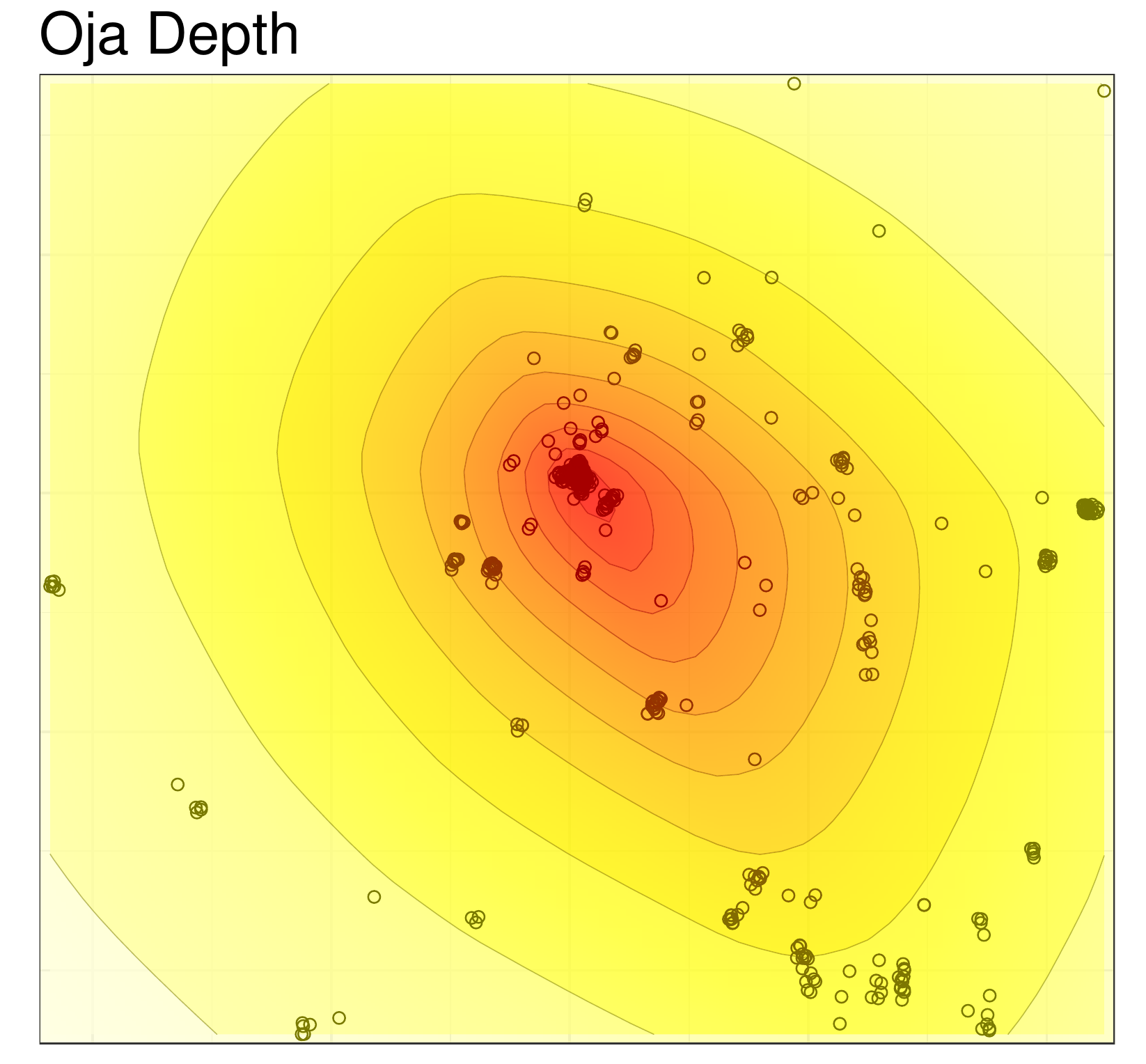}\hfill
    \includegraphics[width=0.2\textwidth]{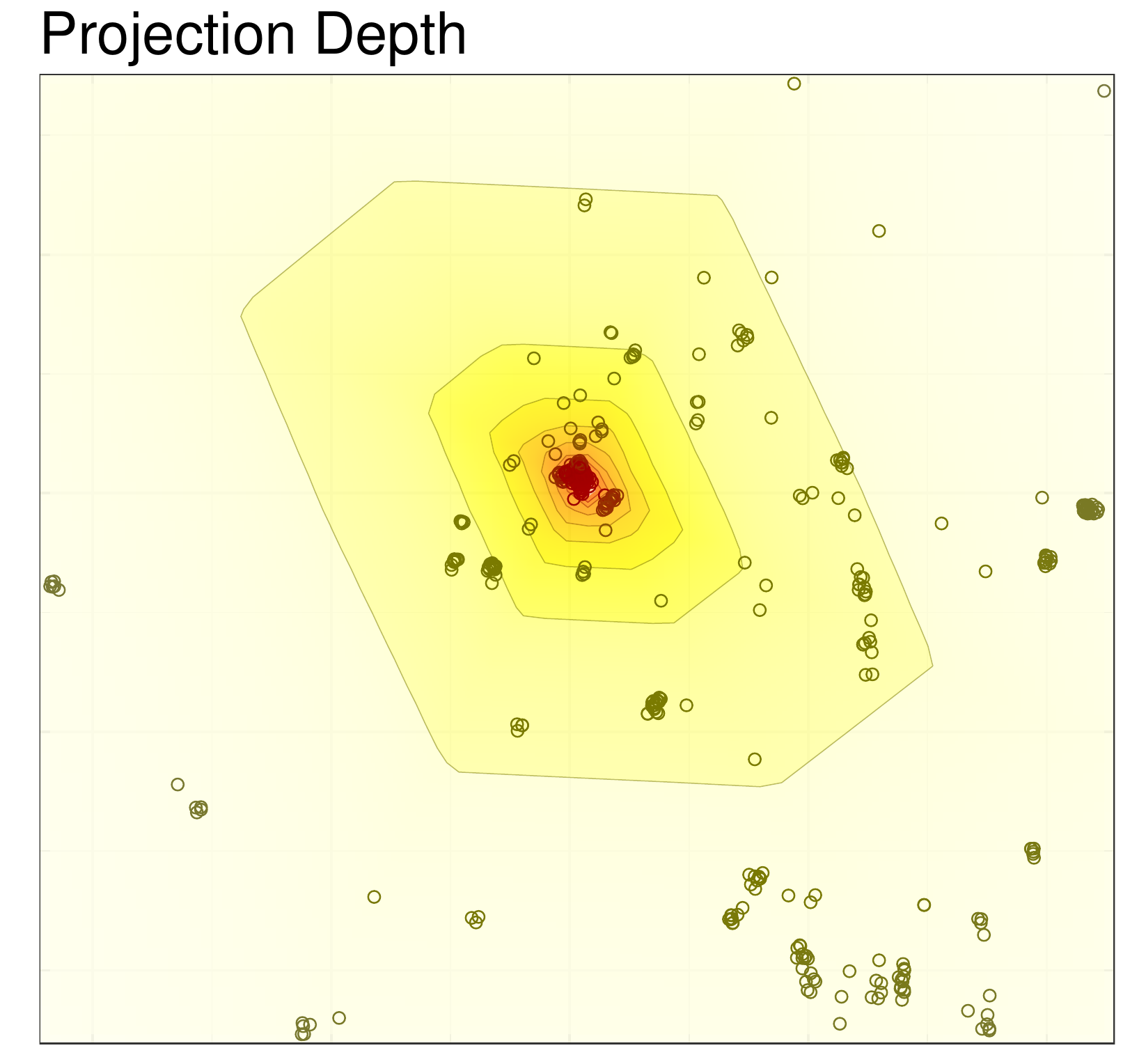}
    \hspace{1cm} 
    \end{minipage}
    
    \begin{minipage}[c]{\textwidth} 
    \hspace{1cm}
    \includegraphics[width=0.2\textwidth]{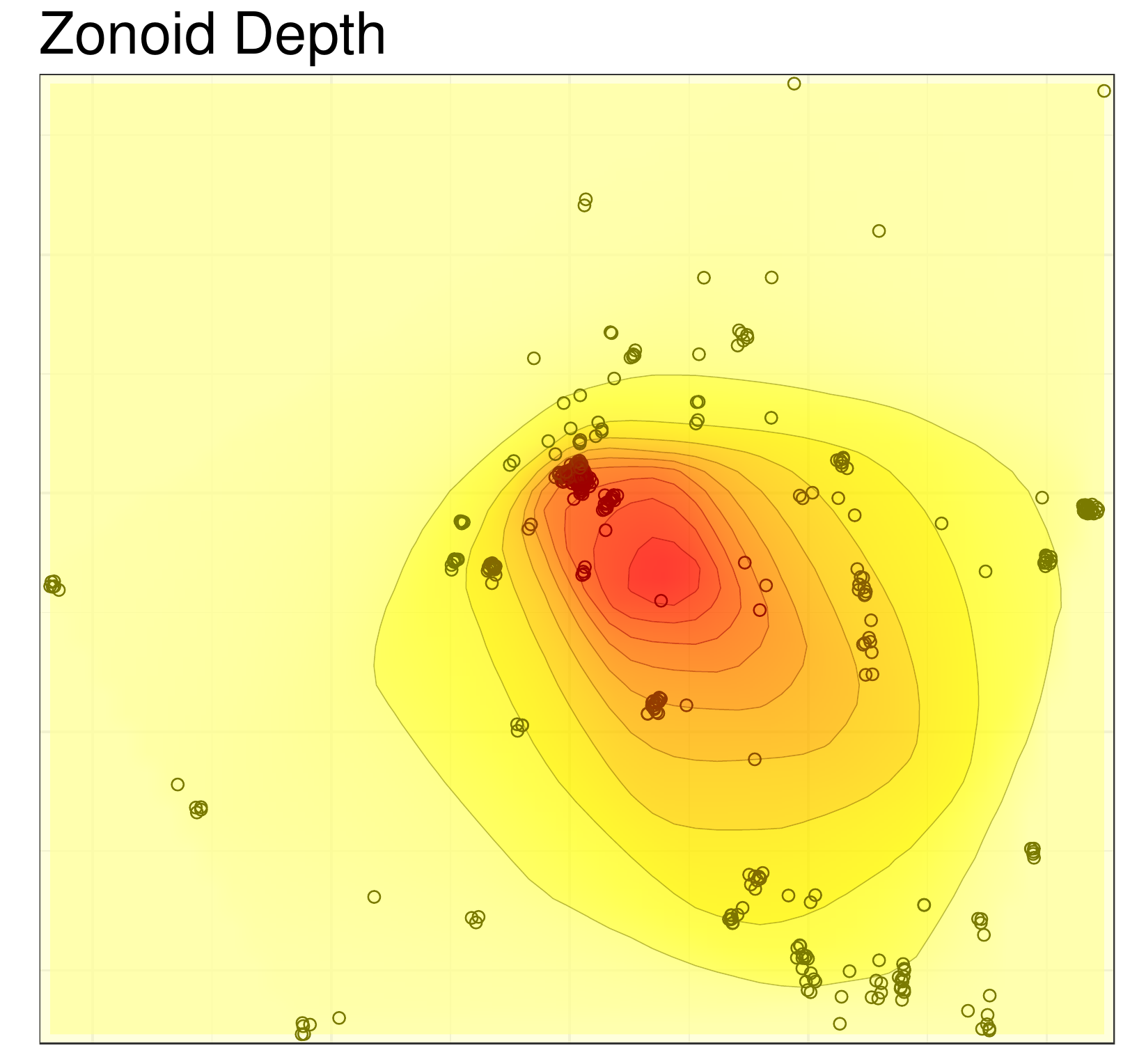}\hfill
    \includegraphics[width=0.2\textwidth]{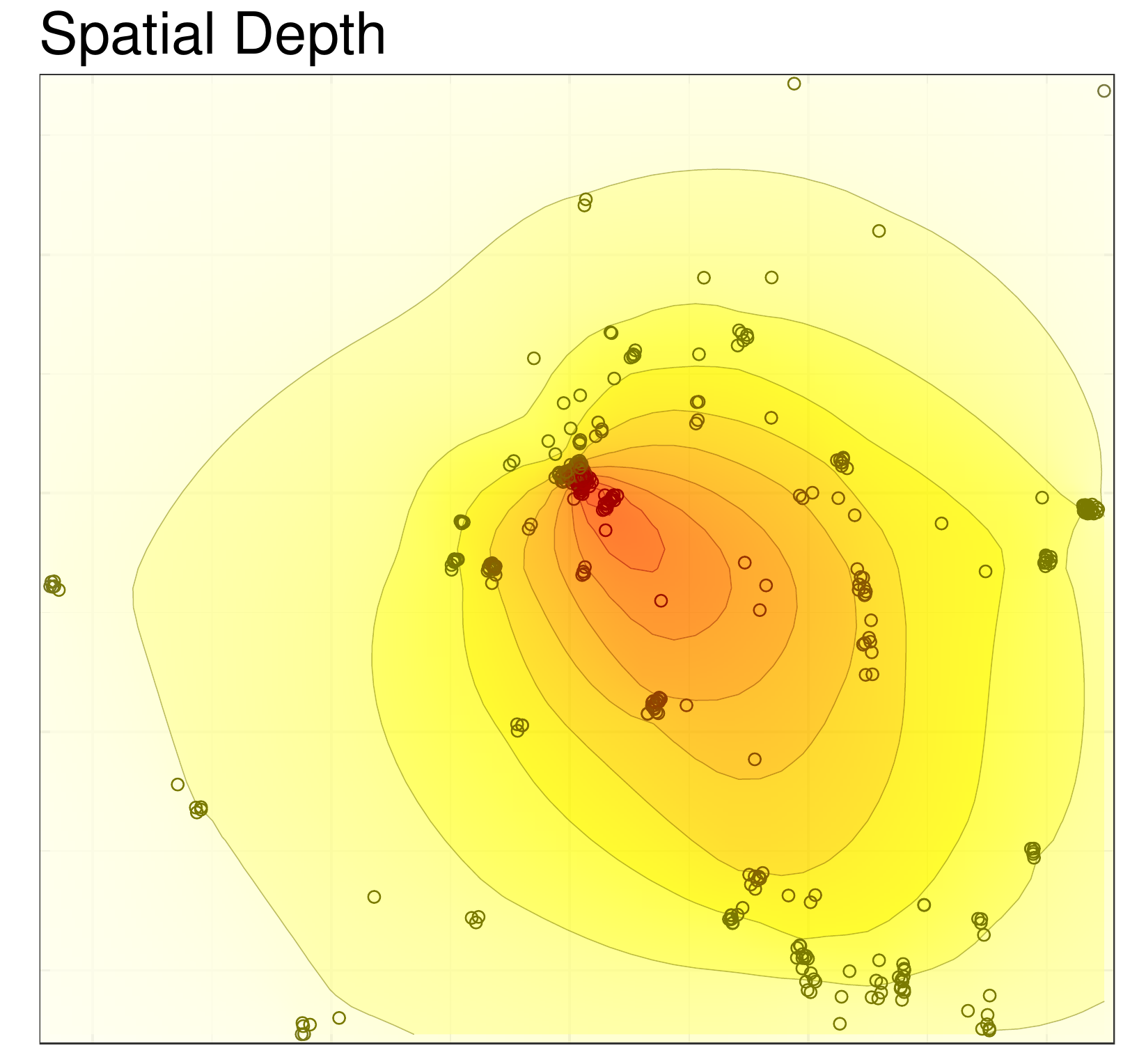}\hfill
    \includegraphics[width=0.2\textwidth]{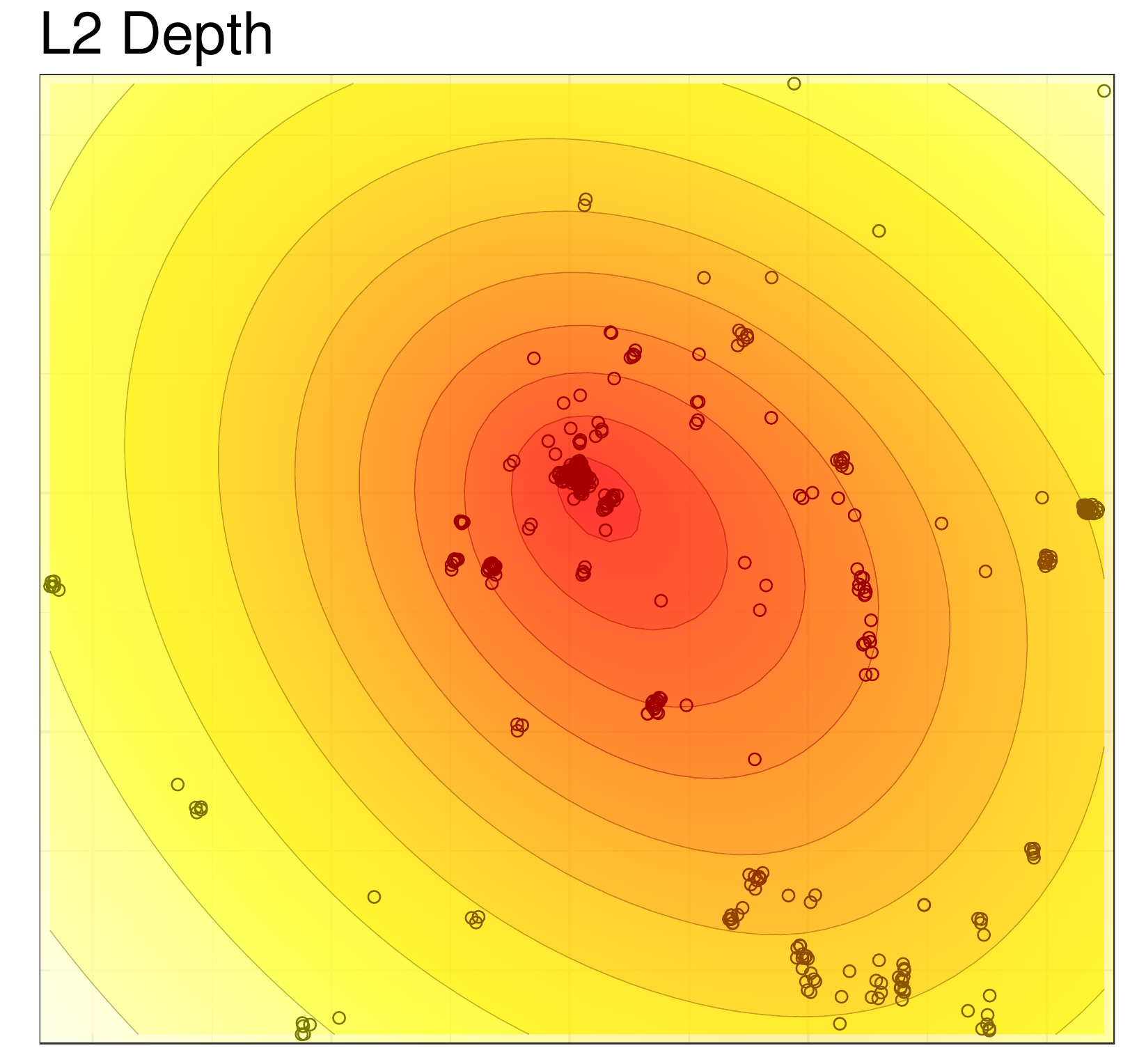}\hfill
    \includegraphics[width=0.2\textwidth]{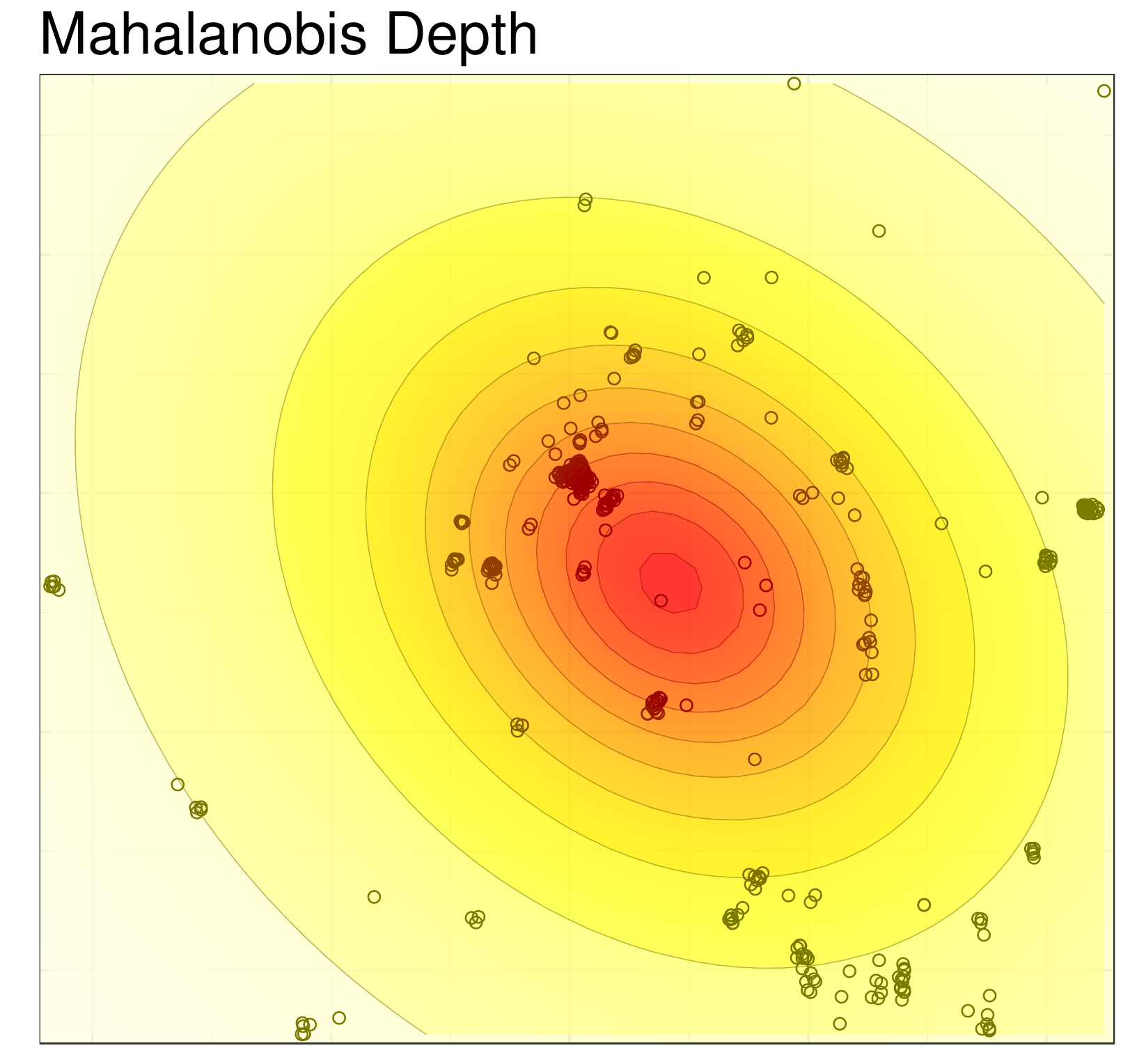} 
    \hspace{1cm}
    \end{minipage}
    
    \caption{Depth maps computed on experimental data collected during an eye-tracking calibration. Colors indicate the depth of each region (the warmer, the deeper).}
    \label{fig:depthRegions}
\end{figure*}

To review the formal mathematical definition of the depth measures listed above, their statistical properties, or their computation, the reader can refer to  \citep{mosler_depth_2012,rousseeuw_computation_2017,small_survey_1990,zuo_general_2000,aloupis_computing_2001,serfling_depth_2006,mosler_depth_2012}. As an illustration of the behaviour exhibited by different depth measures on bivariate multimodal data, we provide in Figure~\ref{fig:depthRegions} examples of several functions applied to one of the calibration datasets from our eye-tracking recordings. 

\subsection{Multivariate Medians} \label{sec:multivariateMedians}

\begin{figure}[htbp]
    \centering
    \includegraphics[width=1\linewidth]{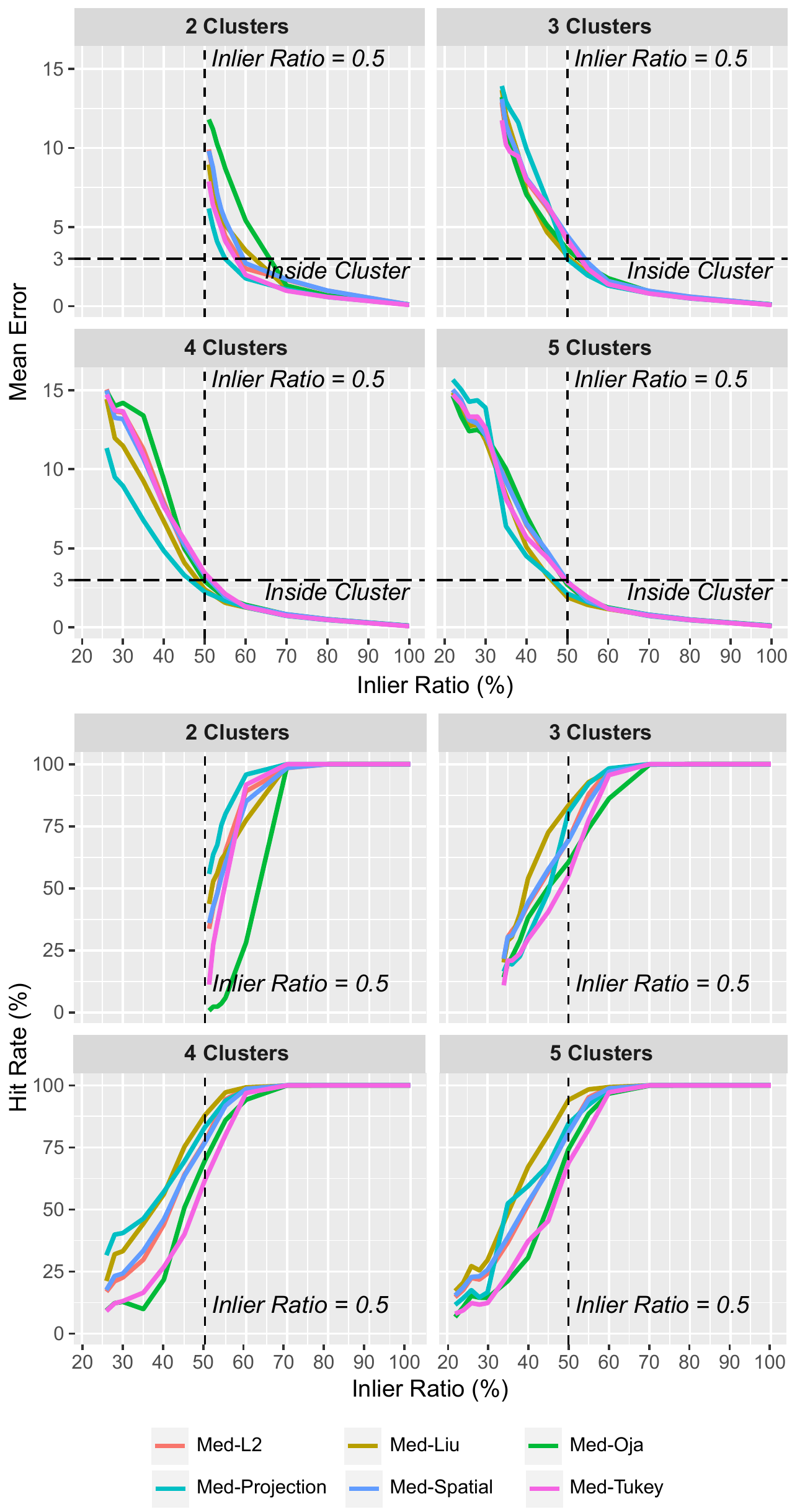}
    \caption{Tolerance of depth medians to clustered outliers. The four top-most figures show the average error to the true center (the variance of the inliers cluster being equal to 1, a distance of 3 corresponds approximately to the cluster outer limit, which includes about 99.7\% of the samples). The four lower plots provide the percentage of successful identification of the main cluster (when the estimate error is inferior to 3). Results are reported for a number of clusters from 2 to 5.}
    \label{fig:curves-BasicDepths}
\end{figure}

\begin{figure}
    \centering
    \includegraphics[width=\linewidth]{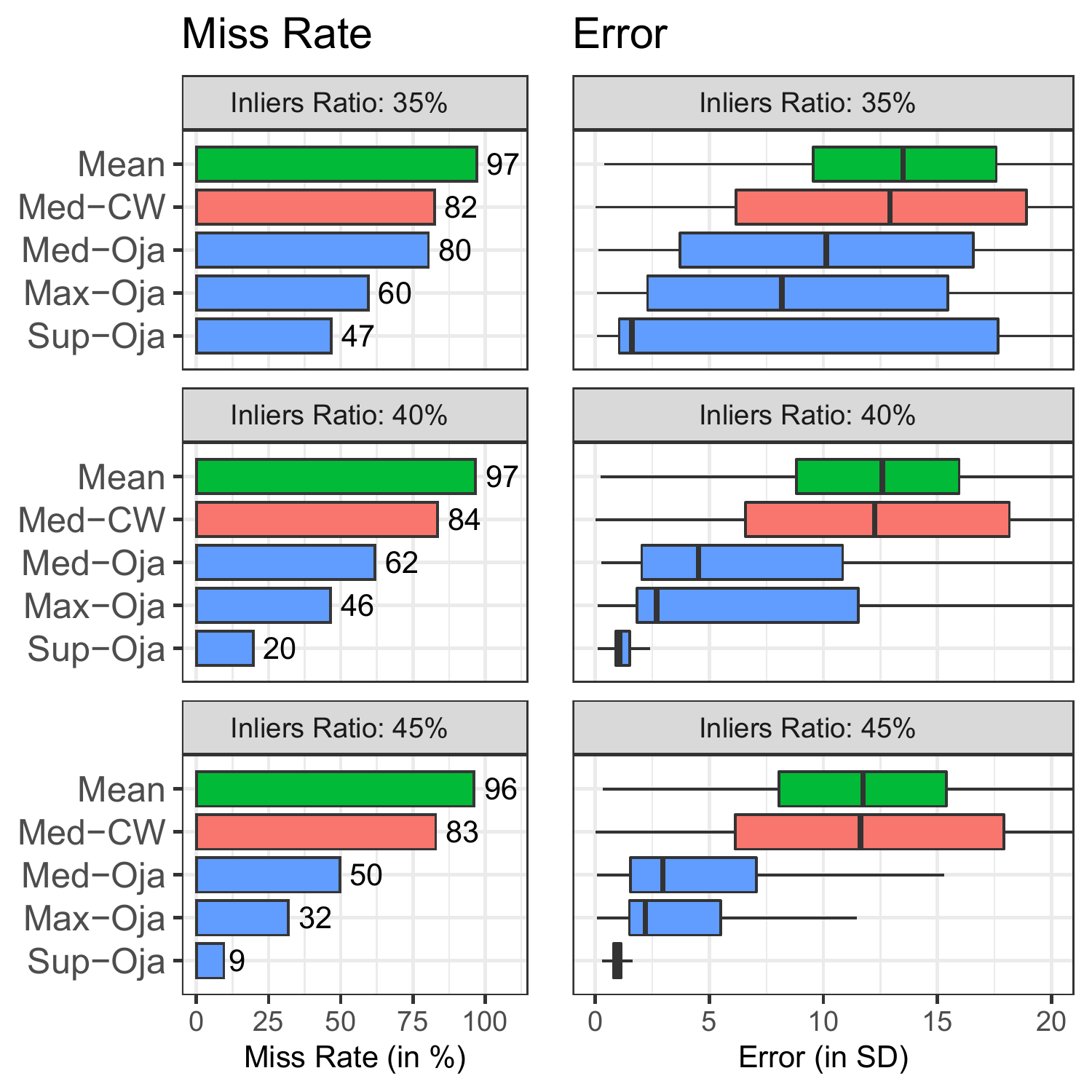}
    \caption{Comparison of the Oja classic multivariate median (\emph{Med-Oja}), sample median (\emph{Max-Oja}) and the two-steps method consisting in computing the sample median from the $10\%$ deepest samples (\emph{Sup-Oja}). The miss rates and errors are provided for different inliers ratios, on distributions composed of 3 clusters, and 0 to 25\% uniform noise. The average of samples (\emph{Mean}) and Coordinate-Wise Median (\emph{Med-CW}) are also reported as baseline.}
    \label{fig:boxplot-MedMaxSup}
\end{figure}

 When outliers are randomly distributed, depth-based MM typically show good tolerance to noise, as illustrated in Table~\ref{table:uniformNoise} (errors ranging from $0.228$ to $0.286\sigma$ for $75\%$ contamination). If outliers are grouped in clusters on the other hand, while multivariate medians derived from depth measures still present better results than the mean or coordinate-wise median, their robustness significantly drops, and they can not sustain the high quantities of outliers encountered in our experimental datasets. As the outliers ratio increase, the medians tend to shift away from the main cluster center, attracted by other high density regions. These errors grow dramatically when the contamination gets past 50\%, as one can observe in Figures~\ref{fig:curves-BasicDepths} and~\ref{fig:boxplot-MedMaxSup}.

\begin{table}

    \caption{Depth Medians average error to the center of a multivariate normal ($\sigma=1$) for an increasing amount of uniform noise. 500 repetitions were performed for each level of noise.}
    
        \centering
            \resizebox{1\linewidth}{!}{
            \begin{tabular}{lrrrrr}
                \toprule
                \multicolumn{1}{r}{\textit{\textbf{Uniform Noise (\%):}}} & \multicolumn{1}{c}{\textit{\textbf{0}}} & \multicolumn{1}{c}{\textit{\textbf{25}}} & \multicolumn{1}{c}{\textit{\textbf{50}}} & \multicolumn{1}{c}{\textit{\textbf{75}}} & \multicolumn{1}{c}{\textit{\textbf{95}}} \\
                \midrule
                Med-L2 & 0,0868 & 0,1097 & 0,1507 & 0,2534 & 0,6345 \\
                Med-Projection & 0,0867 & 0,1122 & 0,1515 & 0,2556 & 0,6417 \\
                Med-Liu & 0,0867 & 0,1100 & 0,1519 & 0,2511 & 0,6341 \\
                Med-Spatial & 0,0866 & 0,1110 & 0,1525 & 0,2546 & 0,6430 \\
                Med-Oja & 0,0850 & 0,1344 & 0,1798 & 0,2859 & 0,7215 \\
                Med-Tukey & 0,0645 & 0,0838 & 0,1284 & 0,2282 & 0,6126 \\
                Mean  & 0,0565 & 0,2370 & 0,3647 & 0,4635 & 0,5886 \\
                \bottomrule
            \end{tabular}
            }

    \label{table:uniformNoise}
\end{table}

The computational complexity of MMs is usually high, involving for instance gradient research in vast regions \citep{aloupis_algorithms_2003}. Furthermore, when facing multimodal distributions, the deepest regions are often located in the empty space in between clusters. For these reasons, an alternative to the standard MM consists in selecting the sample with the highest depth, which not only allows faster computation, but also guarantees falling within one of the clusters in the absence of uniform noise. This estimate is called the sample median \citep{liu_notion_1990}. We suggest an extension of this approach, selecting a small fraction of the samples with the highest depths, rather than the deepest point alone. This subset might still contain outliers (samples from other clusters and/or uniform noise), but usually in much smaller proportions than in the original distribution. From this subset, a new measure of central tendency can therefore be applied. If the samples are averaged, this operator would correspond to a depth-trimmed mean with a very high $\alpha$ parameter (the fraction of extreme values rejected) \citep{masse_monte_2003, donoho_breakdown_1992, rousseeuw_robust_1987}. Instead of the mean, taking the sample median, which require recomputing the depth values with respect to this sub-population and selecting the deepest sample, generally offers more robustness. Note that the concept of trimmed median only applies to the multivariate setting. In a univariate distribution, the median will remain the same regardless of the amount of trimming, as samples are removed from both ends in equal numbers. However, unless the distribution is centrally symmetric, that property does not hold for most multivariate depth functions.

To evaluate this depth-trimmed median, we tested different depth functions with the following two-steps scheme: firstly, selecting the $10\%$ deepest samples, and, then, estimating the depth-based sample median within this group. We show in Figures~\ref{fig:boxplot-MedMaxSup} the results of this method (labelled as \emph{Sup}), compared to the conventional MM (the location with the highest depth, labeled \emph{Med}), the sample median (the deepest sample of the global distribution, labeled \emph{Max}), as well as the arithmetic mean (\emph{Mean}) and Coordinate Wise Median (\emph{CW-Med}) used as baseline. It is worth noticing that with multimodal mixtures, the \emph{Sup} method shows slightly better results than the sample median, which itself offers improvements over the standard MM. Yet, despite these results, none of these methods offer enough robustness for the highest outliers ratios. In the next section we introduce our new approach to overcome the limitations of multivariate depth medians when dealing with highly contaminated multimodal distributions.

\section{Recursive Depth Estimator with Corrected Location Through Filtering}\label{sec:BRIL}

We describe in this section our new approach, named Bootstrap and Refine Iterative Locator (BRIL), which relies on three specific steps to identify the main mode of multivariate distributions: i) \emph{Boostrap\footnote{Note that we use the term \emph{bootstrap} in the traditional sense of an initializing process, a first estimate to be later improved (e.g., a seed). It has no relation with the statistical re-sampling method of the same name.}}: a recursive estimate based on depth measures; ii) \emph{Refine}: a sequential procedure of outlier filtering, through unimodality and normality tests; iii) \emph{Iterate}: an iterative identification of each cluster, in order to select the main mode, i.e., the group with the highest cardinality. Our algorithm is distributed as an R package available at \url{adrienbrilhault.github.io/BRIL/}. The following sections detail every step, with a scrutiny on data to demonstrate the advantage of this method over the classic depth medians, the coordinate-wise median, and other techniques such as clustering.

\subsection{BOOTSRAP: Recursive estimate of location based on depth median} \label{sec:bootsrap}

As shown in section~\ref{sec:multivariateMedians}, classical measures of central tendency based on depth can withstand a certain quantity of contamination (clustered outliers and/or uniform noise), unlike other locators such as the arithmetic mean, which are strongly and regularly affected by outliers, even in small proportions.

However, when the percentage of outliers increases above 50\%, their performances drastically drop, and they fail at identifying the main cluster in most distributions (hit rates can get as low as 15\%, as shown in Figure~\ref{fig:curves-BasicDepths}). We should point out that while most researches in robust statistics consider 50\% as an upper bound for the quantity of outliers, from the assumption that the true distribution should represent the majority of the data, this does not always hold for multimodal distributions. For example, in mixtures of Gaussians, the main sub-population (i.e. the inliers) can be defined as the group with the highest cardinality. In these conditions, the ratio of inliers can therefore reach $\frac{K-1}{K}-1$, where $K$ is the number of components of the compound distribution. When considering mixtures which also includes uniform noise, this ratio might even drop further. 

\begin{figure}[!htbp]
    \centering
    \includegraphics[width=1\linewidth]{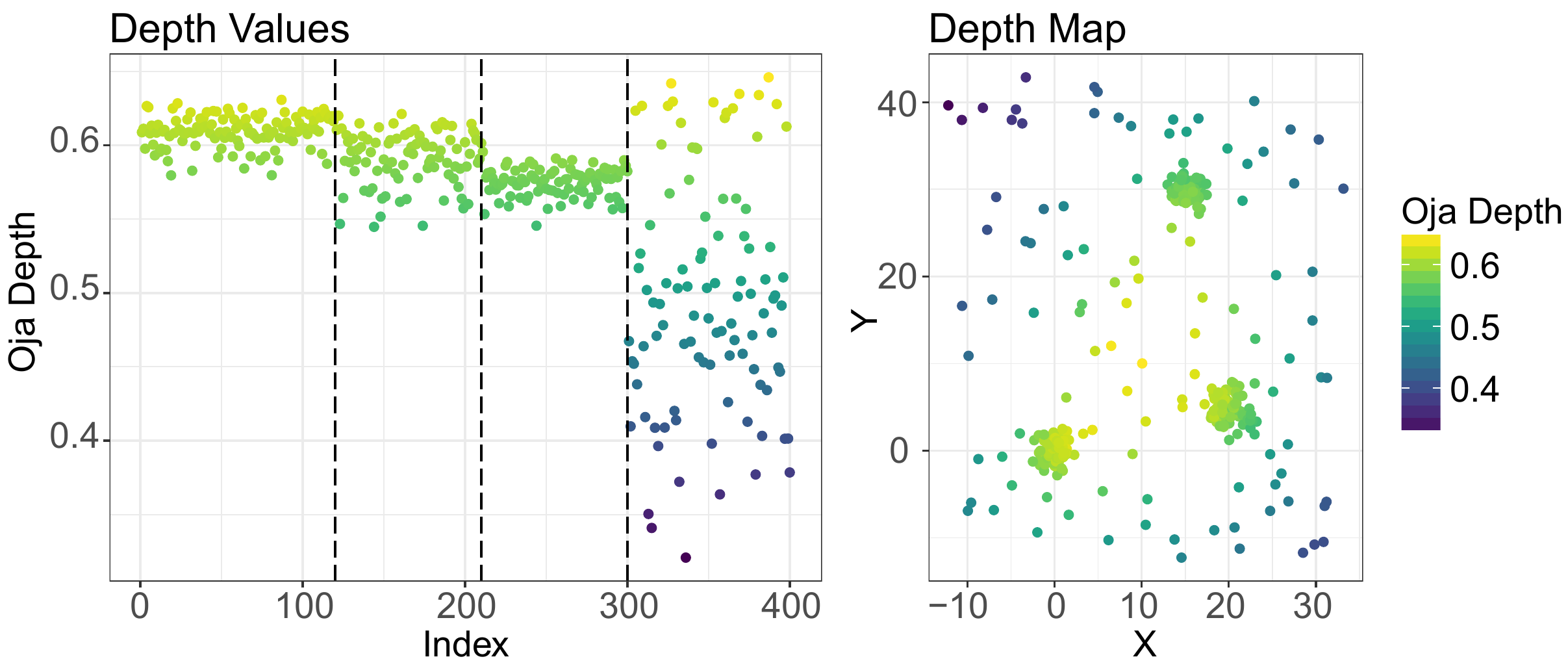}
    \includegraphics[width=1\linewidth]{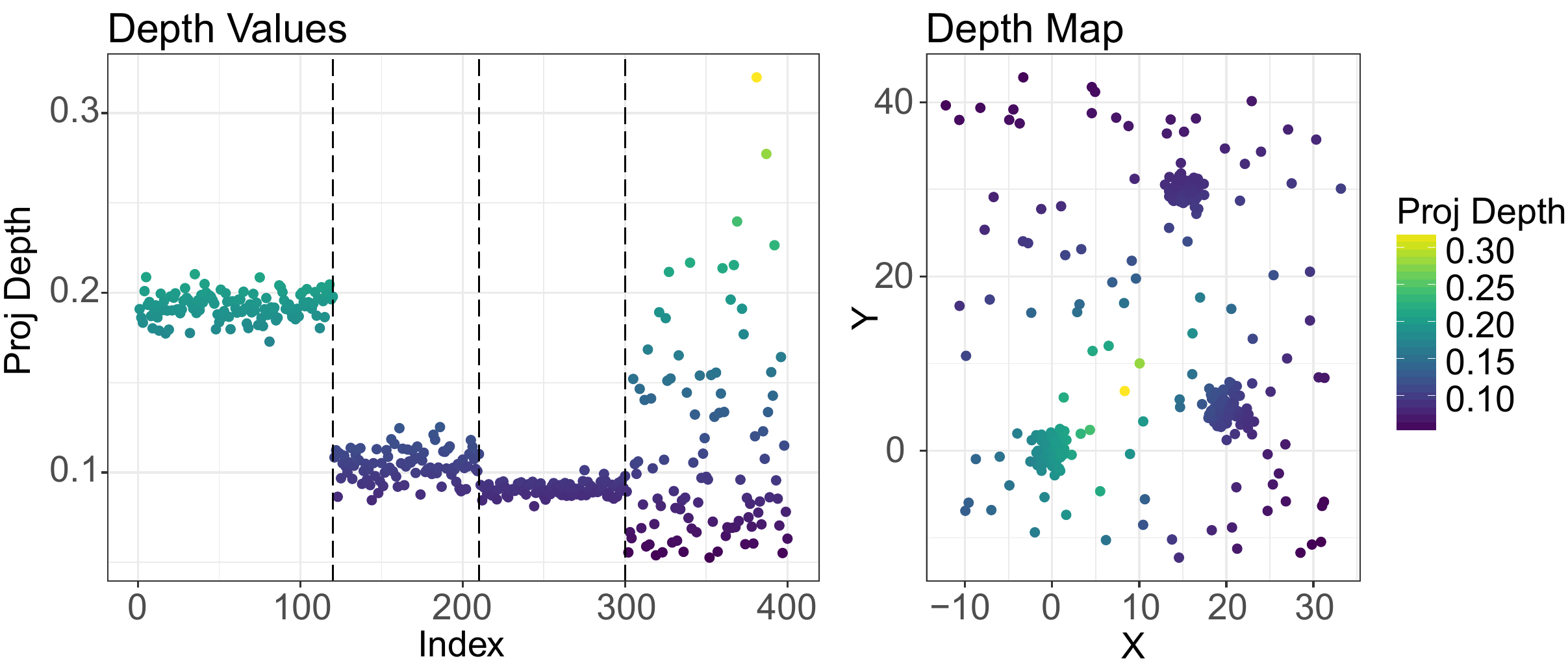}
    \caption{Depth values from the Oja (top-most plots) and Projection depths (bottom-most plots). The left-most figures show the depth values for each of the samples sorted by index (the first 120 points correspond to the main cluster, followed by 2 secondary clusters of 90 samples each, and finally 100 points of random noise). The right-most figures show these same samples in Cartesian coordinates, with their depth values represented on a color scale.}
    \label{fig:depthMaps}
\end{figure}

Examining depth maps from distributions with less than 50\% inliers, we see that the deepest sample generally does not belong to the main cluster, but to the space in between clusters. Nonetheless, there is often a tendency of having higher depth values in this main cluster than in secondary ones. Illustrative examples are provided in Figure~\ref{fig:depthMaps} for Oja and Projection depths. In both cases, the sample with the highest depth is located in between clusters, and, while the depth values of the uniform samples are spread, those from each of the clusters tend to be relatively compact, in a narrow range, with the main cluster showing values in average higher than the other groups.

From these observations, we see that the selection of a fraction of the samples with the highest depth can result in a subset with a relatively low proportion of outliers, from which the central tendency is easier to estimate, as suggested in Section~\ref{sec:multivariateMedians} with the two-steps method named \emph{Sup}. The choice of the optimal size for this subset is, however, sensitive, as it strongly depends on the respective sizes of clusters, their spatial configuration, and the separability of their depth values. By selecting too few samples, there is a risk of leaving out all the inliers, keeping a subset which does not intersect with the main cluster. On the other hand, if the group is too large, the proportion of inliers can be too low to compute a reliable estimate of central tendency. Selecting an optimal value for this parameter would require prior knowledge on the number of clusters, their size, and the quantity of uniform noise.

\begin{figure}[!htbp]
    \centering
    \includegraphics[width=.95\linewidth]{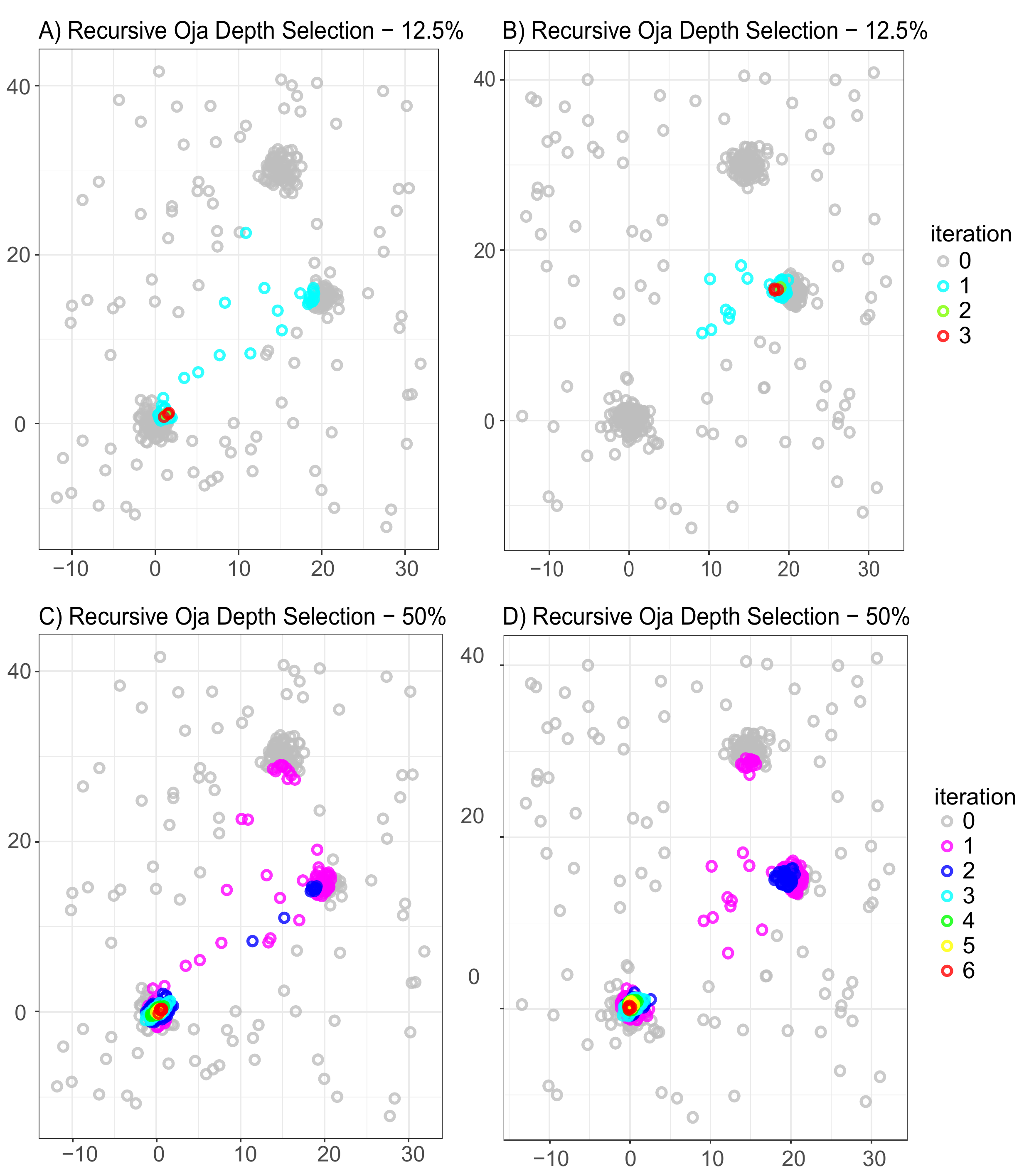}
 
    \caption{Recursive estimates of Oja depth median for two different distributions (the first in Figures A and C, the second in B and D). Both mixtures were drawn from the same clusters coordinates and parameters (30\% inliers, centered at (0,0), 22.5\% of clustered outliers in each of the two other groups, and 25\% uniform noise). The samples selected at each iteration are depicted in different colors, keeping either the 12.5\% (Figures A and B) or the 50\% deepest (Figures C and D). Keeping a small proportion of the samples at each iteration often fails to identify the largest cluster, since the first iteration might not contain any inlier (as in Fig. B).}
    \label{fig:recMedian}
\end{figure}

Instead of this two-step approach, we therefore suggest a more flexible methodology consisting in recursively computing the depth values, discarding a fixed fraction of the samples (those with the lowest depths), and re-iterating this process until reaching a sufficiently small number of samples. Keeping a subset too small at each iteration presents the risk of discarding too many inliers, as previously mentioned. Keeping too many samples leads to an increase in execution times, as more recursions are required. After preliminary simulation studies, we suggest using a 50\% trimming as a trade-off between robustness and computational costs. Figure~\ref{fig:recMedian} exemplifies the execution of this recursive approach using Oja depths, keeping either $12.5\%$ or $50\%$ of the samples at each iteration. Note that when discarding a fraction of the samples with the lowest depths, ties are resolved using the same simplification as in \citep{liu_multivariate_1999,masse_monte_2003}, selecting the first samples encountered (with the smaller indices). Other options would be a random selection, or discarding all the samples sharing the same rank.

\begin{figure}[!htbp]
    \centering
    \includegraphics[width=1\linewidth]{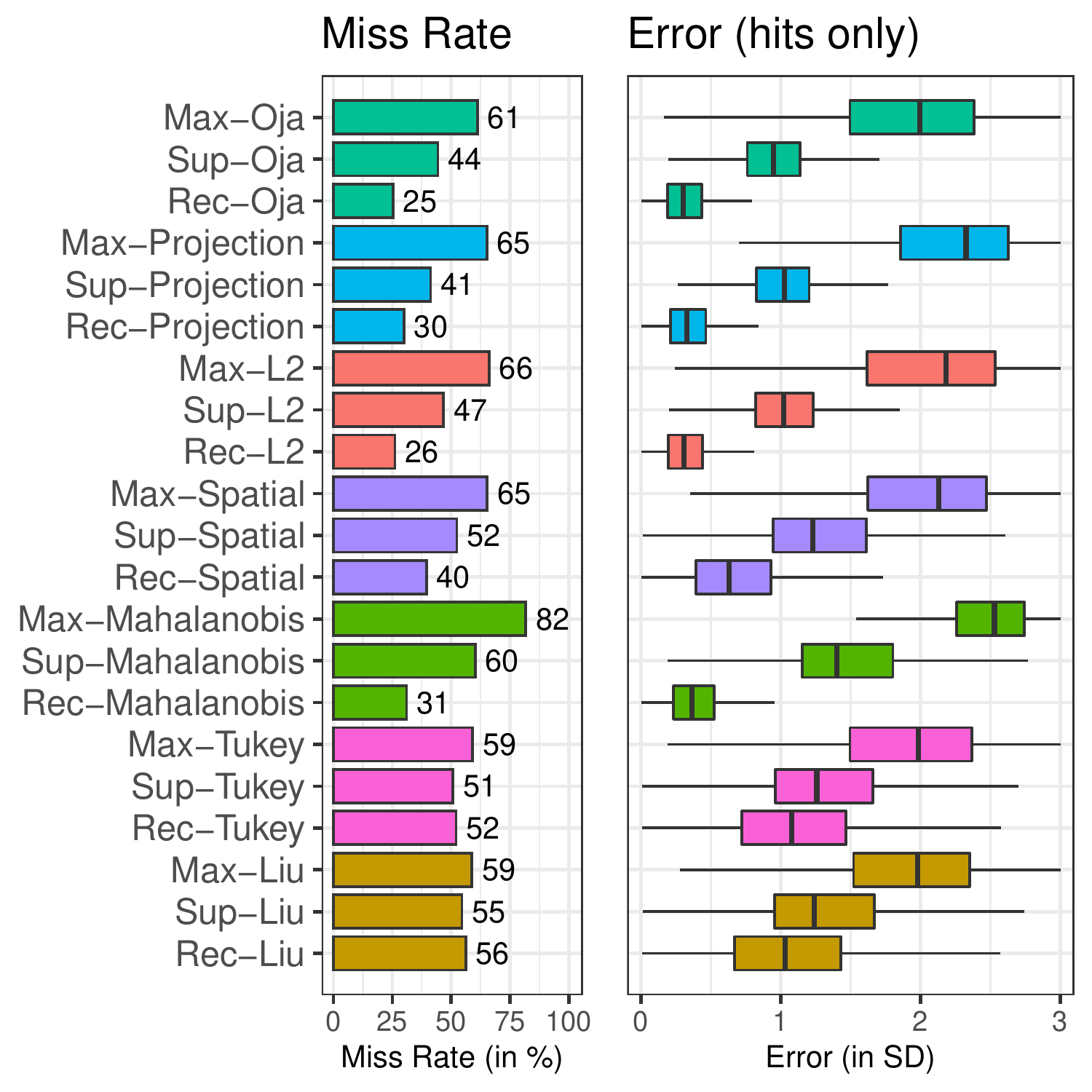}
    \caption{Left figure: proportion of estimates located outside the main cluster (with an error to its center above 3 $SD$), across all the simulations with a ratio of inliers inferior to 50\%, a percentage of uniform noise from 0 to 25\%, and a number of clusters from 3 to 5. Right figure: Errors to the true center, considering solely the hits (estimates falling within the main cluster).}
    \label{fig:recursiveDepthBoxplot}
\end{figure}

The advantages of this recursive framework are shown in Figure~\ref{fig:recursiveDepthBoxplot}, for distributions with less than 50\% inliers. With most depth functions, the sample medians (sample with the highest depth, labelled as \emph{Max}) are located within the main cluster in around 35-40\% of our simulations. The \emph{Sup} method (the two-passes scheme where the top $10\%$ deepest samples are selected and used to re-compute the depth-based sample median) offers more robustness in terms of hit rates and average error. Yet, the recursive version (prefixed \emph{Rec}) significantly outperforms the other approaches with most of the depth measures considered. Among the depths showing the best results, such as Oja, L2, Mahalanobis, and Projection depth, the recursive estimate even reduces by more than twice the miss rate. This recursive locator not only identifies the main cluster more frequently, but also improves the precision of the location among the correct identifications, as shown in the right part of Figure~\ref{fig:recursiveDepthBoxplot}, due to the recursive removal of low depth values which avoids outliers dragging the median away from the cluster center.

\begin{figure}[!hbtp]
    \centering
    \includegraphics[width=1\linewidth]{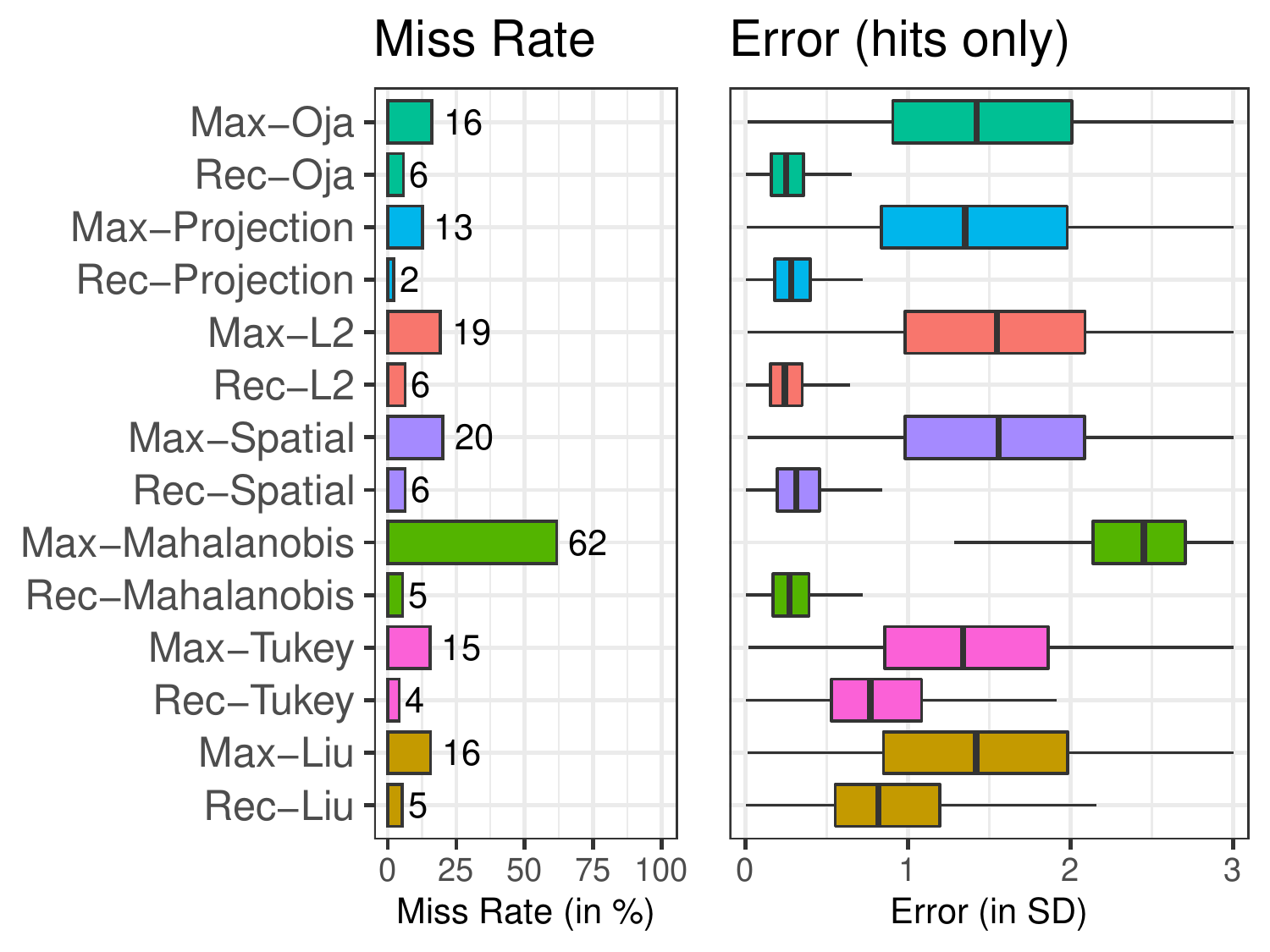}
    \caption{Left figure: proportion of estimates located outside the main cluster, across all the simulations with a ratio of inliers between 50\% and 75\%, a percentage of uniform noise from 0 to 25\%, and a number of clusters from 3 to 5. Right figure: errors to the true center, considering solely the hits.}
\label{fig:recursiveDepthBoxplotAbove50}
\end{figure}

The results presented in Figure~\ref{fig:recursiveDepthBoxplot} focused on highly contaminated data (above $50\%$ outliers), situations in which most traditional methods show poor performances. As previously discussed, depth medians generally offer robust results when the overall quantity of inliers remains above half of the global distribution (see for instance Figure~\ref{fig:curves-BasicDepths}). Analyzing the results in these more favorable conditions, with distributions containing 50 to 75\% inliers, we show in Figure~\ref{fig:recursiveDepthBoxplotAbove50} that most standard depth sample medians do find the main cluster in 80-85\% of our simulations, and present a relatively accurate center estimate. Nonetheless, even with small quantities of outliers, our recursive method still provide better performances. Recursive depth locators not only achieve hit rates close to $100\%$, but also show smaller errors and variability when the main cluster was encountered. As an example, the mean error obtained by the Oja sample median over all simulations with 50 to 75\% inlier was $2.96$, compared to $1.52$ for its recursive version, and the median error dropped from $1.651$ to $0.26$. While at different degrees, all depth measures tested showed a significant improvement for the recursive approach.

In conclusion, considering the results discussed in this section, we demonstrated that the recursive trimming of low depth samples provides great advantages over the conventional MM and sample median, especially when the dataset is strongly contaminated. The recursive Oja and Projection estimates, in particular, have seen their hit rate raise from approximately $35-40\%$ to $70-75\%$, and their mean distance error to the true center halved in case of hits.

\subsection{REFINE: Improving the location by outliers filtering}\label{sec:refine}

Despite the improvements presented in the previous section, the recursive approach still fails to detect the main cluster in about $30\%$ of the simulations with high contamination. When the first group is found, the average distance of the estimate to the true center remains around half a standard deviation, which could also be reduced. These limitations result from the fundamental nature of depth metrics, which were not intended for a multimodal scenario \citep{serfling_depth_2006}. If there are several clusters with large cardinalities, with one is positioned in a central position, it will also generally show higher depths than an outer cluster with a larger count. This problem would arise in a univariate distribution as well: while the median and the mode often present relatively similar results in an unimodal distribution, the median is not suited to locate the dominant mode in the multimodal case.

The other steps of the BRIL algorithm, detailed in these two next sections, offer solutions to overcome these issues. While the recursive depth locator serves as \emph{Bootstrap} of the procedure, providing a first estimate of central location, the \emph{Refine} step objective is to identify the samples belonging to the same cluster as this estimate. This stage serves a dual purpose: firstly, to refine the location of the estimate, by considering solely the samples belonging to the same sub-population, with a minimal number of outliers; and secondly, this partitioning can be used in the next step of the algorithm, the \emph{Iterate} procedure, which recursively removes from the global distribution the samples identified as part of each cluster, in order to select the group with the highest cardinality at the end of the procedure. This will, therefore, avoid selecting a smaller cluster that might have been picked by the first execution of the recursive depth locator due to its higher depth.

In order to refine the location computed in the \emph{Bootstrap} step, we will employ a common outlier filtering methodology, consisting in recursively removing the more distant samples \citep{dempster_new_1981, rousseeuw_robust_1987, caroni_sequential_1992}. This type of sequential procedure is generally recommended when the number of outliers is unknown \citep{thode_testing_2002}. It relies on an outlyingness measure, used to select the most discrepant sample(s) to be removed at each iteration, and on a stopping criterion, frequently under a Gaussian assumption. Our outliers removal strategy consists of two steps. The first based on Euclidean distances and a unimodality test. The second on robust distances and a normality test.

\begin{figure}[!htbp]
    \centering
    \includegraphics[width=1\linewidth]{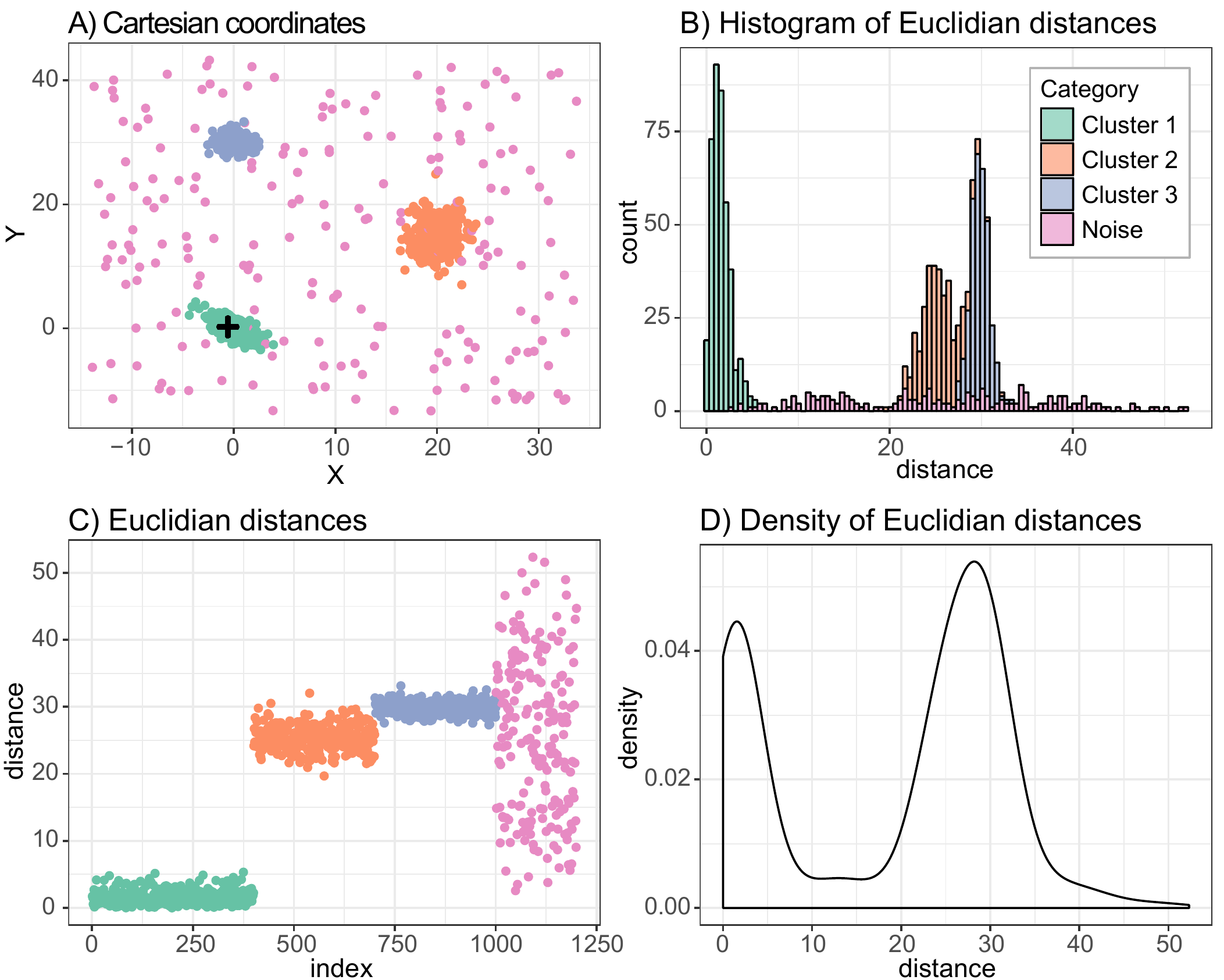}
    \caption{Recursive filtering of outliers. Figure A shows an illustrative multimodal distribution. Euclidean distances from each sample to the estimate computed with the recursive depth method (indicated by the black cross) are presented in the form of histogram in Figure B, density in Figure D, and by index in Figure C (the first 400 samples corresponding to the main cluster, followed by 200 in the second and third clusters, and finally 200 points of uniform noise). The histogram and density function of distances exhibit clear multimodal properties, but will converge to a unimodal setting once most of the outliers have been filtered by the progressive removal of the furthest sample(s).}
    \label{fig:filteringDistances}
\end{figure}

In the first filtering stage, since we do not have prior knowledge on the quantity and nature of the outliers, which can be uniformly distributed, organized in clusters, or both, Euclidean distances to the first estimate were used to select the furthest sample to be discarded at each iteration. If the mixture contains clusters of outliers, the histogram of distances will typically exhibits multimodal properties, with the first peak corresponding to the cluster selected, as illustrated in Figure~\ref{fig:filteringDistances}. In these conditions, progressively removing the most distant samples will eventually lead to a unimodal subset of distances, where the samples from the cluster represent the great majority of the data remaining.

\begin{figure}[!htbp]
    \centering
    \includegraphics[width=1\linewidth]{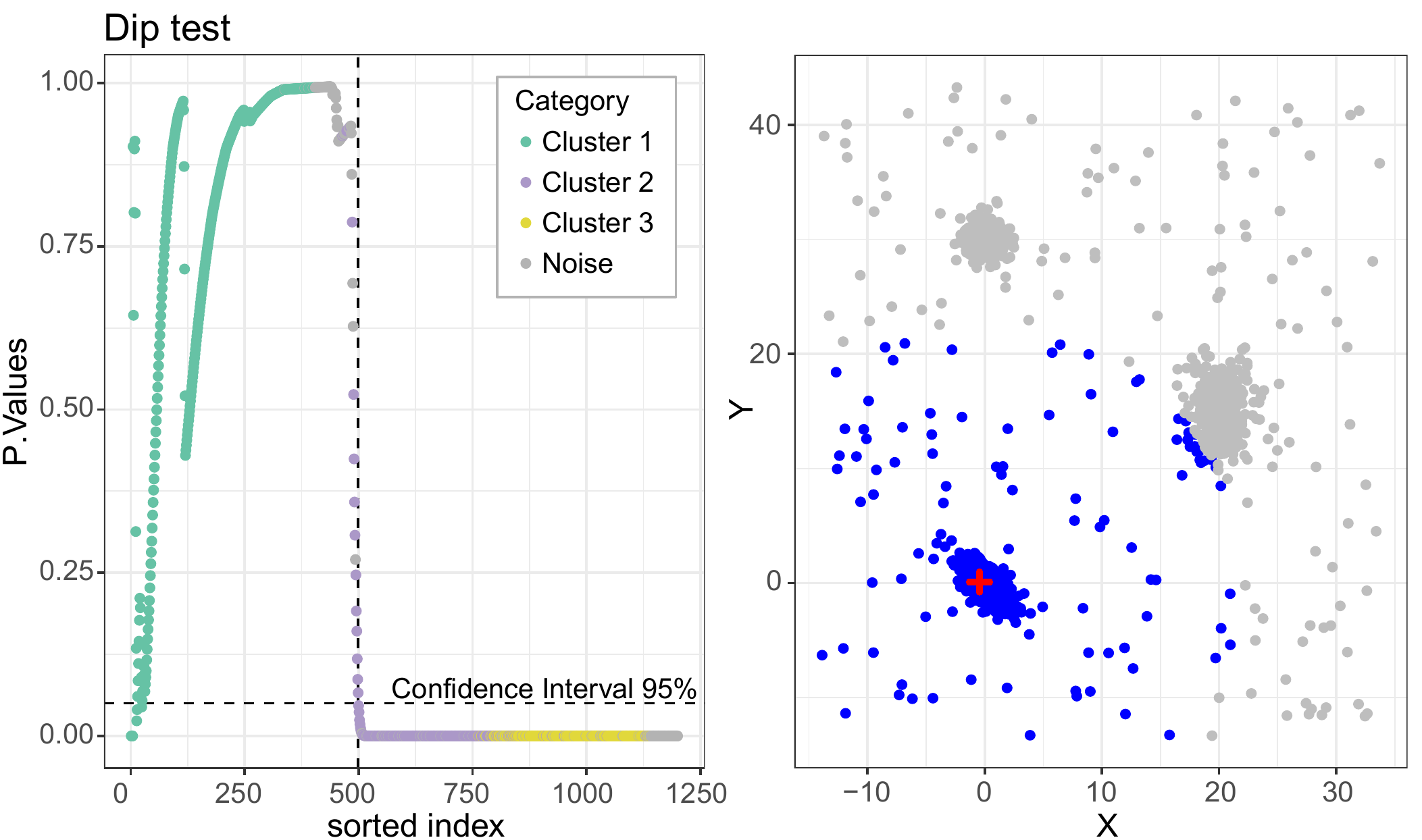}
    \caption{Iterative outliers removal based on Euclidean distances and the Dip test of unimodality. Samples are ordered by their Euclidean distance to the first estimate of location obtained through recursive depth trimming (indicated by the red cross on the right figure). At each iteration, the Dip test is applied to these distances and the furthest sample is removed, until the test fails to rejects the unimodal hypothesis. The $p$-values at each iteration are provided on the left figure, while the right plot shows the samples in Cartesian coordinates, with the unimodal subset selected at the end of the procedure displayed in blue.}
    \label{fig:dipTest}
\end{figure}

Tests for unimodality and multimodality have been extensively studied for univariate distributions \citep{cheng_mode_1999,fischer_testing_1994,hartigan_testing_2000}. Most common techniques include: i) the Silverman bandwidth test \citep{silverman_using_1981}, which uses a normal kernel density estimate with increasing bandwidth; ii) the excess mass test proposed in \citep{muller_excess_1991}; and iii) the Hartigan’s Dip Test \citep{hartigan_dip_1985}, which has been widely adopted due to its low computational complexity, its high statistical power, and the absence of tuning parameter. The extension of these tests to multivariate distributions is not straightforward, and several definitions of unimodality have been suggested for the multidimensional setting \citep{paez_modeling_2018,kouvaras_random_2007}. The Hartigan’s SPAN and RUNT statistic \citep{hartigan_runt_1992}, and the MAP test \citep{rozal_map_1994} are often cited as multivariate alternatives, but these types of procedures are usually far more complex than univariate ones, both conceptually and computationally, relying for instance on the construction of several spamming trees \citep{siffer_are_2018}. They also require a delicate adjustments of tuning parameters, and, for these reasons, most studies resort instead to the projection of multivariate data into one dimension to test for unimodality \citep{johnson_applied_2007}. To transform multivariate samples into univariate values, rather than linear or principal curve projections as in~\citep{ahmed_investigating_2012}, we suggest considering the Euclidean distances of the samples with respect to the current central estimate, in a way that relates to the method from \citet{siffer_are_2018}, who used distances to fold the multivariate distributions around a pivot. Our filtering method consider the previous recursive depth estimate as pivot to compute the distribution of Euclidean distances, to which the DIP test is applied at each iteration, removing the most distant sample and repeating the procedure while the test rejects the hypothesis of unimodality with a sufficient confidence level (we set the significance level to 5\%, i.e., stopping when the test $p$-value exceeds $0.05$). This filtering process is illustrated in Figure~\ref{fig:dipTest}.

Once the unimodal subset is extracted, the second filtering phase relies on a normality test. An outlyingness measure is again required to determine which sample to discard at each iteration, but the Euclidean distances used in the first filtering step are no longer advisable, since the cluster is not necessarily spherical. In such scenario, outliers can present smaller Euclidean distances than samples from the cluster. A common alternative when dealing with ellipsoidal distributions is the Mahalanobis distance. However, this measure is known to be sensitive to the presence of outliers, which strongly affect the data covariance matrix \citep{devlin_robust_1981}. This problem is generally referred to as the masking effect, and addressed by the use of robust estimators of location and scatter. Robust distances are traditionally defined by the Equation~\ref{eq:mahalanobis}  \citep{hadi_identifying_1992,huber_robust_2011,rocke_identification_1996,rousseeuw_clustering_1987}, such that $T(X)$ is a central location estimator and $C(X)$ a matrix of dispersion/scatter. Note that if we choose $T(X)$ as the arithmetic mean and $C(X)$ as the usual sample covariance matrix, this equation becomes the Mahalanobis distance.

\begin{equation}\label{eq:mahalanobis}
    RDi = \sqrt{(x_i - T(X))^t C(X)^{-1} (x_i - T(X))}
\end{equation}

In place of the classic covariance matrix, the scatter parameter provides an estimate less sensitive to outliers. Suitable robust estimators of location and scatter include the Minimum Covariance Determinant (MCD) \citep{rousseeuw_fast_1999}, the Minimum Volume Ellipsoid (MVE) \citep{rousseeuw_multivariate_1985}, the Orthogonalized Gnanadesikan-Kettenring (OGK)  \citep{gnanadesikan_robust_1972,maronna_robust_2002}, M-estimators \citep{maronna_robust_1976,tyler_distribution-free_1987}, S-estimators \citep{lopuhaa_relation_1989,rocke_robustness_1996}, or MM-estimators \citep{tatsuoka_uniqueness_2000}. A survey is available in \citep{rousseeuw_high-breakdown_2013}.

In regard to the location estimator, $T(X)$, one can either use the position computed by the aforementioned estimators, such as MCD or MVE, which not only provide a scatter estimate but as well a center location, or apply the recursive depth estimate presented in Section~\ref{sec:bootsrap}, which provides a robust estimate of central location even in contaminated distributions. To improve the reference location used by robust distance, we recommend to using this depth-based method on the unimodal subset selected in the first filtering step, rather than using the previous estimate which was computed from the global distribution.

We compared three types of robust distances in preliminary studies, based on MCD, MVE, and OGK, respectively. These 3 operators require a single parameter. For MCD and MVE it corresponds to the size of the subset used to compute the robust location and scale. This number was set to $(n+p+1)/2$, where $n$ correspond to the number of samples, and $p$ the dimension of the data, which is the most commonly adopted value in the literature, as it offers good properties in term of breakdown point \citep{lopuhaa_breakdown_1991,hubert_minimum_2017}. OGK, on the other hand, requires the choice of a robust univariate estimate of scale, as, for instance, the median absolute deviation (MAD), the $\tau$ scale \citep{yohai_high_1988}, or $Sn$ and $Qn$ operators \citep{rousseeuw_alternatives_1993}. \citet{maronna_robust_2002} offer a comprehensive review and comparison of these alternatives. Our implementation of OGK follows their recommendations, using the $\tau$ estimate.

\begin{figure}[!htbp]
    \centering
    \includegraphics[width=.49\linewidth]{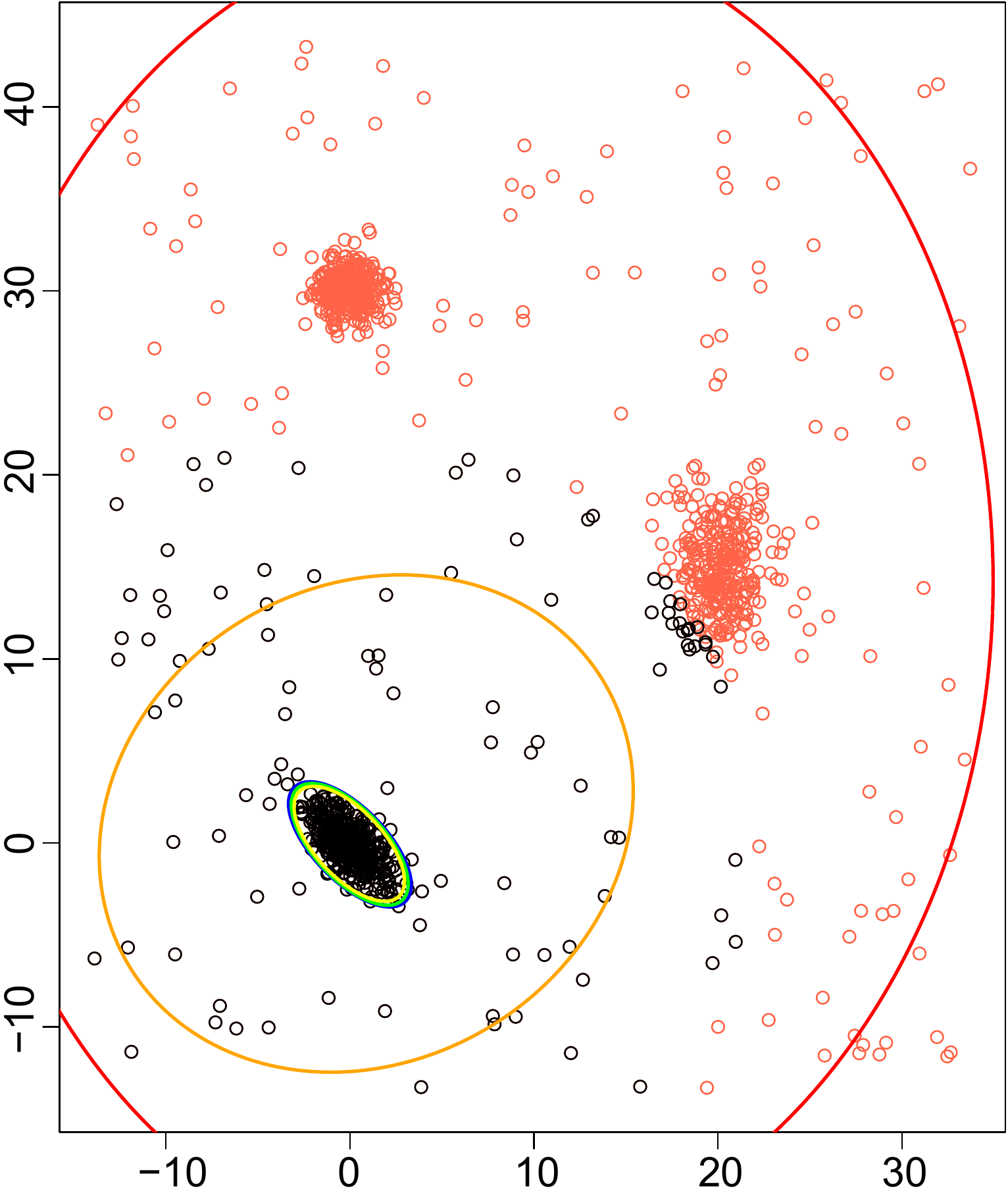}
    \includegraphics[width=.49\linewidth]{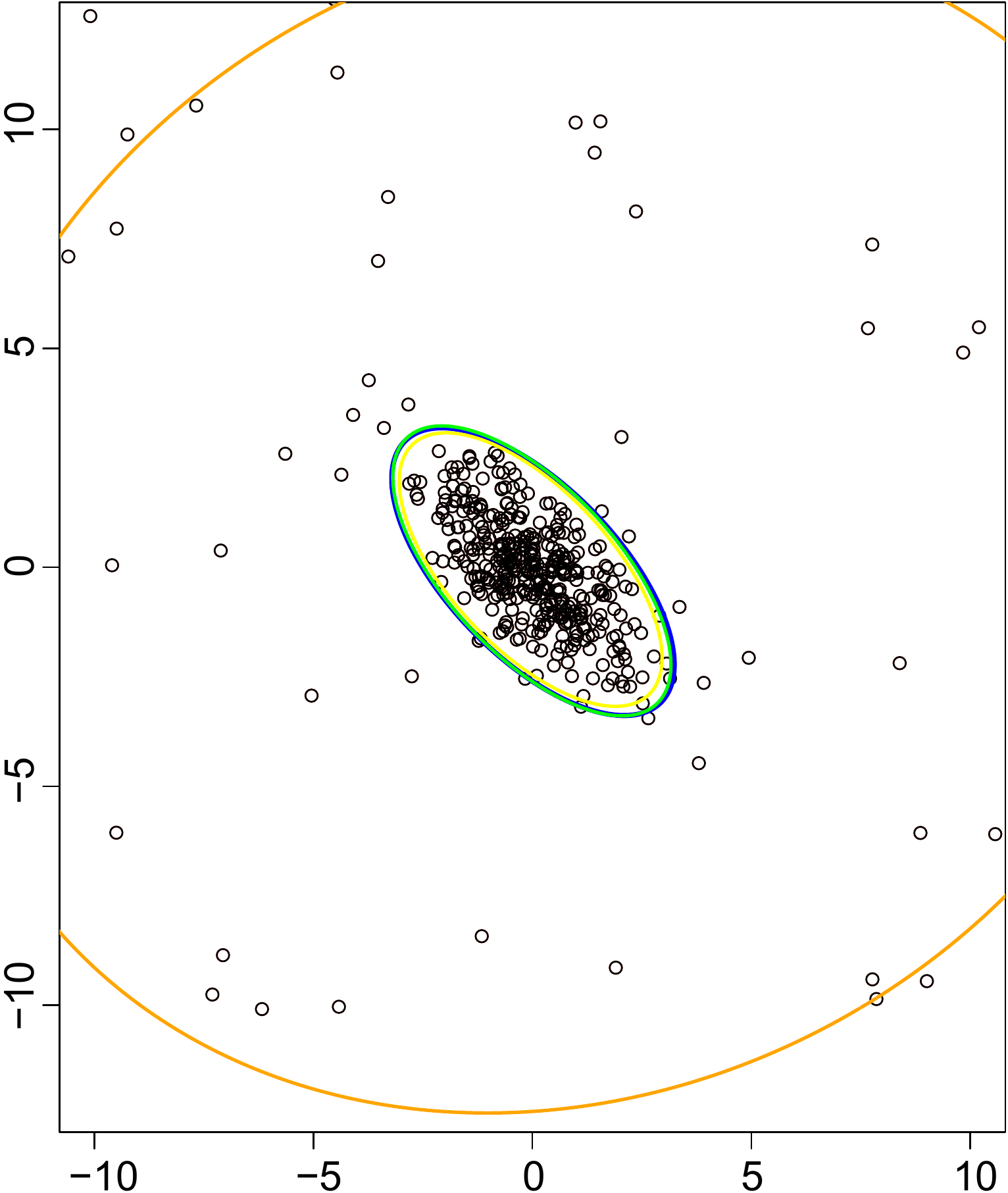}
    \caption{Robust estimates of location and scatter (the covariance confidence ellipses indicate the regions containing 97.5\% of the samples under a normal assumption). The samples rejected after the first step of outlier deletion based on Euclidean distance and unimodality tests are depicted as red points, while black ones constitute the unimodal subset selected. The red ellipse correspond to the Mahalanobis distances computed from the whole distribution, while the orange one was estimated from the unimodal subset only. The blue, green, and yellow ellipses correspond to the robust distances estimated on the unimodal subset, based on the MVE, MCD, and OKG, respectively. The left figure shows the whole distribution, the right one a close-up on the main cluster.}
    \label{fig:robustDistances}
\end{figure}

Simulations over a wide range of mixtures showed similar results for each of these robust distances. An example is provided in Figure~\ref{fig:robustDistances}, using 200 samples in the main cluster, 120 in each of the two secondary clusters, and 200 samples of uniform noise. As one can notice, the classic Mahalanobis distances are not suited to deal with noisy data, even after filtering a great part of the outliers through the unimodal tests. Robust estimates of location and scatter, on the other hand, offer reliable results, as they can withstand up to 50\% outliers and are being applied to the unimodal subset, whose contamination is typically far below.

Based on these evaluations and its shorter computing time, the MCD method was chosen as the scatter estimate. The other estimators would be nonetheless suitable. In term of location parameter, we used the corrected recursive depth estimate presented in Section \ref{sec:bootsrap}, applied to the subset selected in the filtering step based on unimodaility.

From the ordering of the samples provided by these robust distances, the final step of the \emph{Refine} procedure consists in filtering the remaining outliers in the unimodal subset. A possible approach for outlier filtering, detailed for instance in \citep{hadi_identifying_1992,hubert_minimum_2010,rousseeuw_clustering_1987,rousseeuw_unmasking_1990}, consists in removing at once all the outliers with a robust distance superior to a given threshold. Different values have been suggested, such as considering the square root of the $97.5\%$ quantile of the chi-square distribution with $p$ degree of freedom ($p$ being the data dimension) \citep{hubert_minimum_2010,korkmaz_mvn:_2014,etherington_mahalanobis_2019}. The choice of this cut-off value remains however very sensitive, as it depends on the nature of the distribution, on the quantity of noise, and on the robust distances parameters \citep{filzmoser_multivariate_2005}. For these reasons, generic thresholds recommended in the literature can often lead to large errors, either false positives (outliers selected as part of the cluster) when set too high for the dataset considered, or false negatives (inliers rejected) when too low. Therefore, we suggest instead an approach which does not require prior knowledge on the quantity of outliers, and consists in an iterative removal of the most distant samples, as in our first filtering step, until reaching a stopping criterion. Note that the opposite procedure is possible as well, starting with a small subset and adding samples at each iteration, as in \citep{atkinson_fast_1994,hadi_identifying_1992}. In regard to the stopping criterion, different methods have been suggested, relying for instance on tests of normality, on the distribution of residuals, or on the convergence of parameters such as the energy or the dispersion matrix.

In order to search for clusters or modes in multivariate distributions, one must first define what is considered a group, a sub-population. While some researches have been investigating atypical shapes such as rings, stars, or non-parametric models, most studies usually characterize clusters as groups of samples following a normal law, or presenting an ellipsoidal shape. As noted in \citep{burman_multivariate_2009}, the local distribution around a mode can be modeled as a multivariate normal under some smoothness assumptions. Adopting this widely used definition, which is consistent with our recordings of ocular fixations and with a great number of other measurements of natural phenomenons, we based our outliers filtering step on a Gaussian assumption.

\sloppy 
When considering multivariate samples, a rejection of the normality hypothesis in one of the components, by an univariate procedure, is enough to rule out multivariate normality (MVN), as all marginal distributions and linear transformations of a multivariate normal are themselves normal \citep{mardia_tests_1980, looney_how_1995}. However, testing each variable separately is not sufficient to conclude on normality, as non-normal multivariate distributions can also have normally distributed marginals~\citep{thode_testing_2002, mecklin_appraisal_2004}. Therefore, MVN tests require the use of tailored statistics, accounting for the multivariate structure of the data. This topic was the object of numerous studies over the last decades, leading to the creation of more than 50 procedures \citep{mecklin_monte_2005, looney_how_1995}. Reviews of some of these tests, with discussions on their consistency, power, affine invariance, and other properties, are available in \citep{mecklin_appraisal_2004, looney_how_1995, andrews_methods_1973, gnanadesikan_methods_1977, cox_testing_1978, henze_invariant_2002, romeu_comparative_1993}. Among the methods most frequently cited, we find: i) tests based on the empirical characteristic function, such as the Henze-Zikler test \citep{henze_class_1990} or the BHEP family \citep{baringhaus_consistent_1988, epps_test_1983}; ii) the Royston extension of the Shapiro-Wilk univariate test \citep{royston_techniques_1983, royston_approximating_1992}; iii) tests based on skewness and kurtosis, for instance those from Mardia \citep{mardia_measures_1970}, Kankainen \citep{kankainen_tests_2007}, or the Doornik-Hansen omnibus test \citep{doornik_omnibus_1994}; iv) the energy test from \citep{szekely_new_2005}.

There is no clear consensus on the best way to assess normality in the multivariate scenario, each method presenting advantages and drawbacks depending on the number of samples, the parameters of the tests, or the alternative distributions considered. Power studies of these tests often report inconsistent results, and usually conclude that there is no method universally superior to the others in all the situations evaluated \citep{thulin_tests_2012,bogdan_data_1999,joenssen_power_2014,farrell_tests_2007,mecklin_monte_2005,horswell_comparison_1992,cardoso_de_oliveira_multivariate_2010,hanusz_monte_2017,naczk_assessing_2005}. Nonetheless, methods based on multivariate skewness and kurtosis have grown very popular and are often recommended, as they can detect various types of departures from normality, and generally present good properties such as affine invariance. In this work, we adopted the Mardia tests, widely used in the literature, which rely on the extension of the third and fourth moments to multivariate data \citep{mardia_measures_1970}. There are, in fact, two Mardia tests, based on kurtosis and skewness, respectively. Each of these moments can characterize and diagnose different departures from normality, and it is therefore usually advised to use both tests, as discussed in \citep{kankainen_tests_2007,thode_testing_2002,zhou_powerful_2014}. While some authors suggest a direct combination of both statistics, through single omnibus tests \citep{mardia_omnibus_1983,doornik_omnibus_1994,hanusz_monte_2017}, this approach was questioned in several studies \citep{horswell_comparison_1992}. We preferred a simpler approach: the agreement of both measures, that is, rejecting the null hypothesis when any of the two tests indicates non-normality, as in \citep{cain_univariate_2017,korkmaz_mvn:_2014,looney_how_1995}. We also included another method in our experiments, that has been suggested for MVM testing, and consists in a comparison of the Mahalanobis distances and the chi-square distribution with $p$ degree of freedom through a Kolmogorov-Smirnov test  \citep{malkovich_tests_1973,joenssen_power_2014,etherington_mahalanobis_2019,thode_testing_2002,brereton_chi_2015}.

\begin{figure}[!htbp]
    \centering
    \includegraphics[width=\linewidth]{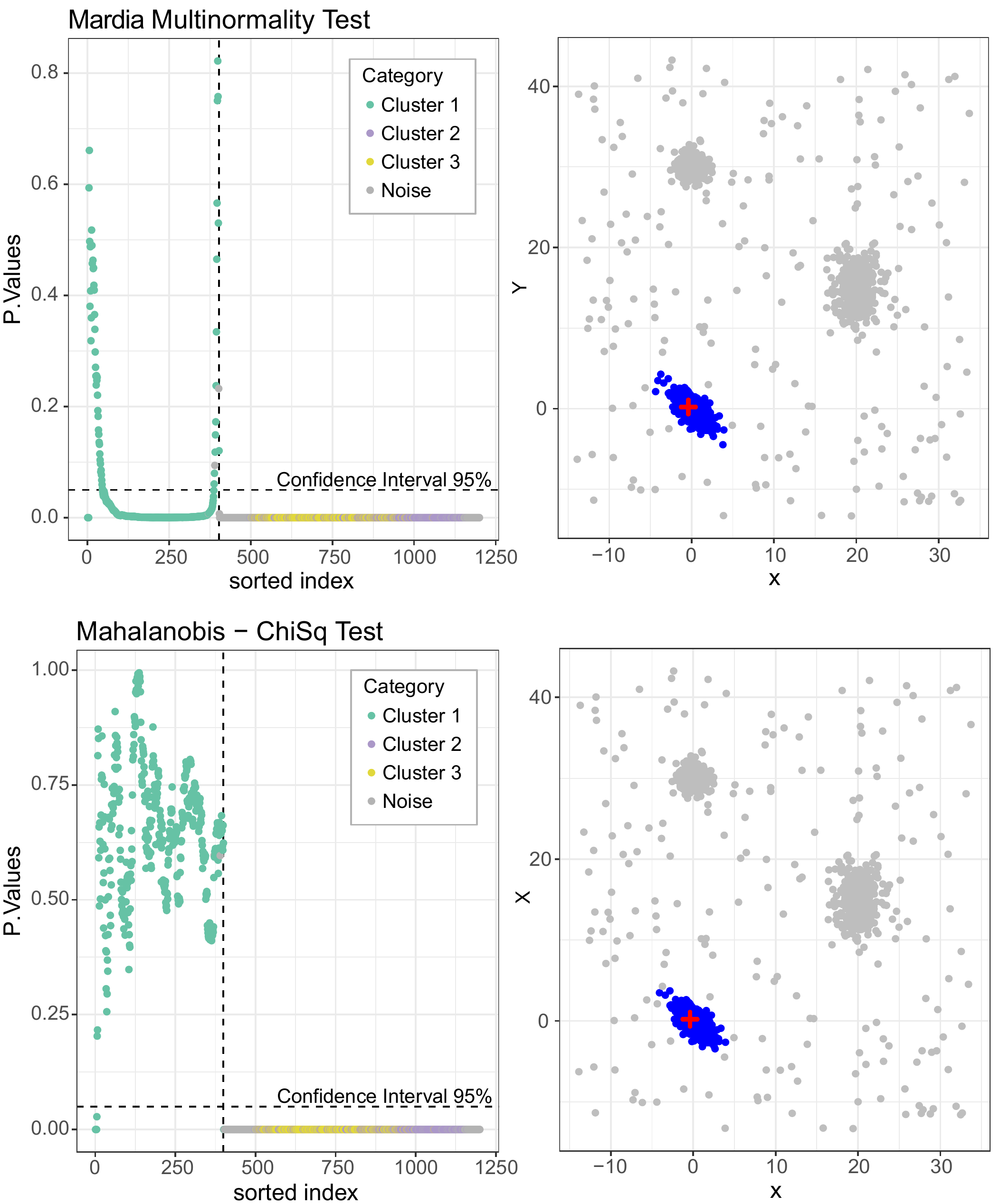}
    \caption{Iterative outlier filtering based on normality tests. Left-most figures: $p$-values of the tests after each outlier removal (ordered by their robust distance with respect to the unimodal subset). Right-most figures: The whole distribution in Cartesian coordinates, with the samples selected depicted in blue. The top-most figures show the filtering based on the Mardia test, whereas the bottom-most ones were obtained by comparing the Mahalanobis distances to the Chi-Squared distribution.}
    \label{fig:NormalityFiltering}
\end{figure}

For both of these methods, similarly to the first step of the \emph{Refine} procedure, the filtering was performed through a sequential procedure. At each iteration, the most distant outlier, based on the robust distances, is removed, until the MVN test fails to reject the normality hypothesis (we used a significance level of $0.05$). Figure~\ref{fig:NormalityFiltering} shows the results obtained on an illustrative distribution, presenting the $p$-values computed at each iteration and the final set of samples considered part of the cluster under this normality assumption. The two tests considered showed satisfying results in both our experimental and synthetic datasets, and appear suitable for the outlier filtering procedure.

Once the subset of samples belonging to the cluster has been identified through the successive normality tests, its center is computed using the arithmetic mean, which constitutes our final location for this mode. The results from this estimate, that we call ``refined location'', are referred to as \emph{``BRL''} (Bootstrap and Refine Locator).

\begin{figure}[!htbp]
    \centering
    \includegraphics[width=1\linewidth]{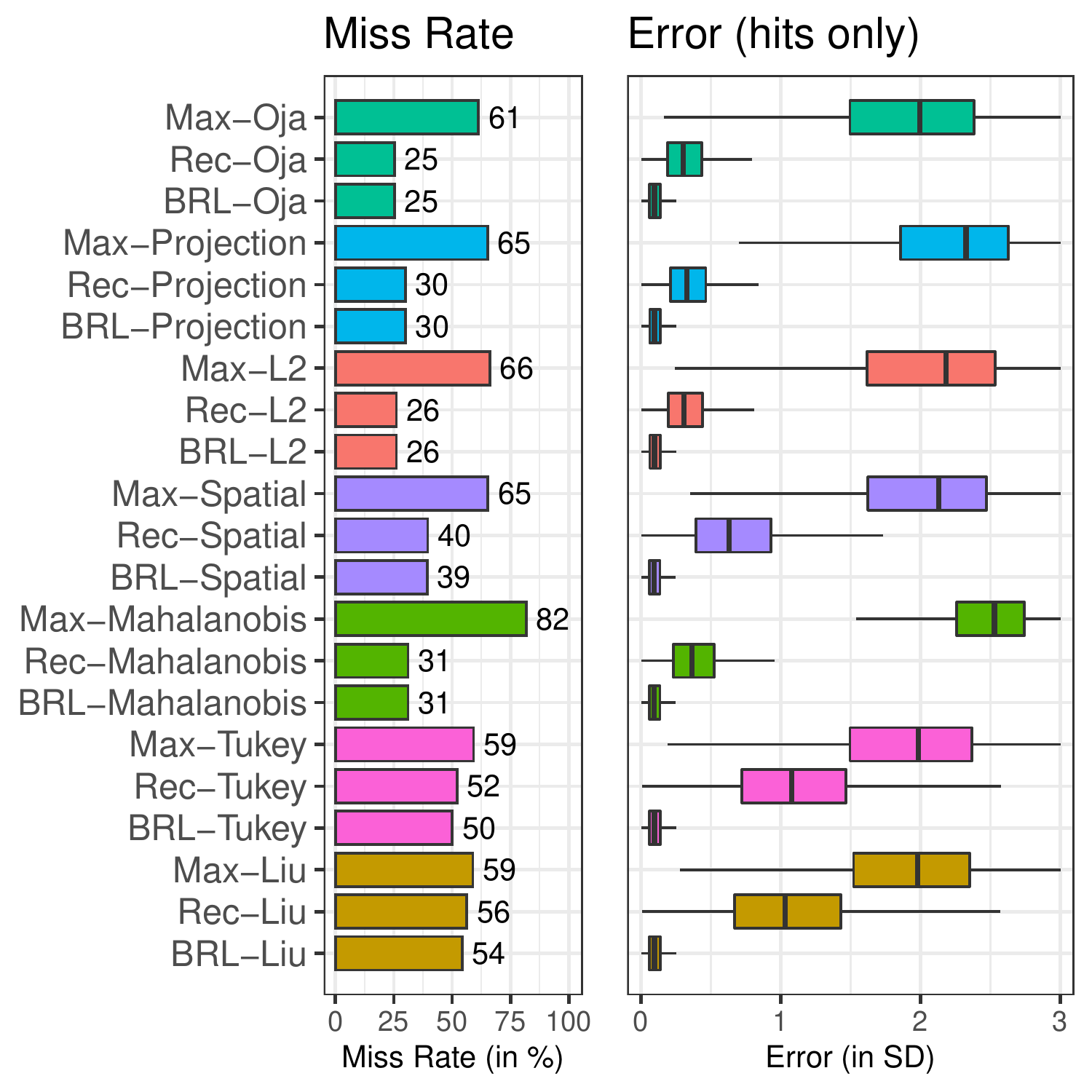}
    \caption{Left figure: proportion of estimates located outside the main cluster, across all the simulations with a ratio of inliers below 50\%, a percentage of uniform noise from 0 to 25\%, and a number of clusters from 3 to 5. Right figure: Errors to the true center, considering solely the hits.}
    \label{fig:boxplotRefine}
\end{figure}

In Figure~\ref{fig:boxplotRefine}, we present the results from this method in the synthetic dataset, with the same measures as those used in Figures~\ref{fig:recursiveDepthBoxplot} and \ref{fig:recursiveDepthBoxplotAbove50}. These values were computed for all the simulations with a quantity of inliers inferior to 50\%.

Analyzing the performances from the refined locations, we notice a substantial improvement when compared to the first estimate computed by the recursive depth method (\emph{REC}). From the results presented in Figure \ref{fig:boxplotRefine} (right), one can see that the errors were considerably reduced, with an average error around $0.1$ standard deviation away from the true center in case of hits. In turn, the percentages of estimates located outside the main cluster (i.e., the miss rates) were not statistically different between the \emph{BRL} and \emph{REC} methods. This finding was expected, as the refined location rely on the estimate from the recursive depth method. Therefore, if the first recursive estimator selected a secondary cluster instead of the main one, the following refining steps using unimodality and normality tests will still be applied to the same incorrect cluster.
The slight improvements observed in some of these miss rates, while marginals, can be explained by the adjustment step applied after reaching an unimodal subset. When the initial estimate is located on the edge or close to the main cluster, as it sometimes happens with high outliers ratios, the location is considered a miss, since its distance to the true center is superior to $3\sigma$. The application of the first filtering process using Euclidean distances and unimodality tests is usually sufficient to clear the initial distribution of most of the outliers, as long as the first estimate was closer to the main cluster than to secondary ones. Therefore, after reaching unimodality, the corrected center estimate will likely fall inside this main cluster, and be considered a hit.

\subsection{ITERATE: iterative identification and removal of clusters} \label{sec:iterate}

With the refining step presented in the previous section, the precision of the center location reached highly satisfying levels, with mean errors inferior to $0.1$ standard deviations, almost identically to the average computed from the ground-truth labels. However, there are still a number of simulations for which it fails to identity the main cluster and, instead, select a secondary one of smaller cardinality, usually due to the central preference of depth measures.

To overcome this issue, we designed an iterative procedure, removing the samples identified in the \emph{Refine} step from the whole distribution, in order to search for new clusters with the same methodology: applying the \emph{Bootstrap} and \emph{Refine} approach again on the remaining samples. This design relates to search and selection strategies proposed, for instance, in \citep{greene_selective_2018}. At each iteration, a new cluster will be selected, labeled and removed, until all samples are assigned to clusters, or until the remaining distribution is already considered unimodal before any outlier filtering. In the latter scenario, the \emph{Refine} procedure will still be applied, but when it ends, the global iterative search will terminate, even if samples remain unassigned, in order to avoid creating several small spurious clusters from unimodal data (generally the uniform noise remaining after having removed all the clustered samples). Once all iterations have been concluded, the main mode location is simply selected by choosing the center associated to the group with the highest count among all clusters encountered but the the last. If there was only one iteration, then the center of this group is selected. Removing the last cluster from the candidates avoid situations where the uniform samples remaining in the final iteration would fail to reject the normality hypothesis during the \emph{Refine} step, and, therefore, would be considered as a cluster. If the quantity of uniform noise is superior to the cardinality of the main cluster, then this uniform data could be erroneously considered as the main mode. Note that when there is no uniform noise, the depth sample median in the penultimate iteration of the procedure will guarantee finding the largest group out of the two remaining, since it'll represent more than 50\% of the distribution. In conclusion, no matter the scenario considered, the last group encountered will not be the main cluster and can be left out from the candidates.

\begin{figure*}[!htbp]
    \centering
    \includegraphics[width=0.9\textwidth]{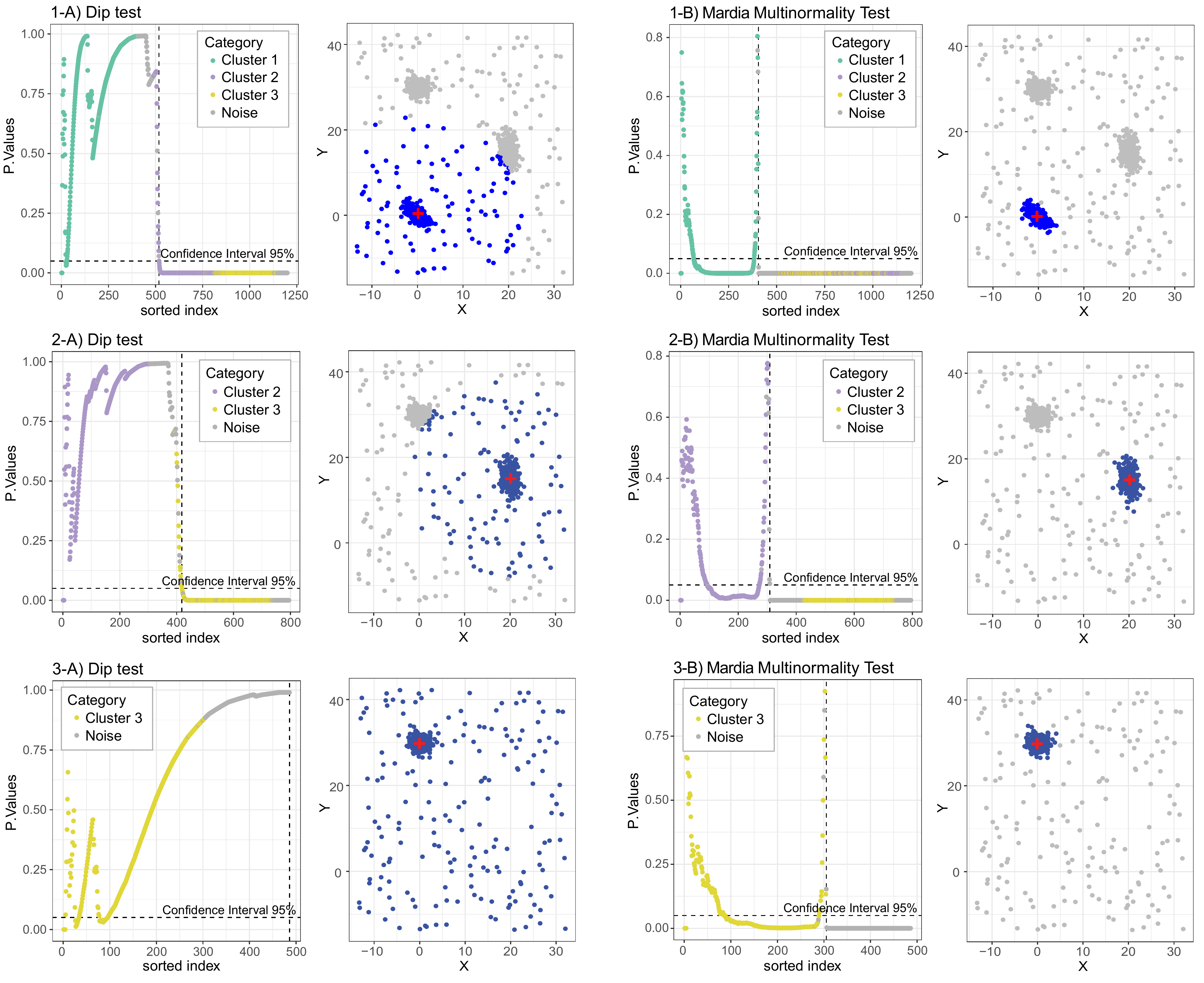}
    \caption{Iterative identification of modes. Each iteration consist in finding an initial seed though a recursive depth estimator (\emph{Bootstrap}), then applying two successive filtering processes (\emph{Refine}). The first is based on Dip tests computed from the Euclidean distances to the \emph{Bootstrap} estimate, in order to extract an unimodal subset (as shown in the first two columns). The second rely on robust distances and a multivariate test of normality (Mardia Test), as displayed in the last two columns. The samples selected at the end of the filtering are removed from the global distribution and the whole procedure is re-iterated. Each row of figures corresponds to an iteration. The stopping criteria was met when the remaining data showed unimodal properties at the beginning of the third iteration.}
    \label{fig:iterationsBril}
\end{figure*}

\begin{figure}[!htbp]
    \centering
    \includegraphics[width=0.8\linewidth]{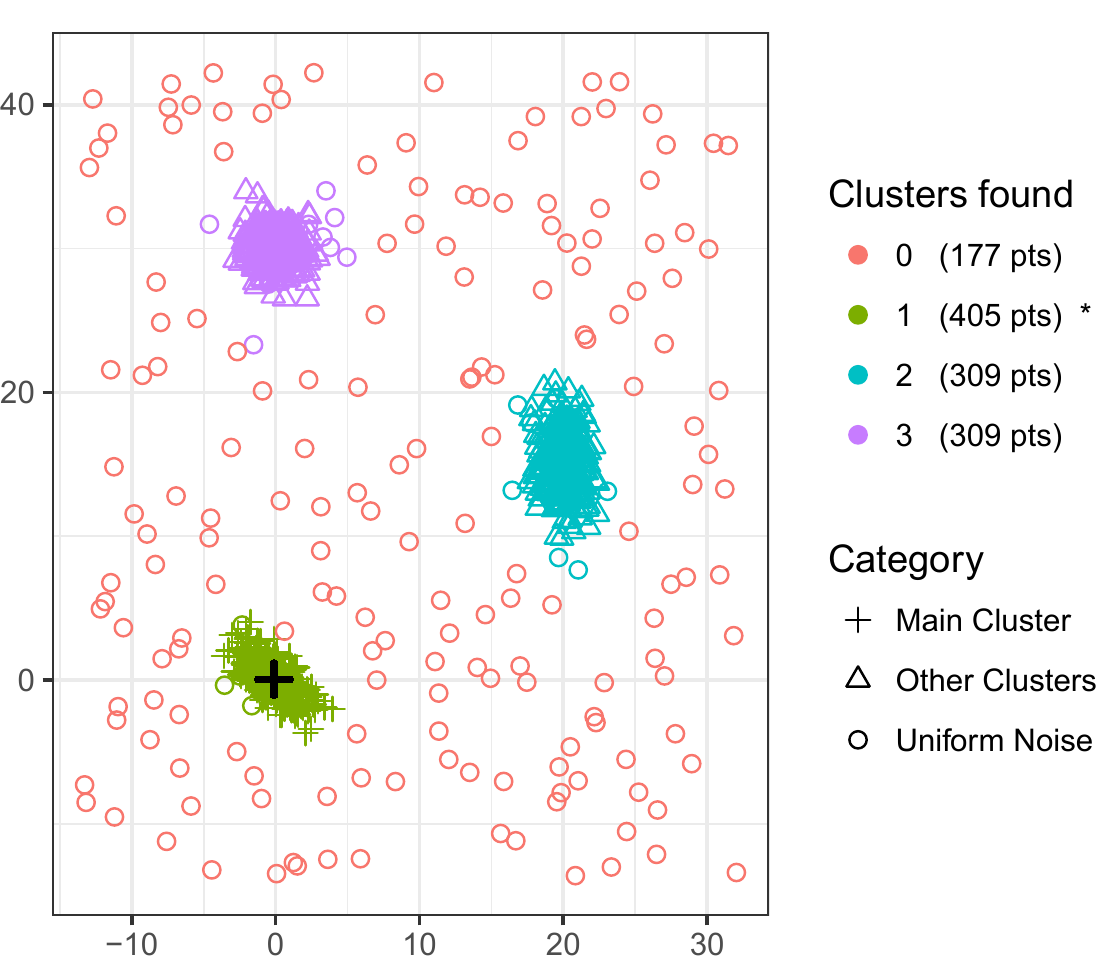}
    \caption{Groups identified after 3 iterations of the \emph{BRIL-Projection} method. The estimate of the main mode, represented by a black cross, correspond to the center of the group with the highest cardinality (405 samples in this example).}
    \label{fig:clustersBril}
\end{figure}

This procedure is exemplified in Figures~\ref{fig:iterationsBril} and~\ref{fig:clustersBril}. Figure~\ref{fig:iterationsBril} shows the selections performed in the different iterations, leading to the identification of $3$ groups in a sample distribution with an high quantity of outliers, both as clusters and uniform noise. Each row corresponds to an iteration of the algorithm, and is decomposed in several successive steps. First, an estimate of the mode is computed through the depth recursive method (\emph{Bootstrap}). Then, outliers are filtered with a recursive removal based on the Euclidean distance to this first estimate and the DIP test of unimodality. Once the unimodal subset has been identified, the center is corrected through the same recursive depth method as in \emph{Bootstrap}, and robust distances to this location are computed with the MCD estimator. These distances are used to order samples in the second phase of outliers filtering, progressively removing the most distant samples until the Mardia test points to normality. The final estimate for each group is obtained by averaging the positions of this last subset. These samples are then removed from the global distribution, and the whole procedure is repeated. In this illustrative distribution, the procedure stops at the end of the third iteration as the remaining samples are already unimodal before any filtering. Figure~\ref{fig:clustersBril} shows all the groups identified at the end of the procedure. The largest one, depicted in green, is selected as the main mode due to its higher cardinality.

\begin{figure}[!htbp]
    \centering
    \includegraphics[width=\linewidth]{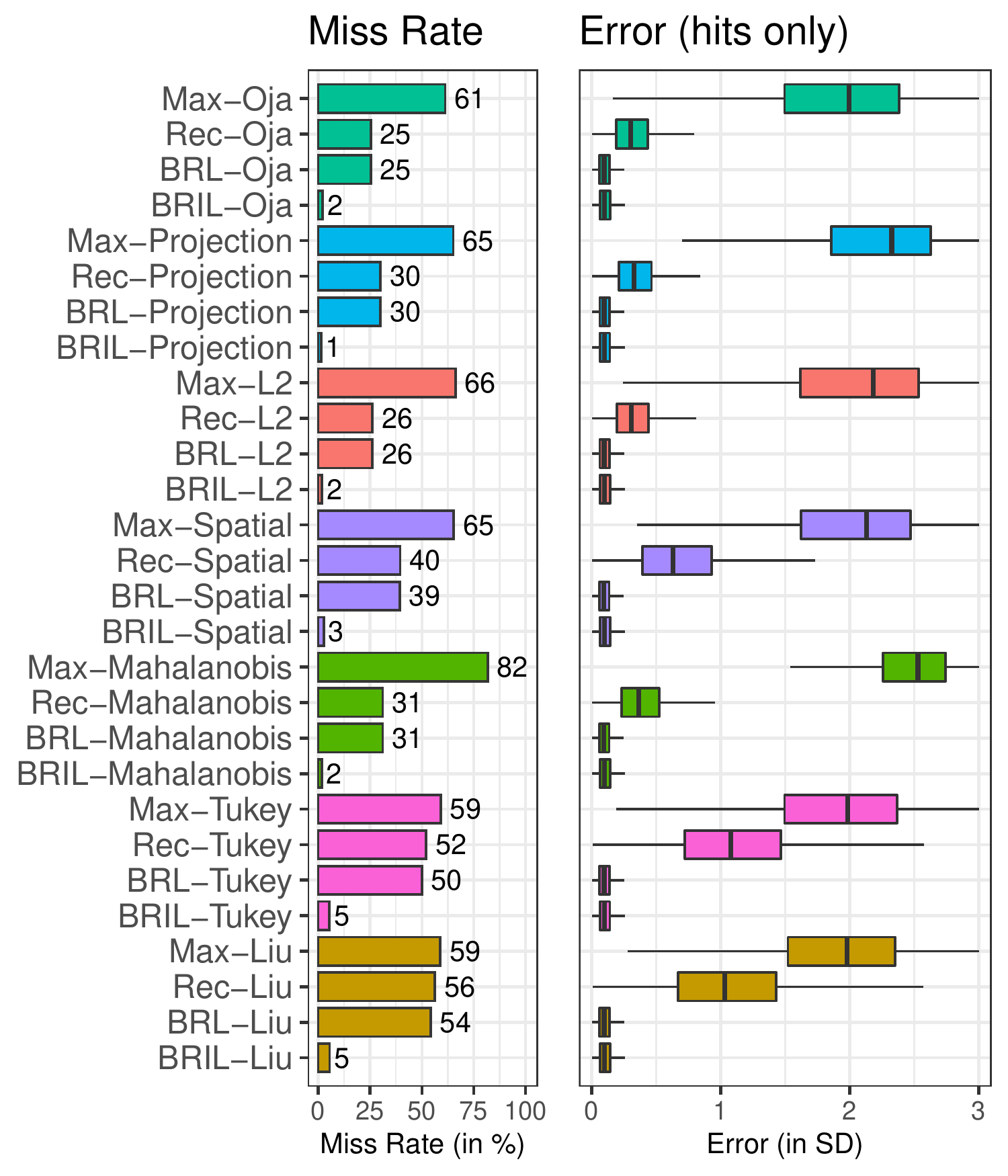}
    \caption{Comparison of our final method (\emph{BRIL}) with the sample median (\emph{Max}), its recursive version (\emph{Rec}), and Refined one (\emph{BRL}), for different depth measure. The left figure shows the proportion of estimates located outside the main cluster, across all the simulations with a ratio of inliers below 50\%, a percentage of uniform noise from 0 to 25\%, and a number of clusters from 3 to 5. The right figure the errors to the true center, considering solely the hits.}
    \label{fig:resultsBRIL-boxplot}
\end{figure}

Similarly to the experiments presented in the previous sections, we analyzed the gains obtained on the synthetic dataset, which are presented in Figure~\ref{fig:resultsBRIL-boxplot}. This final method, including the \emph{Iterate} step, is referred to as \emph{BRIL}, and presents outstanding results. Besides keeping the average error as good as with the \emph{BRL} version in case of hits, this iterative mechanism significantly reduces miss rates (i.e., the number of simulations where the main cluster was not encountered). These miss rates showed values close to zero for all of the depth metrics tested. While a few samples can be miss-classified in the \emph{Refine} step, our method shows robust results and is able to identify all the actual clusters in the great majority of our simulations, regardless of their position, proximity, variance, or the presence of uniform noise. Moreover, the center locations are particularly accurate due to the low quantity of noise remaining after the outliers filtering procedures.

\section{Recursive MVE / MCD }\label{sec:RecMCD}

Minimum Volume Ellipsoid (MVE) and Minimum Covariance Determinant (MCD) are popular robust estimators of location and scatter, due to their high breakdown point, and fast computation \citep{rousseeuw_clustering_1987,rousseeuw_fast_1999,rousseeuw_multivariate_1985}. They also belong to the class of convex body minimizers. Their estimates of location and scatter are defined as the mean and covariance matrix of the $h$ observations among $n$ presenting the smallest volume, in the case of MVE, or the minimal determinant of the sample covariance matrix in the case of MCD, with $h$ usually chosen as $(n+d+1)/2$, such that $n$ is the number of samples and $d$ the number of dimensions \citep{becker_mve_2004}. These methods are often encountered in the literature of robust statistics alongside depth measures as they are as well suited to deal with contaminated multivariate data, for purposes as diverse as noise filtering, classification, clustering, regression, fitting, etc. In the late 1970s, Sager also suggested the use of convex body minimizers to estimate the mode of multivariate distributions, through a recursive procedure selecting at each iteration the subset of samples of a given size with the smallest volume, and discarding the others, until reaching a group containing less than $d+1$ elements \citep{sager_estimation_1978,sager_iterative_1979}. The estimate of the main mode is, then, obtained by averaging the samples from the last iteration. Unfortunately, the limited computational processing power at the time, combined with the absence of optimized algorithm to compute minimum volume subsets, kept this proposal only hypothetical. Recently, a new study revived Sager’s idea, and compared three different implementations of this procedure based on MCD, MVE, and Minimum Volume Convex Hull (MVCH) \citep{kirschstein_minimum_2016}. 

\begin{figure}[!htbp]
    \centering
    \includegraphics[width=1\linewidth]{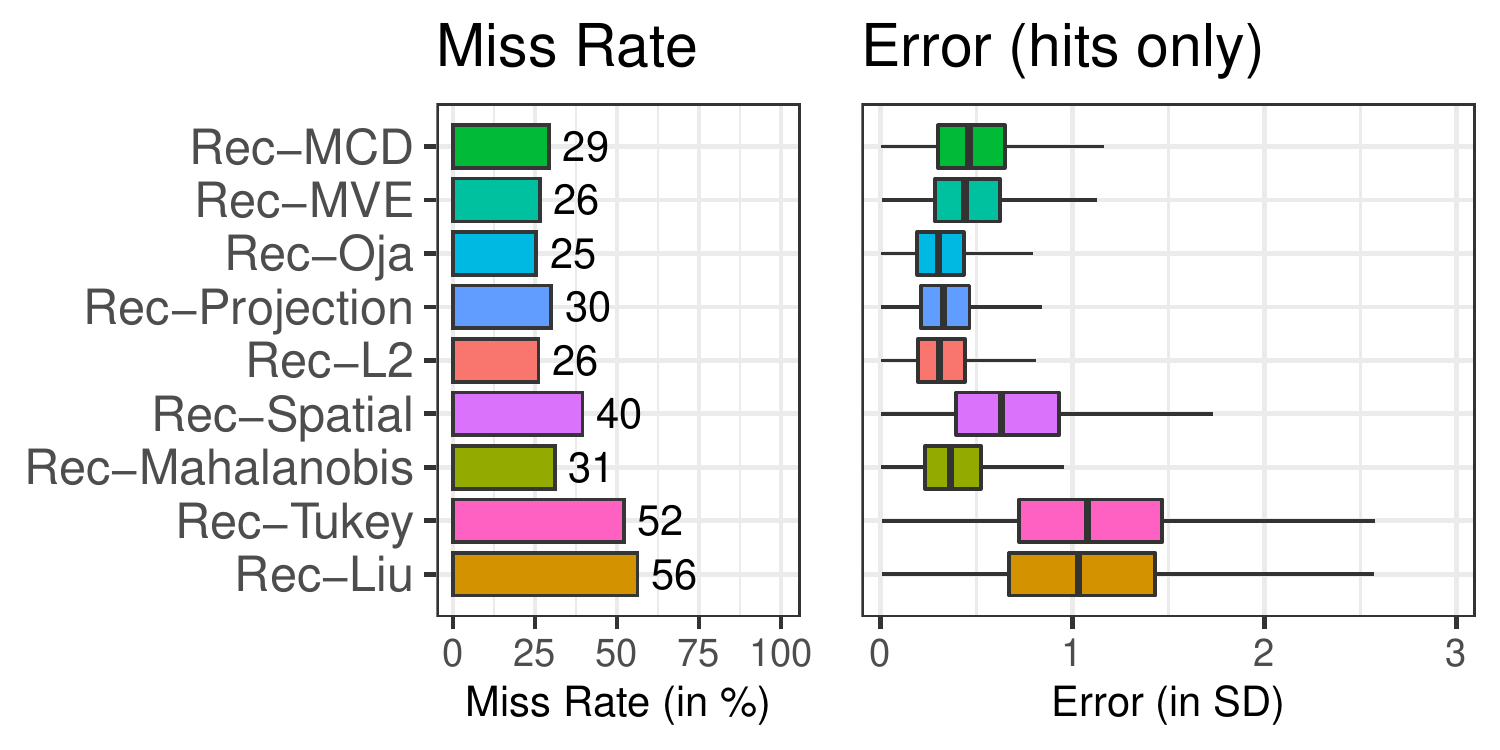}
    \caption{Comparison of recursive locators based either on convex body minizers (MCD and MVE) or depth measures across all the simulations with a ratio of inliers below 50\%, a percentage of uniform noise from 0 to 25\%, and a number of clusters from 3 to 5.}
    \label{fig:boxplotsRecRCD-MVE}
\end{figure}

The methodology we introduced in section~\ref{sec:bootsrap}, estimating the mode through recursive filtering of a fixed fraction of the samples of lowest depth, presents similarities with the procedure theorized by Sager with convex body minimizers. We compare in Figure~\ref{fig:boxplotsRecRCD-MVE} the results of the recursive MCD and MVE methods (\emph{Rec-MCD} and \emph{Rec-MVE}), with those from our recursive estimates based on depth measures. \emph{Rec-MCD} and \emph{Rec-MVE} show overall performances relatively similar to those from locators based on Oja, L2, Projection or Mahalanobis depths. While their miss rate are very close to each other (from 26 to 31\%), it is interesting to notice a slight advantage for the top depth-based estimators in term of precision of localization.

\subsection{Limitations} \label{sec:limitations}

The intrinsic nature of depth measures and convex body minimizers differ on several aspects. Unlike depth functions, MCD and MVE operators do not directly provide an ordering of the samples constituting the global distribution, but instead simply select a given fraction of the samples presenting the minimum volume or covariance determinant. It is nonetheless possible to derive outlyingness values from these subsets, as detailed in Section~\ref{sec:refine}, by computing robust distances based on an unique iteration of MCD/MVE (using the estimates of location and scatter from the samples selected). However, convex body minimizers being combinatorial methods, changes in the size parameter can lead to entirely different subsets, which will, in turn, radically change the ordering and outlyingness values. On the other hand, the subset obtained from selecting the $n^{th}$ samples with the highest depths will always be included in any subset of deepest samples of a size superior to $n$.

\begin{figure}[!htbp]
    \centering
    \includegraphics[height=5cm]{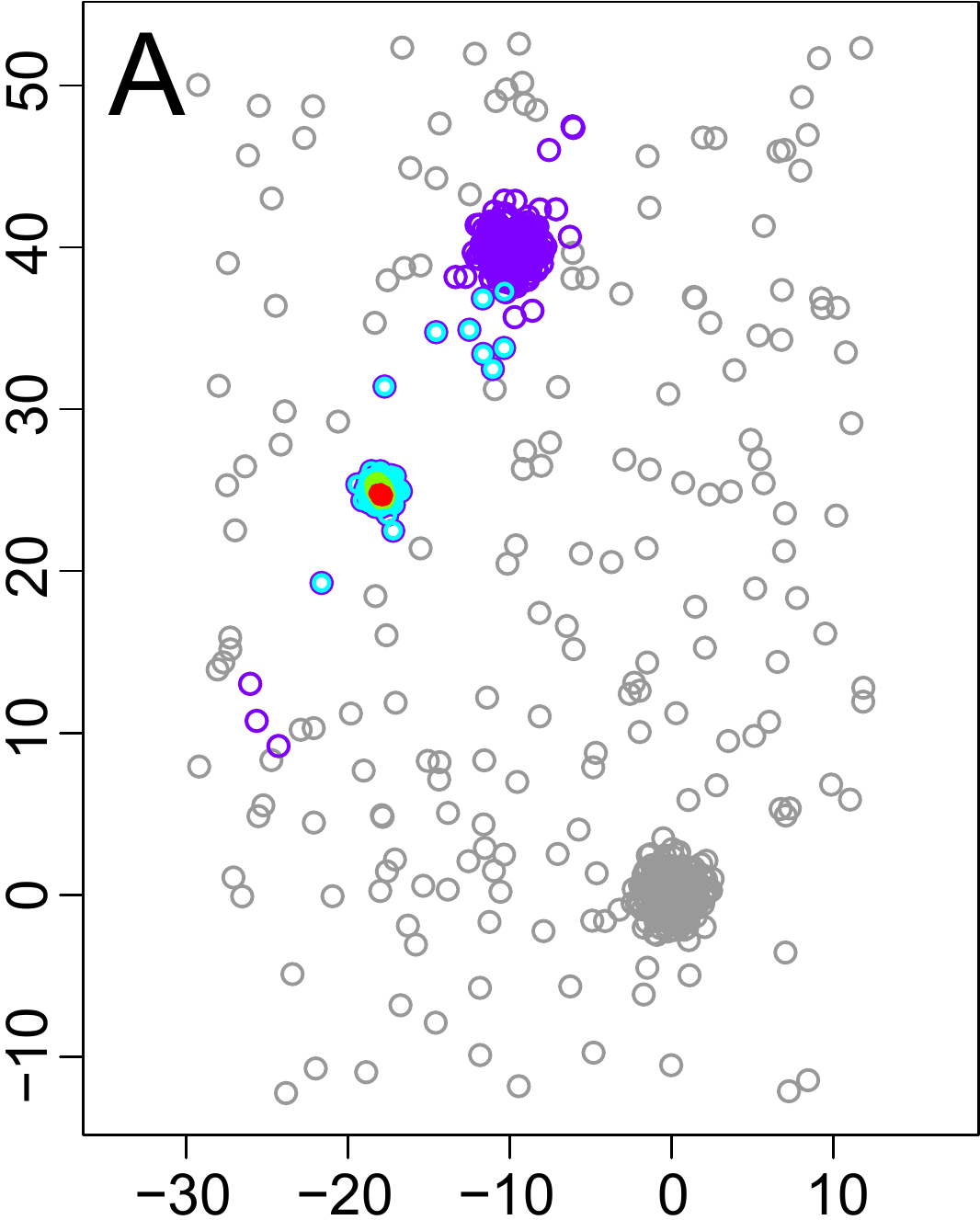}
    \includegraphics[height=5cm]{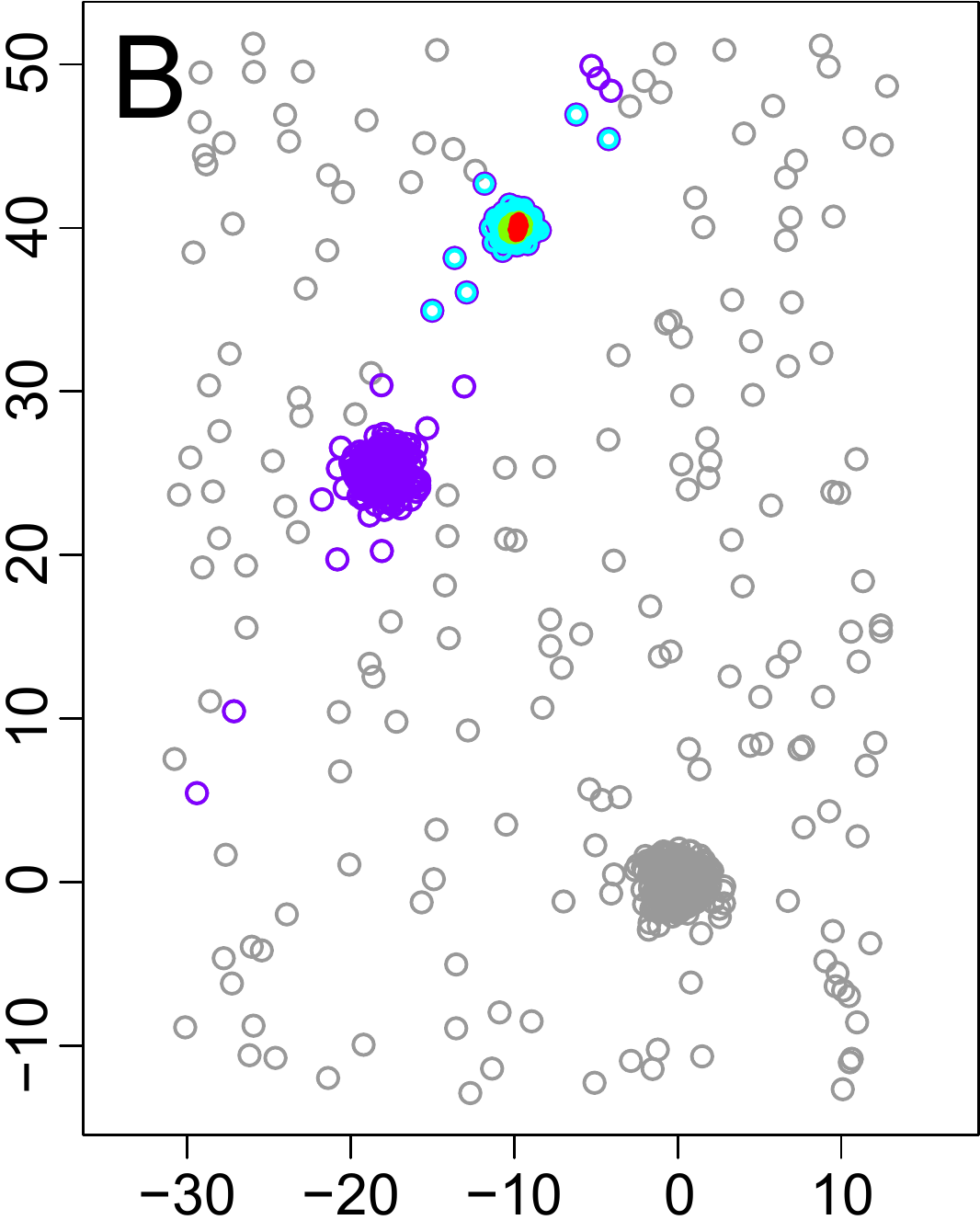}
    \includegraphics[height=5cm]{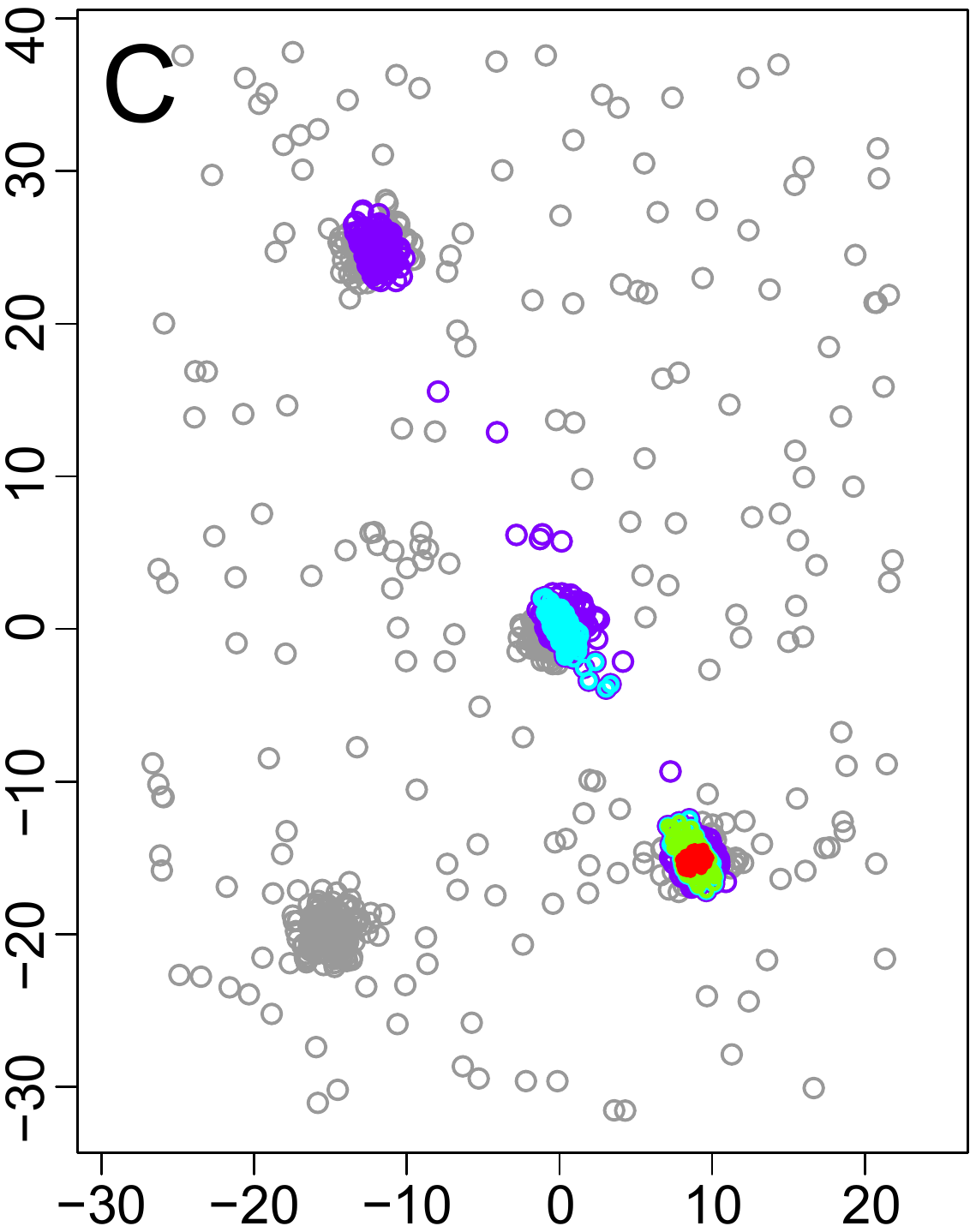}
    \includegraphics[height=5cm]{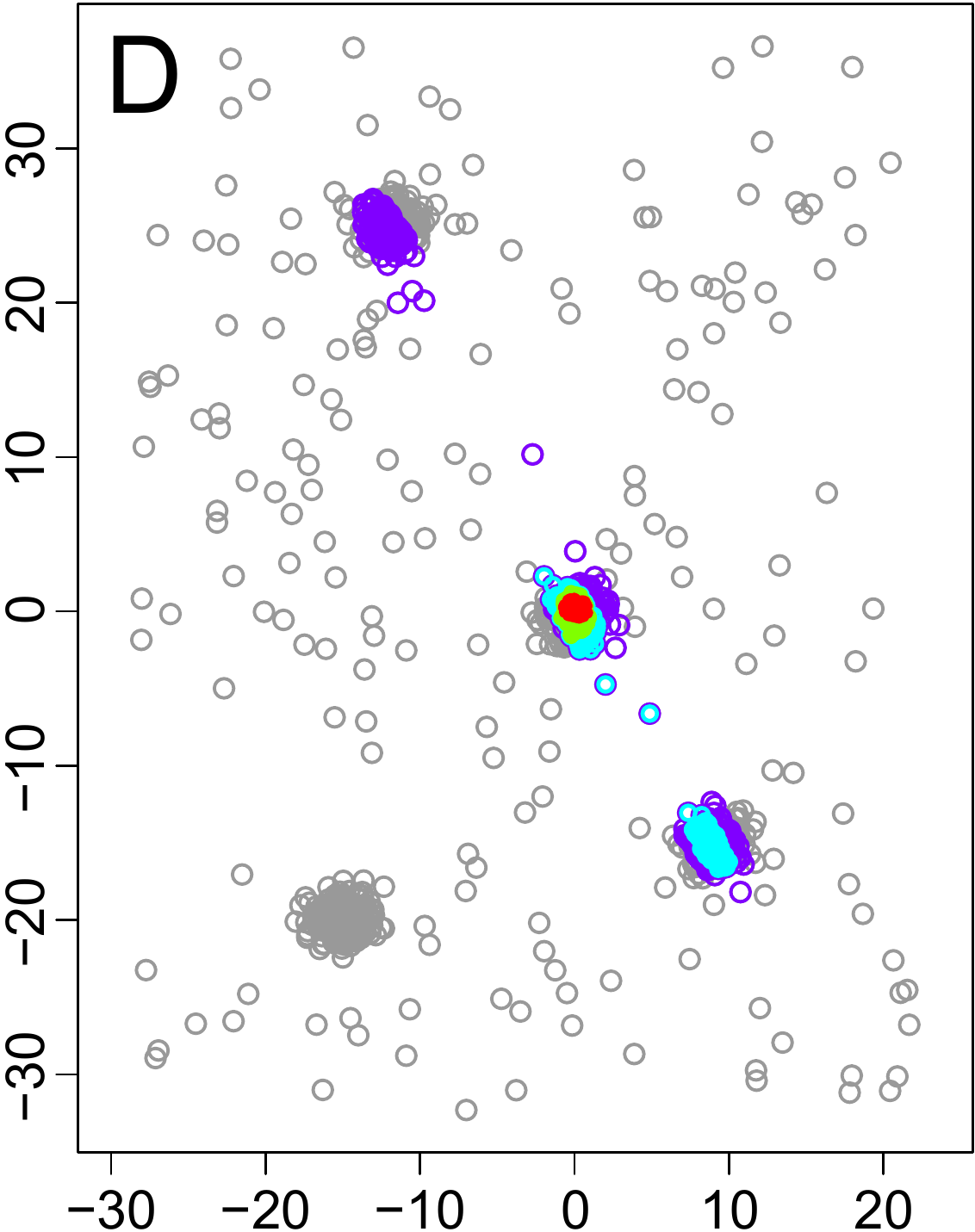}

    \caption{Recursive MVE applied to four illustrative datasets. The colors represent the samples selected at each iteration (with a trimming factor of 0.5). The true mode (cluster with the highest cardinality) is centered at coordinates (0,0).}
    \label{fig:errorsRecMVE}
\end{figure}

For these reasons, the choice of the number of samples kept at each iteration of the recursive procedure is particularly important with convex body minimizers, as it can greatly affect the robustness to outliers \citep{kirschstein_minimum_2016}. The commonly recommended value of $(n+d+1)/2$ only guarantees the convergence to the true mode under the assumption that the number of outliers remains as well below $(n+d+1)/2$. However, in the situations considered in this work, where the global distribution contains several clusters, this assumption does not always hold. When the principal cluster account for less than 50\% of the whole distribution, the first iteration often selects a group including several secondary clusters, without samples from the main cluster, resulting in an incorrect mode at the end of the successive recursions, as illustrated in Figure \ref{fig:errorsRecMVE}-A and B. Even when samples from the main cluster are selected, not all of them might be kept in the first iteration, depending on the spatial configuration of the different clusters. This can lead to two sorts of error. First, selecting a secondary cluster in the next iteration(s), when too many samples from the principal cluster were discarded in the first step (see Figure \ref{fig:errorsRecMVE}-C). Secondly, even if the correct cluster is chosen, its final center estimate (the result of the last iteration) might be corrupted when it was selected from a non-centered subset of the cluster (see Figure~\ref{fig:errorsRecMVE}-D).

\subsubsection{Application of the BRIL algorithm to Rec-MCD and Rec-MVE}

Given the promising results of \emph{Rec-MCD} and \emph{Rec-MVE} observed in Figure~\ref{fig:boxplotsRecRCD-MVE}, showing performances similar to our recursive estimates based on depth measures, we suggest the use of these convex body minimizers as bootstrap estimates in our algorithm described in Sections~\ref{sec:refine} and~\ref{sec:iterate}. These additional steps aim at overcoming the limitations previously described, when inliers represent less than half of the distribution. First, by refining the location estimate through recursive trimming based on unimodality and normality tests. Then, by iterating the global procedure of identification and removal of each cluster, in order to select the cluster with the highest cardinality as first mode, in the exact same way as we did with the depth-based estimates.

\begin{figure}[!htbp]
    \centering
    \includegraphics[width=1\linewidth]{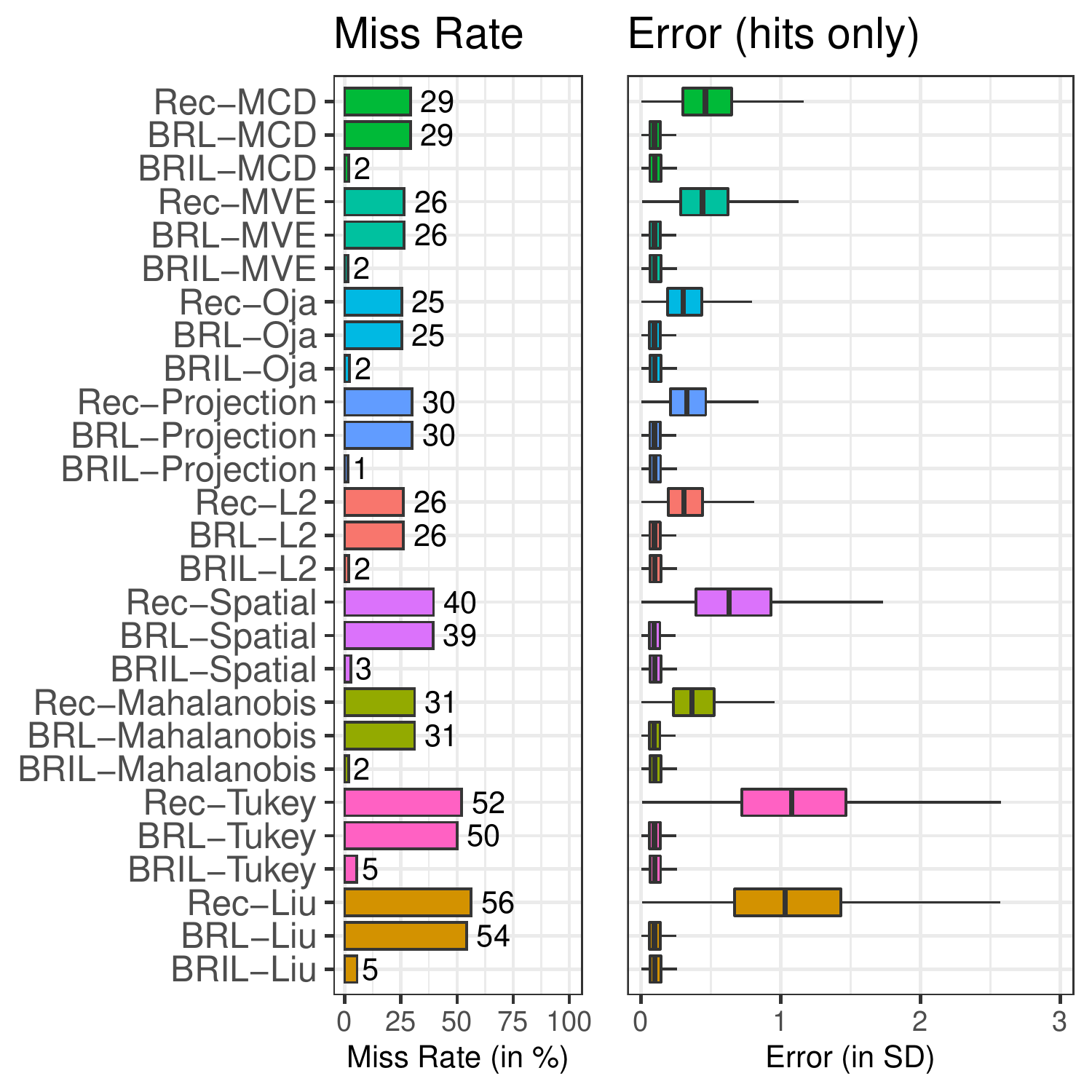}
    \caption{Comparison of \emph{BRL} and \emph{BRIL} procedures applied to recursive estimators based either on depth measures or on convex body minimizers.}\label{fig:boxplotsBrilMCD}
\end{figure}

The results are presented in Figure~\ref{fig:boxplotsBrilMCD}, using the same nomenclature as in the previous sections, the prefixes \emph{BRL} and \emph{BRIL} referring to the refined location and to the complete iterative procedure, respectively. Both of these convex body minimizers methods showed very similar behaviour as depth-based measures when integrated into the BRIL algorithm. We see, indeed, that the refining step increases the precision of the central location when the main cluster was correctly identified by the \emph{Rec-MCD} and \emph{Rec-MVE} first estimate. Whereas the iterative procedure allows to overcome the issues encountered in distributions where an outlier cluster was selected in the first iteration, through the search of the different other clusters and the selection of the largest.

Our simulations showed no significant difference between MCD and MVE performances, consistently with the findings from \citep{kirschstein_minimum_2016}. The reader can refer to the following studies discussing their differences and properties \citep{becker_largest_2001,becker_mve_2004,kirschstein_minimum_2016,woodruff_computable_1994}.

\newpage
\section{Results}\label{sec:results}

\subsection{General Results}

We summarize in Table~\ref{tab:globalsynthresult} the performances of our approach measured on the synthetic dataset. For all the depth functions and convex body minimizers considered, the results presented include: the standard multivariate median (\emph{Med}, only applicable to depth measures); its recursive version (\emph{Rec}); the \emph{BRL} estimate (the recursive locator followed by filtering and correction, as presented in Section~\ref{sec:refine}); and finally the complete \emph{BRIL} procedure (with the selection of the iteration with the highest count, see Section~\ref{sec:iterate}). These results are compared with those from the clustering algorithms showing the best estimates in the study from \citep{brilhault_eyetracking_2019}, i.e., \emph{PAM} (partitional clustering), \emph{MClust} (model-based), and \emph{DBSCAN} (density-based), as well as the average position (\emph{Mean}) and coordinate-wise median (\emph{Med-CW}) as baselines. The data included all simulations with $k=2$ to $5$ clusters, 0 to 25\% uniform noise, and an inliers ratio (within the clustered samples) set between $1/k$ and $0.75$. A total of 15900 different distributions were tested.

\begin{figure}
\centering
    \includegraphics[width=1\linewidth]{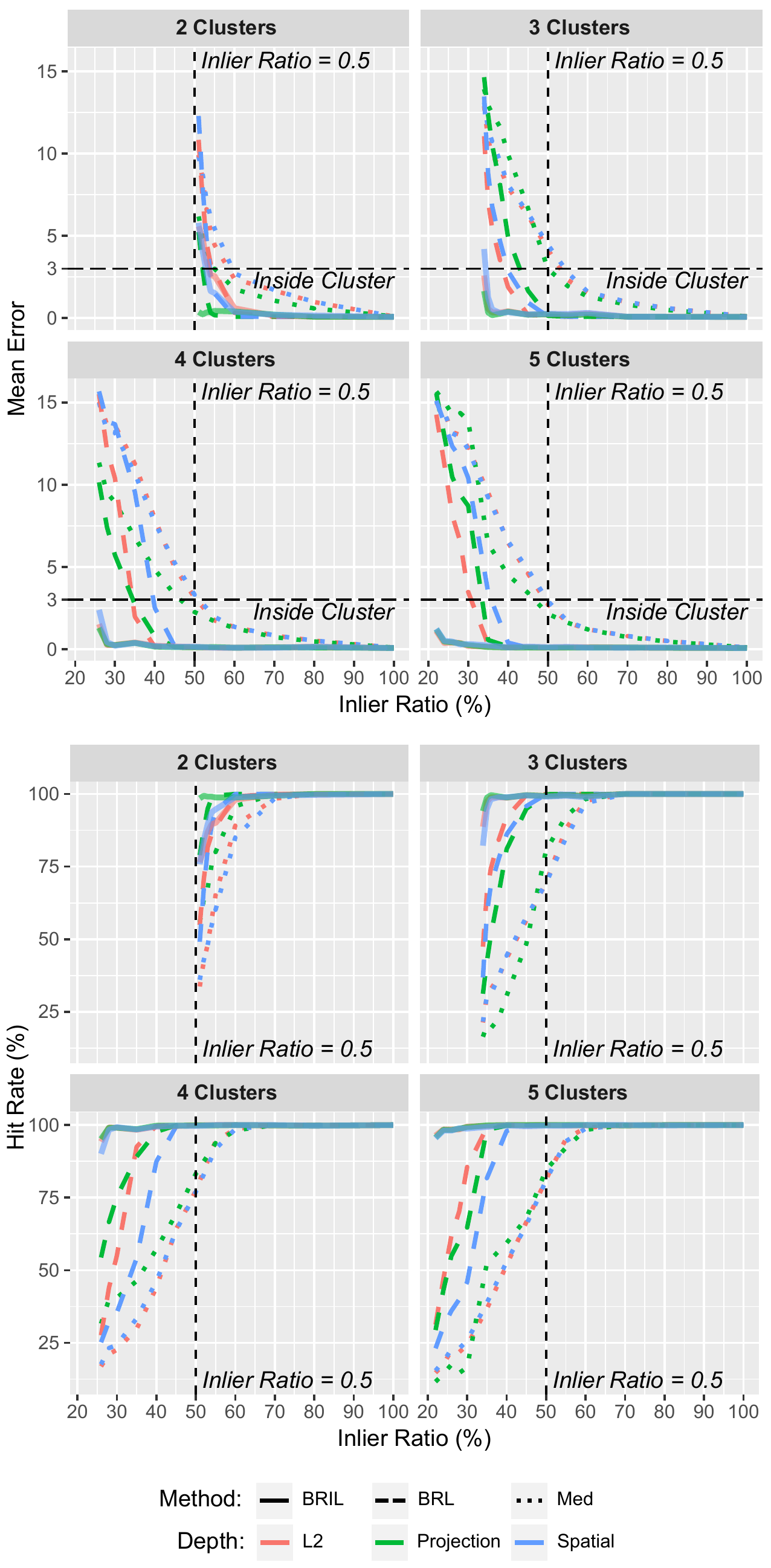}
    \caption{Overall results of our approach with multimodal distributions, as a function of inliers ratio. Dotted lines correspond to the standard MM median, dashed ones to the recursive estimate with correction (\emph{BRL}), and finally the solid lines to the \emph{BRIL} version (including both the \emph{Refine} and \emph{Iterate} steps). For clarity, the results are provided for only three depth measures (L2, Projection and Spatial). The four top-most figures present the mean error across all Monte-Carlo repetitions for different numbers of cluster, while the bottom-most ones show the percentages of estimates successfully falling within the main cluster. The final BRIL algorithm, regardless of the depth measure considered, reaches close to 100\% hit rate as soon as the main cluster cardinality shows a number of samples slightly superior to the second biggest cluster count.}
    \label{fig:curvesMedBrlBril}
\end{figure}

\begin{figure}
\centering
    \includegraphics[width=1\linewidth]{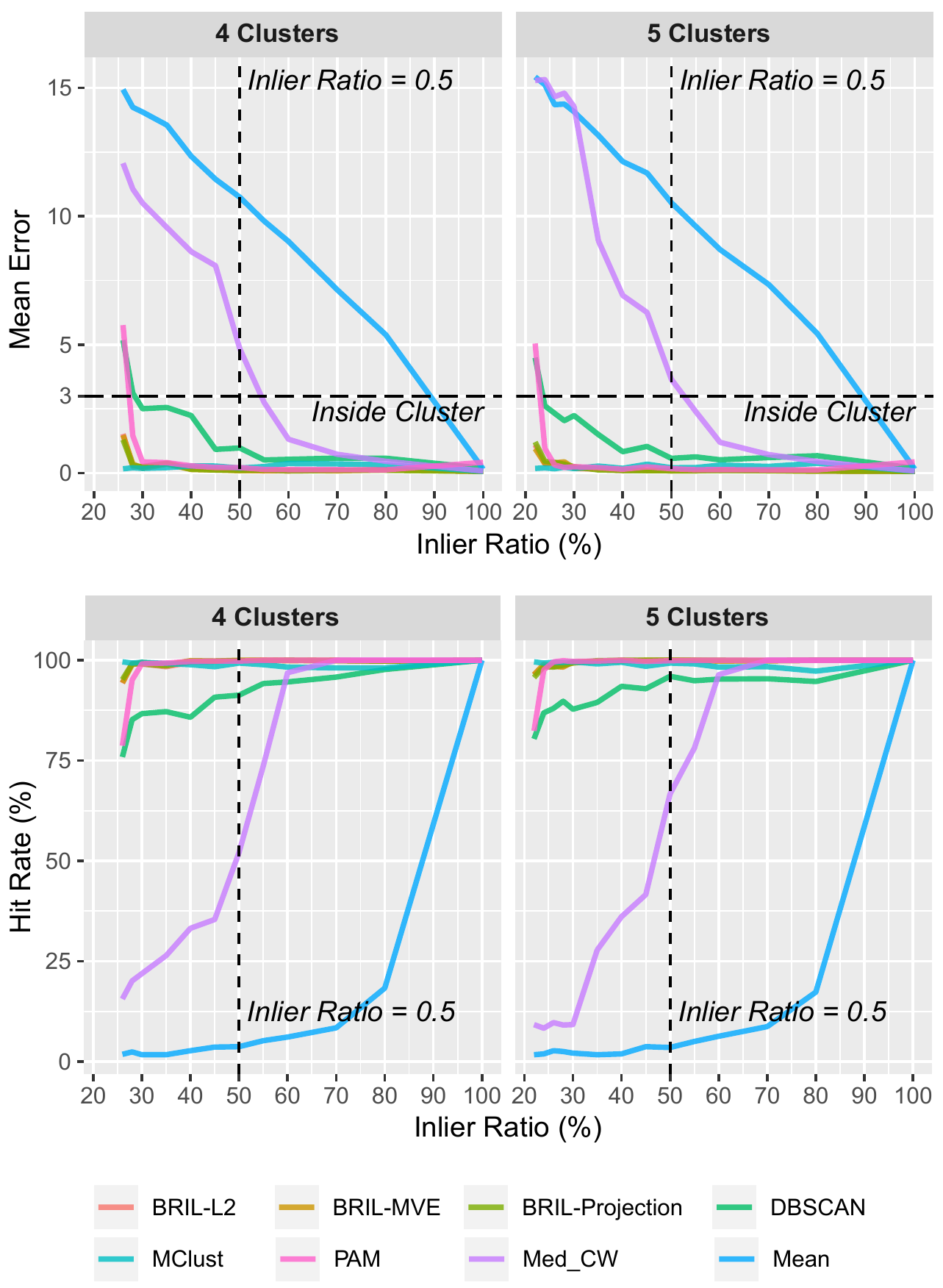}
    \caption{Average error and percentage of hits, similarly to Figure~\ref{fig:curvesMedBrlBril}, for the BRIL algorithm based on L2 and Projection depth, as well as on MVE convex body minimizer, compared with the three clustering algorithms, the coordinate-wise median, and the mean.}
    \label{fig:curvesFinalAll}
\end{figure}

Consistently with the results presented in the previous sections, we see that for every depth function, the \emph{Rec} estimate does increase the percentage of \emph{hits} over the standard MM, and also reduces the distance to the true center when the cluster was encountered (e.g., reaching 0.329 for \emph{Rec-Oja}, 0.337 for \emph{Rec-Projection}, or 0.332 for \emph{Rec-L2}). In turn, the refined locator, while not affecting the hit rates significantly, offers a much more accurate estimate (with average errors as low as 0.097 for \emph{BRL-Oja}, or 0.096 for \emph{BRL-Projection} and \emph{BRL-L2}). Finally, The \emph{BRIL} version show the most reliable estimates, with hit rates ranging from 95.56\% to 98.93\%, and average errors from 0.100 to 0.101 when the main cluster was found, as accurately as Model-Based Clustering (0.103), and better than \emph{PAM} (0.164). Figures~\ref{fig:curvesMedBrlBril} and \ref{fig:curvesFinalAll} illustrate these remarkable performances for different numbers of clusters, as a function of inliers ratio.

\begin{table*}[thb]
  \centering

  \caption{Performance on all simulations with $2$ to $5$ clusters, 0 to 25\% uniform noise, and an inliers ratio, among the clustered samples, set between $1/k$ and $0.75$. The table lists the percentage of hits (i.e., simulations with an error to the center of the main cluster inferior to 3), as well as the average error (with standard deviation) and sum of squared errors for either all simulations (in the third and fourth columns), or only those where the main cluster was found (fifth and sixth columns).}
  
    \begin{tabular}{lcr@{\hskip 4pt}>{\footnotesize}lrr@{\hskip 4pt}>{\footnotesize}lr}
          &       & \multicolumn{3}{c}{\textbf{All Simulations}} & \multicolumn{3}{c}{\textbf{Hits Only}} \\
    \midrule
    \multicolumn{1}{c}{\textbf{Method}} & \textbf{HitRate} & \multicolumn{2}{c}{\textbf{MeanError (sd)}} & \multicolumn{1}{c}{\textbf{SSE}} & \multicolumn{2}{c}{\textbf{MeanError (sd)}} & \multicolumn{1}{c}{\textbf{SSE}} \\

    \midrule
    BRIL-Projection & 98,93\% & 0,341 & (2,518) & 374053 & 0,100 & (0,067) & 828 \\
    BRIL-MVE & 98,56\% & 0,413 & (2,827) & 472790 & 0,101 & (0,069) & 852 \\
    BRIL-MCD & 98,48\% & 0,441 & (2,981) & 526183 & 0,101 & (0,067) & 839 \\
    MClust & 98,37\% & 0,310 & (1,781) & 189373 & 0,103 & (0,096) & 1132 \\
    BRIL-Mahalanobis & 96,73\% & 0,891 & (4,671) & 1309804 & 0,101 & (0,069) & 831 \\
    BRIL-Oja & 96,38\% & 0,950 & (4,775) & 1373181 & 0,101 & (0,069) & 830 \\
    BRIL-L2 & 96,27\% & 1,002 & (4,970) & 1488954 & 0,100 & (0,067) & 813 \\
    BRIL-Spatial & 96,07\% & 1,022 & (4,980) & 1497402 & 0,100 & (0,070) & 832 \\
    PAM   & 95,99\% & 1,242 & (5,652) & 1939812 & 0,164 & (0,170) & 3109 \\
    BRIL-Tukey & 95,64\% & 1,027 & (4,780) & 1384993 & 0,101 & (0,078) & 893 \\
    BRIL-Liu & 95,56\% & 1,047 & (4,850) & 1426258 & 0,101 & (0,079) & 906 \\
    DBSCAN & 89,06\% & 2,123 & (6,906) & 3024259 & 0,253 & (0,388) & 11070 \\
    Rec-MVE & 78,90\% & 5,590 & (10,925) & 8724733 & 0,475 & (0,246) & 13060 \\
    BRL-Oja & 78,86\% & 4,695 & (9,768) & 6805002 & 0,097 & (0,059) & 584 \\
    Rec-Oja & 78,86\% & 4,864 & (9,663) & 6780732 & 0,329 & (0,188) & 6557 \\
    BRL-MVE & 78,62\% & 5,354 & (11,106) & 8806845 & 0,098 & (0,068) & 648 \\
    BRL-L2 & 78,21\% & 4,851 & (9,913) & 7055741 & 0,096 & (0,058) & 571 \\
    Rec-L2 & 78,21\% & 5,019 & (9,804) & 7027744 & 0,332 & (0,197) & 6729 \\
    BRL-Projection & 77,87\% & 5,063 & (10,225) & 7541606 & 0,096 & (0,060) & 574 \\
    Rec-Projection & 77,87\% & 5,246 & (10,124) & 7533022 & 0,337 & (0,181) & 6618 \\
    Rec-MCD & 76,43\% & 6,131 & (11,232) & 9486547 & 0,501 & (0,257) & 14061 \\
    BRL-MCD & 76,39\% & 5,829 & (11,386) & 9478656 & 0,098 & (0,069) & 638 \\
    BRL-Mahalanobis & 75,22\% & 5,317 & (10,030) & 7466731 & 0,095 & (0,058) & 540 \\
    Rec-Mahalanobis & 75,22\% & 5,498 & (9,878) & 7404242 & 0,376 & (0,224) & 8337 \\
    BRL-Spatial & 68,77\% & 6,498 & (10,558) & 8904351 & 0,093 & (0,055) & 466 \\
    Rec-Spatial & 68,68\% & 6,711 & (10,243) & 8687638 & 0,537 & (0,396) & 17716 \\
    BRL-Tukey & 64,15\% & 7,236 & (10,553) & 9485632 & 0,094 & (0,071) & 520 \\
    Rec-Tukey & 62,59\% & 7,598 & (9,682) & 8775283 & 0,994 & (0,556) & 47081 \\
    BRL-Liu & 61,45\% & 7,762 & (10,742) & 10175610 & 0,094 & (0,071) & 498 \\
    Rec-Liu & 59,95\% & 8,093 & (9,859) & 9425306 & 1,016 & (0,576) & 47327 \\
    Med-Liu & 52,93\% & 7,996 & (8,555) & 7944596 & 1,573 & (0,778) & 94444 \\
    Med-Projection & 51,52\% & 7,557 & (8,335) & 7333785 & 1,611 & (0,835) & 98310 \\
    Med-Spatial & 47,28\% & 8,572 & (8,473) & 8416410 & 1,648 & (0,790) & 91473 \\
    Med-L2 & 46,90\% & 8,439 & (8,417) & 8230309 & 1,650 & (0,803) & 91498 \\
    Med-CW & 38,22\% & 8,786 & (8,191) & 8358878 & 1,316 & (0,848) & 54253 \\
    Med-Tukey & 36,26\% & 8,084 & (7,539) & 7078699 & 1,369 & (0,856) & 54792 \\
    Med-Oja & 31,80\% & 9,061 & (7,681) & 8174602 & 1,234 & (0,756) & 38566 \\
    Med-Mahalanobis & 26,77\% & 10,670 & (7,873) & 10187020 & 2,371 & (0,447) & 90322 \\
    Mean  & 3,32\% & 12,137 & (5,732) & 10438196 & 2,016 & (0,721) & 8827 \\
    \bottomrule
    \end{tabular} 

  \label{tab:globalsynthresult}
\end{table*}

\subsection{Noise influence}\label{sec:noise-effect}

\begin{figure}
    \centering
    \includegraphics[width=\linewidth]{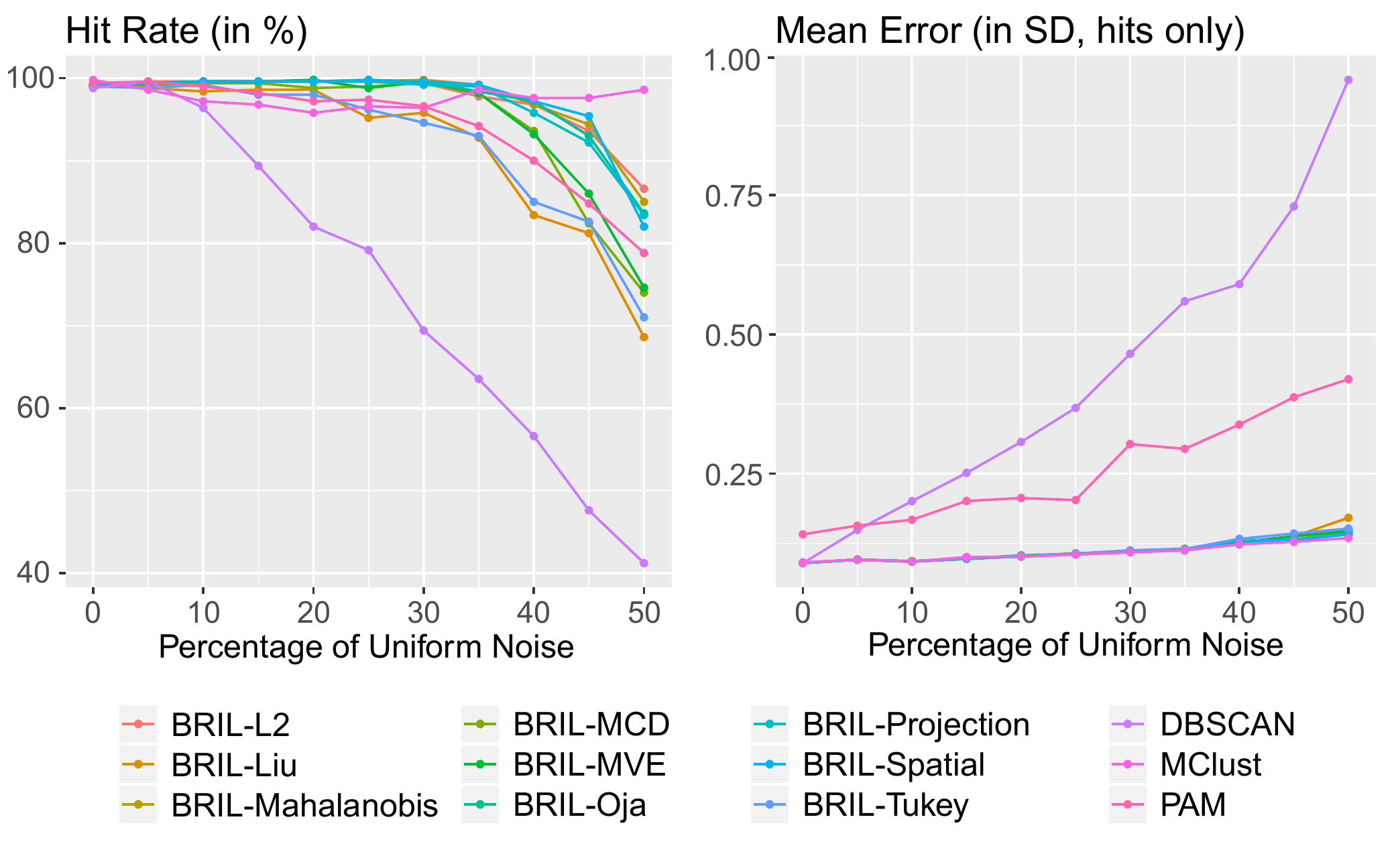}
    \caption{Effect of the quantity of uniform noise (from 0 to 250 samples, corresponding to 0 to $50\%$ of the total mixture) on the average error (when the cluster was encountered) and global hit rate (percentage of simulations where the main cluster was found). Distributions were drawn with $3$ cluster, $500$ samples, and a proportion of inliers equal to $40\%$. Note that the proportion of inliers being relative to the clustered samples only, 40\% inliers under a 50\% noise setting actually means that inliers represent only 20\% of the global mixture.}
    \label{fig:noiseEffect}
\end{figure}

The tolerance to uniform noise of the three standard clustering methods as well as of our algorithm are presented in Figure~\ref{fig:noiseEffect}. The clustering algorithms results are consistent with the findings from~\citep{brilhault_eyetracking_2019}. While \emph{DBSCAN} result significantly deteriorates with increasing contamination, model-based clustering and \emph{PAM} successfully maintain an accurate estimate of the main mode.

Other methods of clustering present much higher sensitivities to uniform noise, but were discarded from the present study. Interestingly, when encountering the main cluster, the BRIL estimates show an error of localization as low as model-based clustering, a technique specifically designed for the type of synthetic distributions tested in our simulations. We see from \ref{fig:noiseEffect} that our algorithm, no matter the depth measure or the convex body minimizer used, can withstand very high high quantity of noise, most of them showing a successful identification in more than 90\% of the simulation even when the uniform noise reaches 40\% (corresponding to an overall percentage of inliers of only 24\%, since the inliers ratio among clustered samples was 40\% as well).

\subsection{Effect of sample size}\label{sec:size-effect}

\begin{figure}
    \centering
    \includegraphics[width=\linewidth]{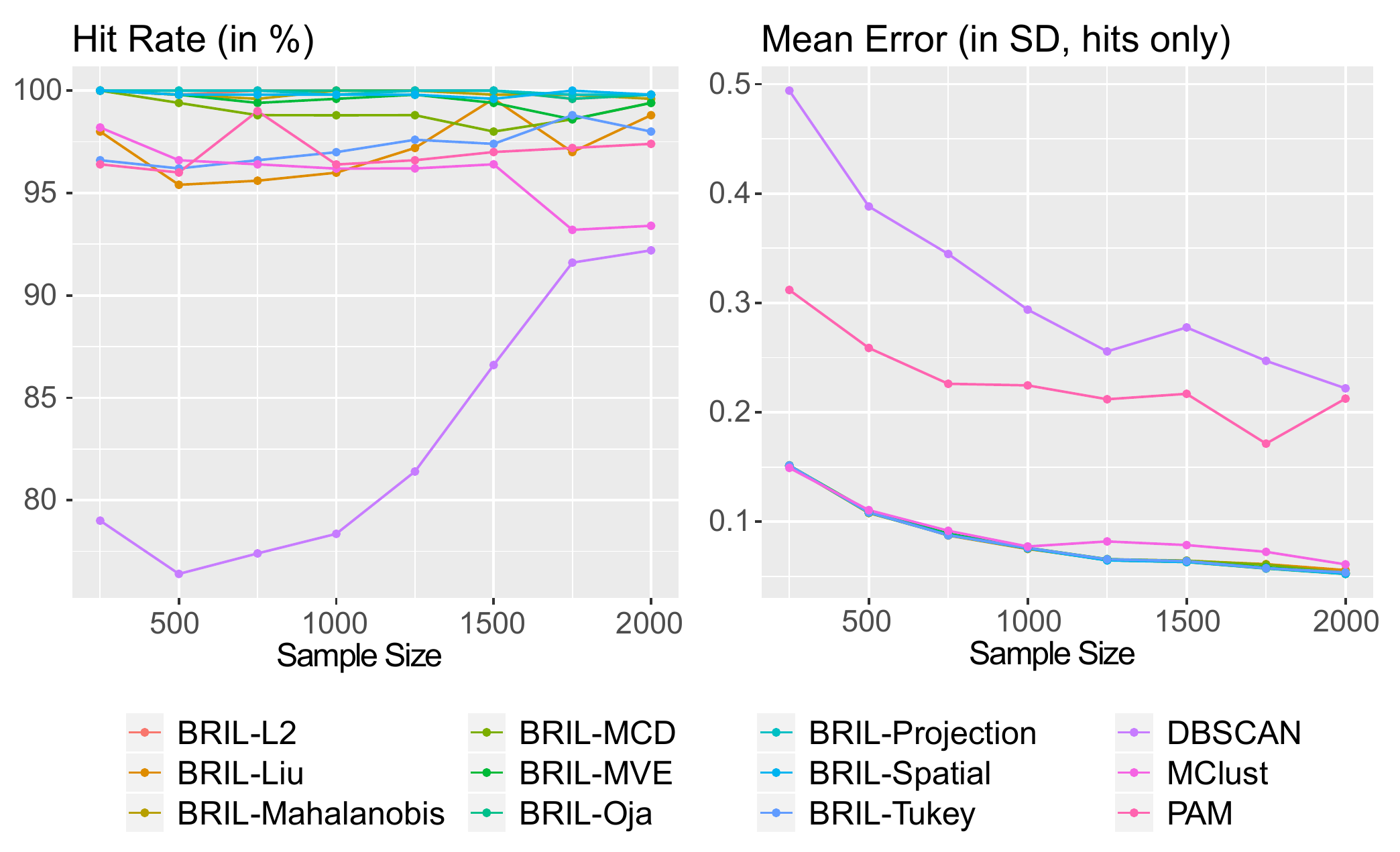}
    \caption{Effect of sample size on the average error and hit rate. Distributions were drawn with 25\% uniform noise, 3 clusters, and 40\% of the clustered samples belonging to the main cluster (i.e. 30\% of the global mixture). Sample size ranged from 250 to 2000 samples.}
    \label{fig:sizeEffect}
\end{figure}

In Figure~\ref{fig:sizeEffect} we report the results of the different algorithms for sample sizes ranging from 250 to 2000. The simulations were run with mixtures composed of 3 clusters (30\% of the samples in the main clusters, 22.5\% in each of the two others), and 25\% uniform noise. Regarding BRIL performances, we see that the sample size does not affect significantly the results of the algorithm in hit rates for all convex body minizers and depth measures tested. In terms of precision of localization when the main cluster was encountered, we can notice a slight improvement with higher samples sizes, yet of low magnitude (the maximum difference observed from 250 samples to 2000 was inferior to $0.1\sigma$). 

\emph{PAM} and \emph{MClust} clustering algorithms provide high invariance to sample size as well. \emph{DBSCAN}, on the other hand,  suffered from a mild deterioration of its performances as the sample size decreases (with an hit rate going from 79\% for 250 samples, to 92\% for 2000, and a mean error dropping from 0.494 to 0.221 in case of hits). This sample size size effect with \emph{DBSCAN} could be explained, as noted in \citep{brilhault_eyetracking_2019}, by: (i) the parameter related to the minimum number of points that, in our experiment, was constant, regardless of the sample size; and (ii) the decrease of the relative difference of densities between the clusters and the uniform noise regions, which were randomly drawn in a spatial area of fixed size.

\subsection{Execution Time}

\begin{figure*}
    \centering
    \subfigure{
        \includegraphics[width=.45\linewidth]{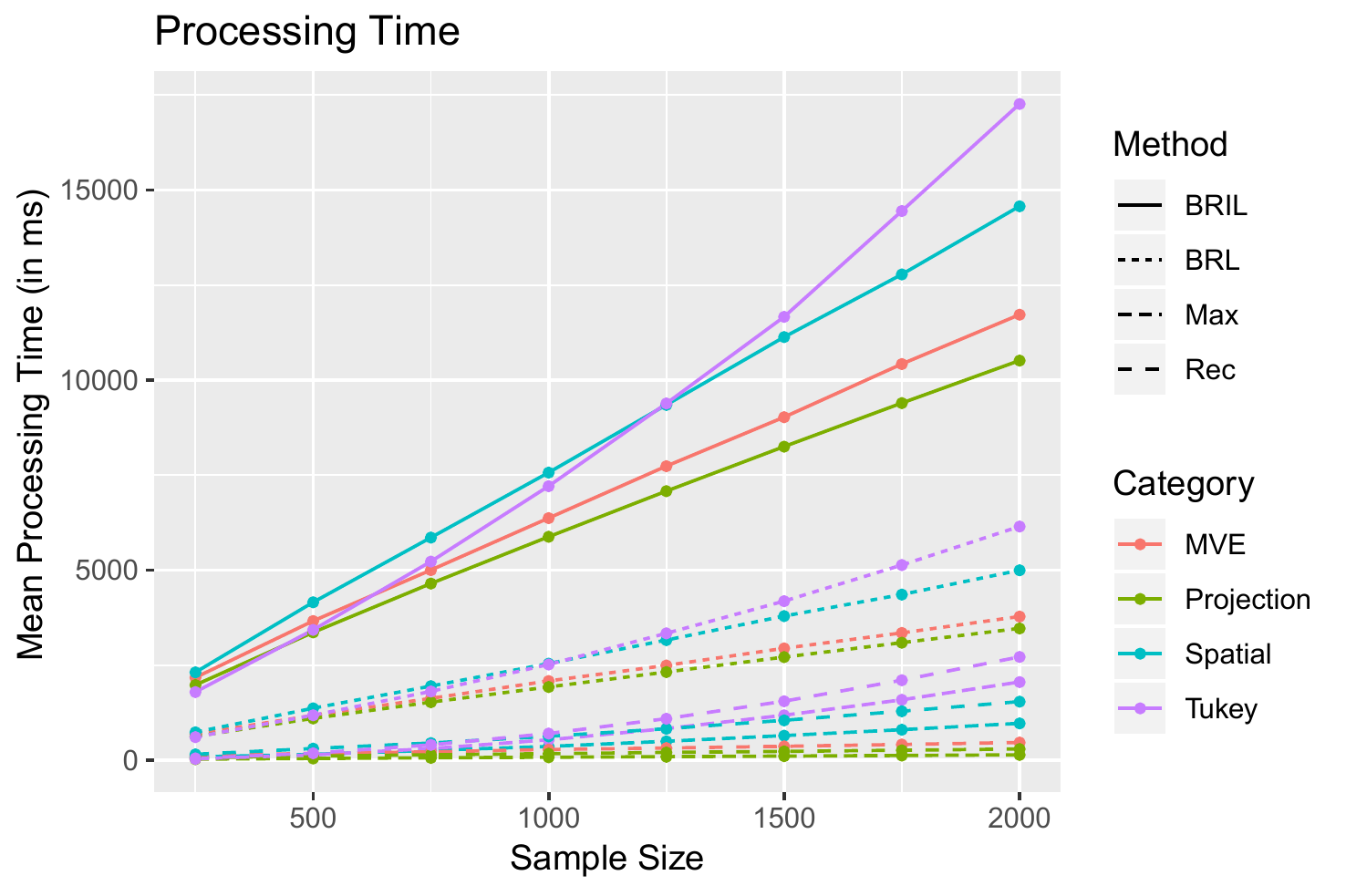}
    }
        \subfigure{
        \includegraphics[width=.45\linewidth]{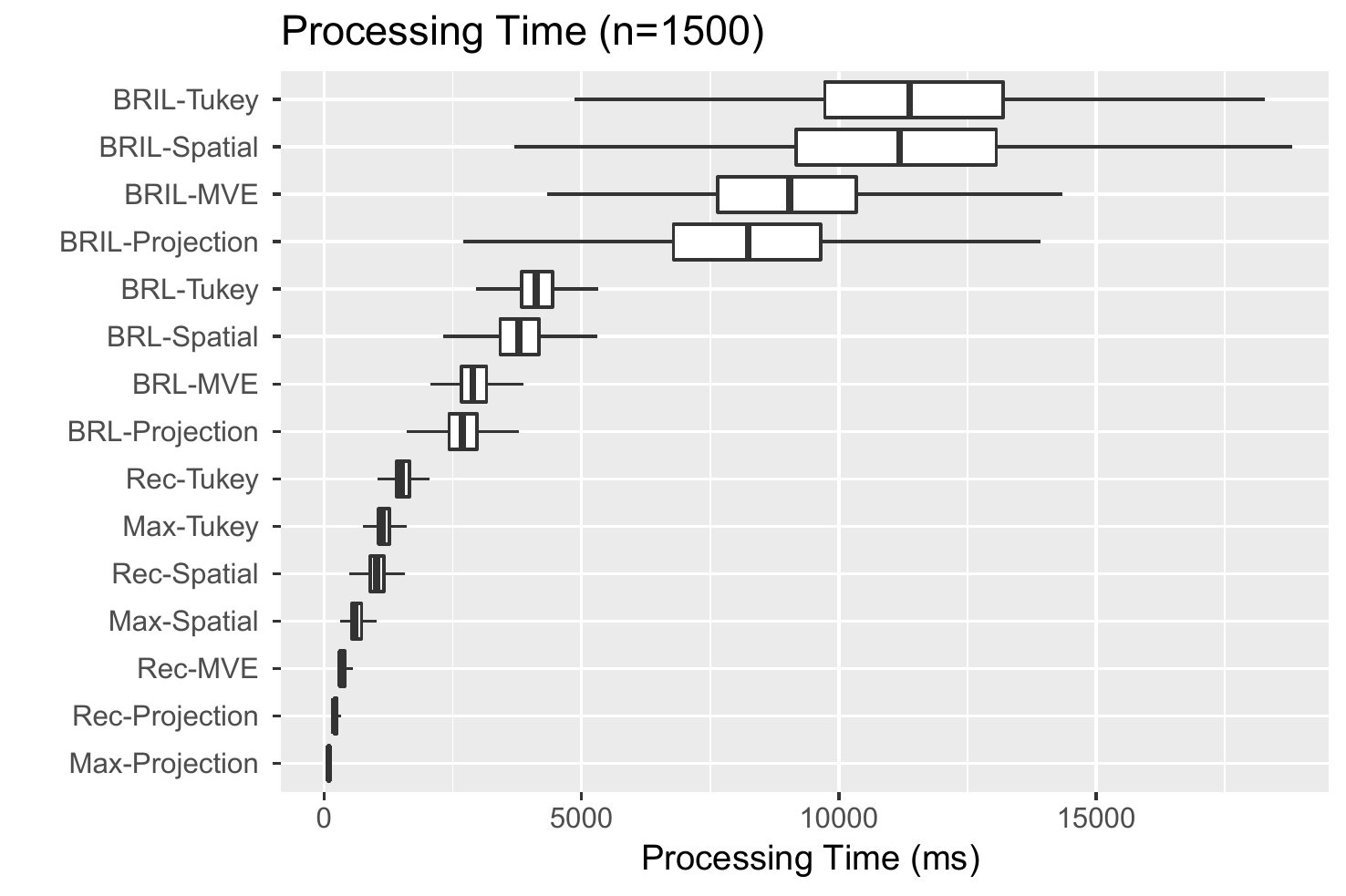}
    }
    \subfigure{
        \includegraphics[width=.45\linewidth]{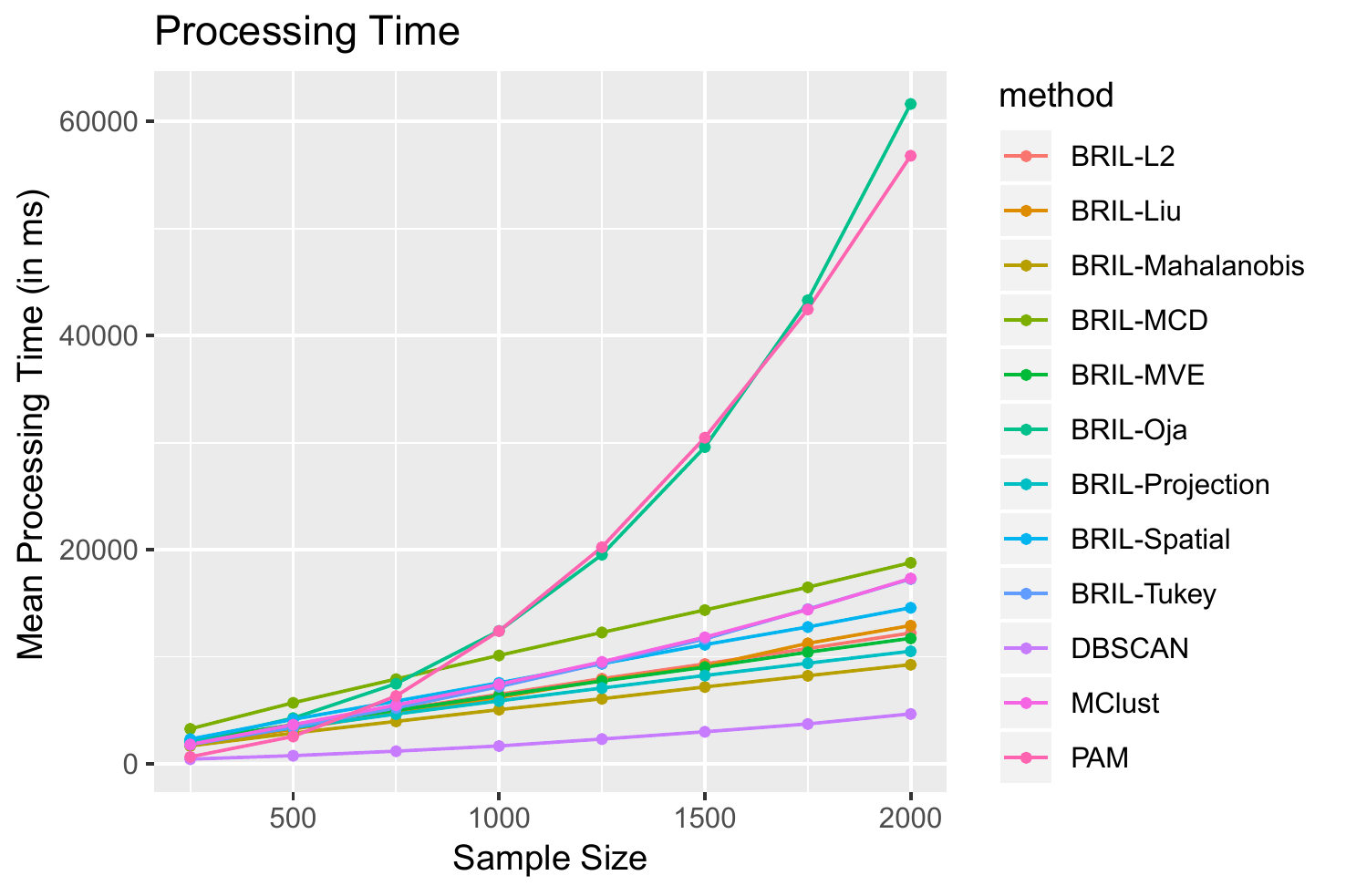}
    }
        \subfigure{
        \includegraphics[width=.45\linewidth]{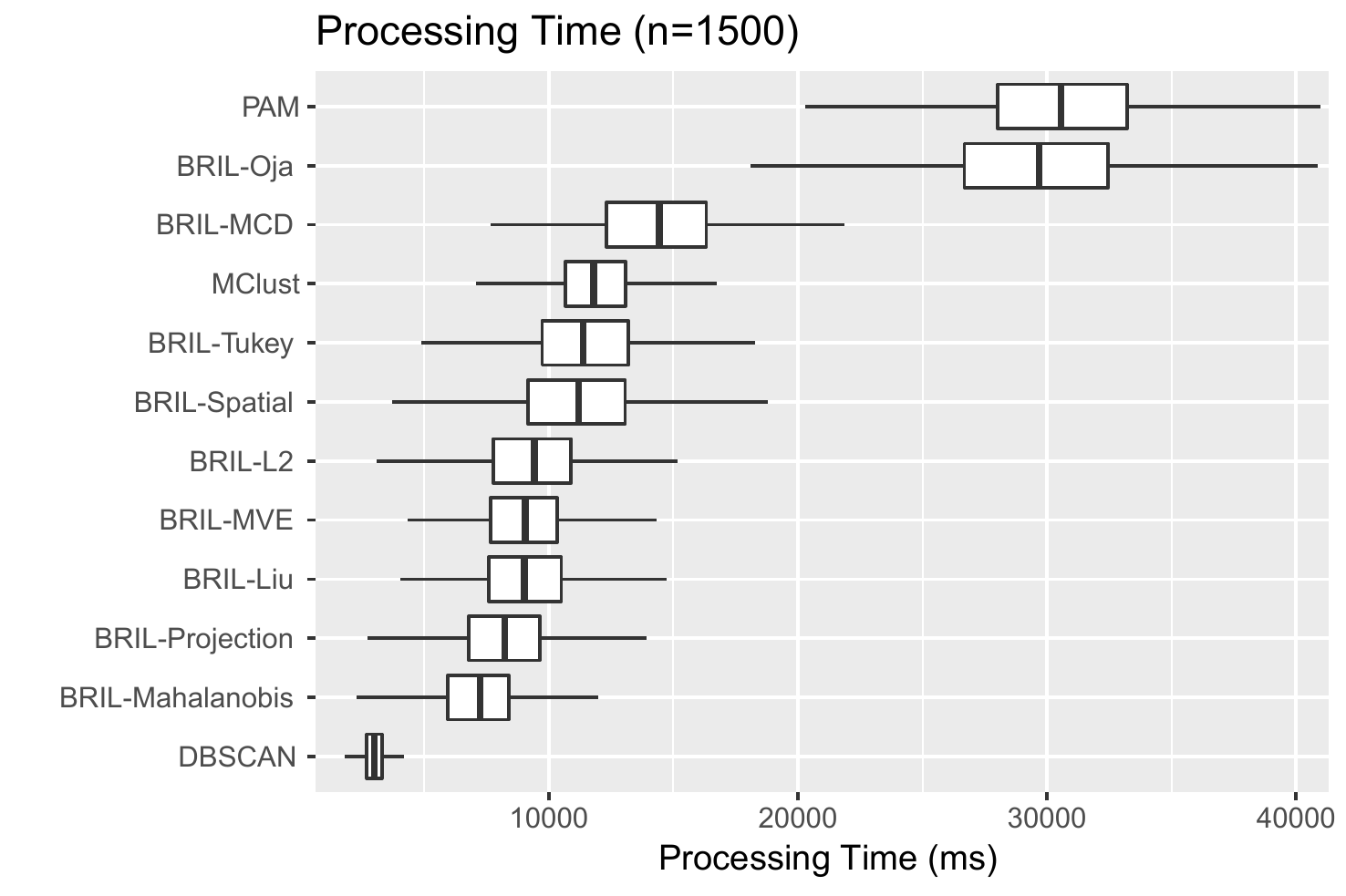}
    }
    \caption{Average computing times. In the first row we present the cumulative impact in processing time of the different steps of our algorithm, namely the recursive estimate (\emph{REC}), the refined version (\emph{BRL}) and the full iterative procedure (\emph{BRIL}). These results are provided for 4 illustrative methods (Tukey depth, Spatial depth, Projection depth and MVE). In the second row, we report the average processing time for all \emph{BRIL} methods, along with the clustering algorithms. The mixtures tested included from 3 to 5 clusters and 25\% uniform noise. 500 repetitions were ran for each sample size, which ranged from 250 to 2000.}
    \label{fig:computingTime}
\end{figure*}

The average execution times measured on our synthetic dataset are presented in Figure~\ref{fig:computingTime}. Our tests were conducted on a personal workstation using the R environment. These measurements are only indicative, as they obviously depends on the computer's hardware, as well as on each algorithm implementation, optimization, sets of parameters and on the distributions considered (some of these techniques having a higher computing cost when processing a larger number of clusters for instance, despite the same total sample size). Yet, these processing times still provide insights on the potential application of our algorithm in a real environment, and how it positions itself in regard to other methods.

One can observe that the differences between the sample median and the recursive version are relatively small. The recursive method, although improving significantly the precision and percentage of hits, have a negligible impact on computing times for the sample sizes tested. The refined locator does increase consequently these processing times, but remain acceptable (under one second for 1000 samples with our experimental setup), and the final iterative version, identifying the other modes/clusters of the distribution, is in average two to three times more costly than the refined estimate. Results obtained with convex body minimizers present similar ranges as those from depth measures, with the exception of estimates based on Oja depth, which have significantly longer processing times. Finally, looking at clustering methods, \emph{DBSCAN} is by far the fastest to compute, but shows poor performances in contaminated distributions. \emph{PAM} and \emph{MClust} algorithms, showing better precision of localization than \emph{DBSCAN} in our synthetic dataset, have in turn much longer processing times. All factors being equal, and leaving out Oja depth, our method is generally slighly faster than \emph{MClust}, signficantly quicker than \emph{PAM}, but much longer than \emph{DBSCAN}.

\subsection{Real-world Application}

As mentioned in the introduction, eye-tracking research with primates, or non-cooperative subjects such as babies or young children, can be quite challenging. In order to compute the user's gaze accurately, a calibration procedure is generally required to estimate individual parameters and the mapping function to the screen coordinates. To compute this projection, a set of reference points is necessary. When working with cooperative subjects, this data is generally acquired through a calibration routine where several calibration targets are presented successively on the screen while the user has to manually confirm his fixations. However, with non-cooperative subjects, the main issue resides in the fact that there is no direct way to assess the intervals during which they were actually looking at the calibration points. Therefore, among the full set of gaze coordinates collected during the presentation of each target, only a part of these samples will present valid positions, the rest corresponding to times the subject was attending other parts of the screen. These situations will typically lead to multimodal distributions, composed of several clusters, corresponding to the different screen locations that the subject fixated, and of random samples which result from ocular saccades (when the eye is moving from a position to another), eye blinks, or artefacts from the recording equipment.

\begin{figure*}
    \centering
    \includegraphics[height=6cm,keepaspectratio]{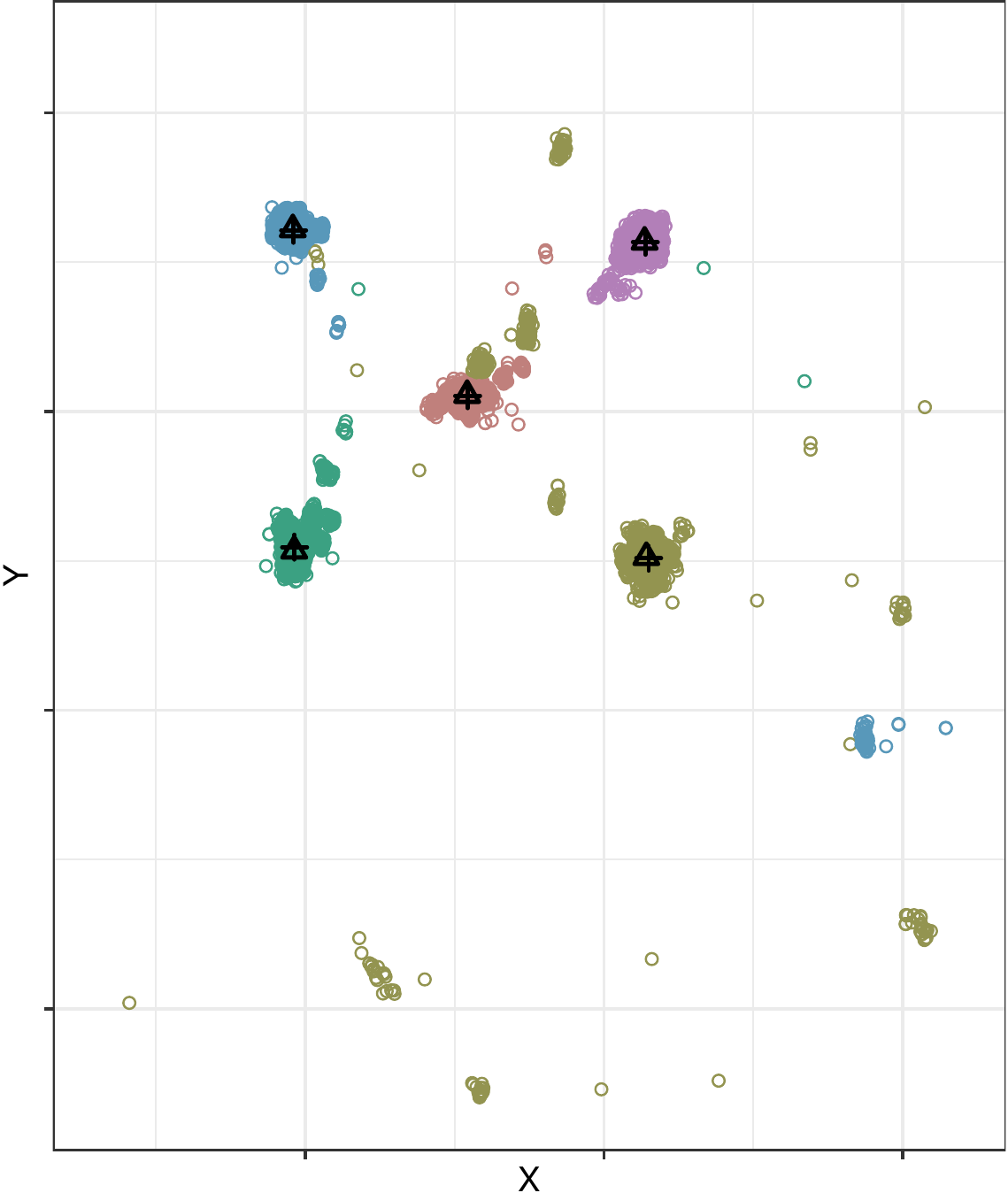} \hfill
    \includegraphics[height=6cm,keepaspectratio]{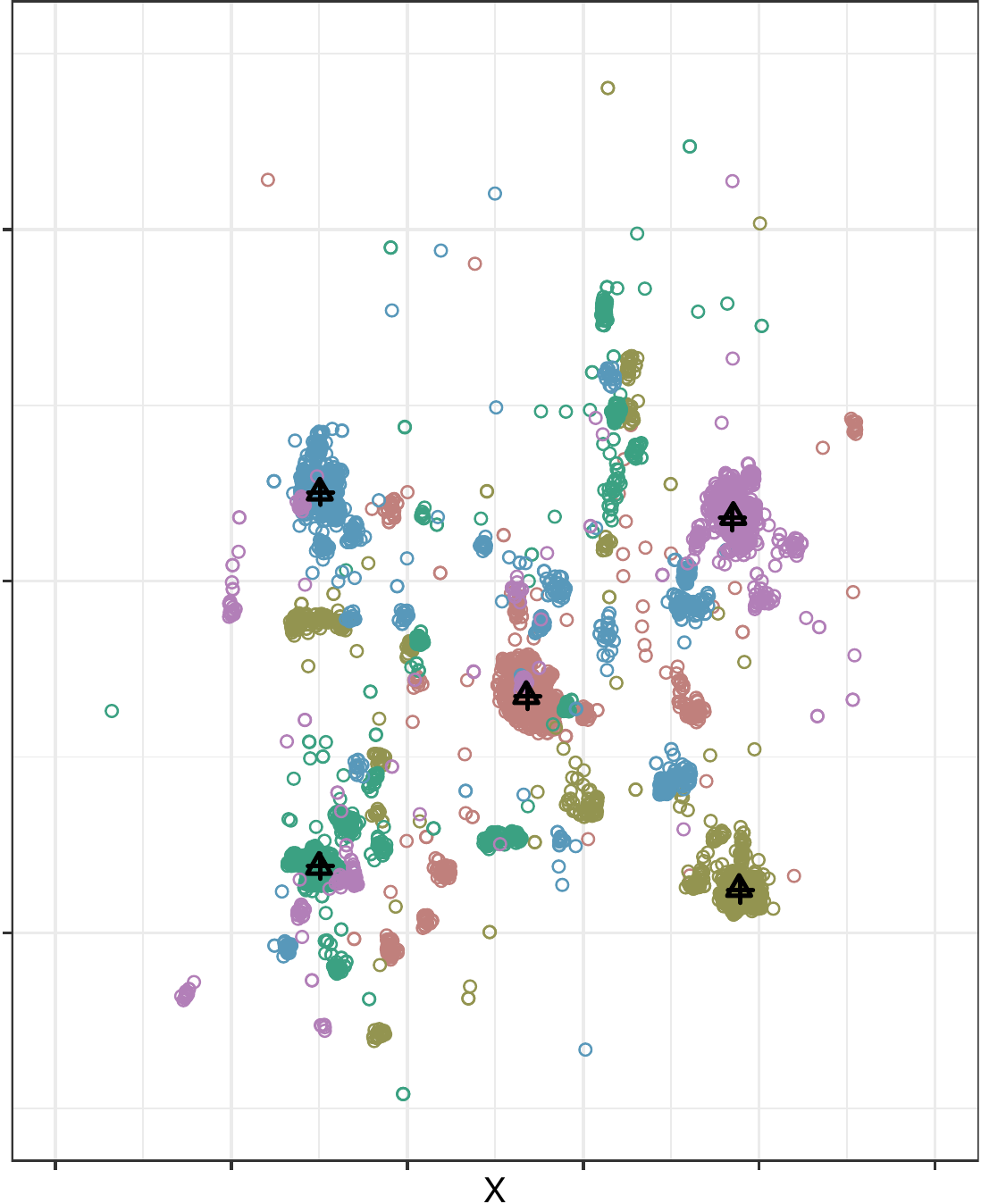}\hfill
    \includegraphics[height=6cm,keepaspectratio]{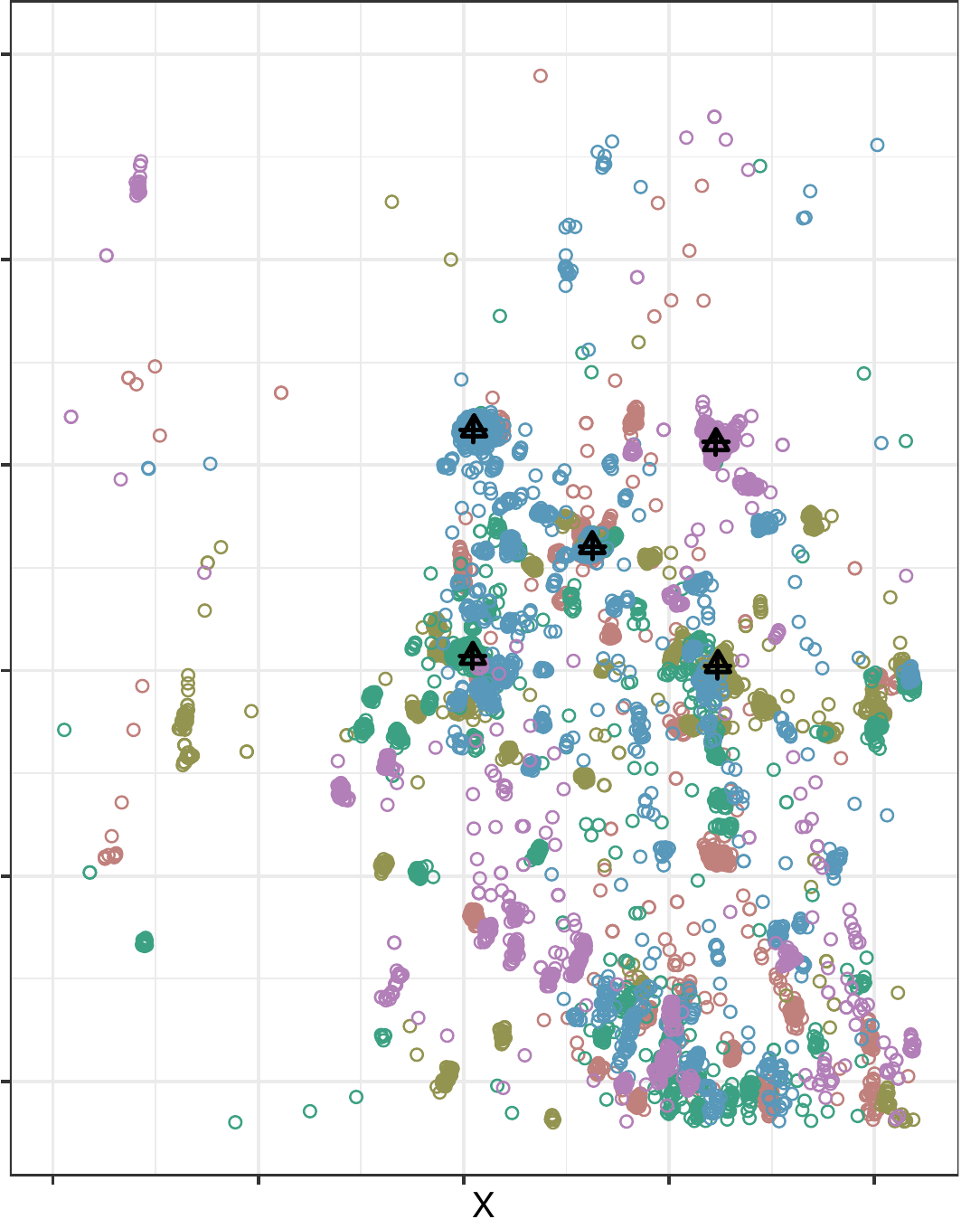}
    \includegraphics[height=6cm,keepaspectratio]{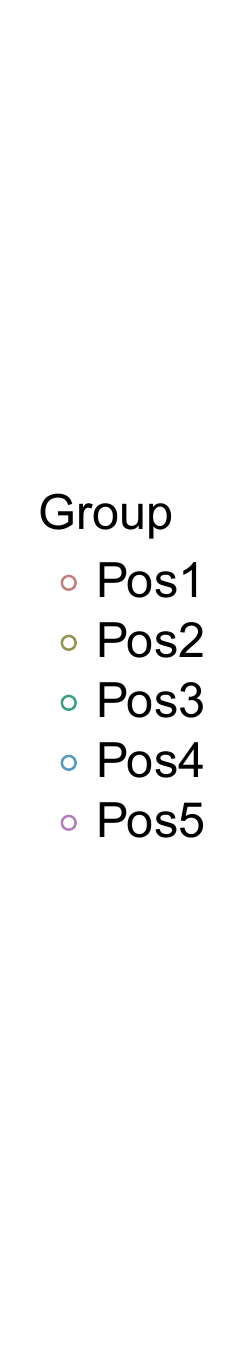}
    \caption{Eye-tracking recordings of three calibration procedures with Capuchin monkeys. Five targets were successively presented on the screen at coordinates (0,0), (100,-100), (-100,-100), and (100,100), for 1200 to 3000ms, with multiple repetitions of each target in a randomized manner. The gaze coordinates collected by the eye-tracker in these intervals are displayed in different colors, based on the target displayed at the time. Black crosses correspond to the ground-truth coordinates of each target, while black triangles indicate the positions computed by the BRIL algorithm using projection depth. Our method is able to accurately locate the main fixation in each dataset, even when it reaches very high levels of noise as in the third example.}
    \label{fig:resultCalibration}
\end{figure*}

We show in Figure~\ref{fig:resultCalibration} three experimental datasets, consisting of eye-tracking calibrations performed by Capuchin monkeys, with increasing levels of contamination. Each procedure included five calibration targets, presented for 1200 to 3000ms, with multiple repetitions of each target in a randomized manner. The gaze coordinates from all the repetitions of a specific target in a given calibration were pooled, and filtered (see Table~\ref{table:tableDatasets}). Out of the remaining samples, 25 random draws of 1000 data points were performed to compute the average, deviation, and sum of squared errors of each algorithm for every target, which were then grouped by session.

The estimates for each of the calibration targets computed with the BRIL algorithm based on spatial depth are indicated by triangle markers in Figure~\ref{fig:resultCalibration}, along with the reference positions considered as ground-truth, represented by the black cross-hairs, which were manually annotated. The complete results, for each of the methods, including the clustering algorithms, are provided in Table~\ref{tab:resultCalibGeneral}. Errors are given in screen pixels, which based on the monitor size, resolution and distance to the viewer, can be converted in visual field angles, 1 degree corresponding to 25 pixels in our experimental setup. While the mean position, commonly used with calibrations realized by cooperating subjects, performs poorly in all situations, we see that the coordinate-wise median, related to the procedure from \citep{model_probabilistic_2012}, is able to provide reliable results under low contamination, but is not suitable as the quantity of outliers increases. In agreement with the conclusions from \citep{brilhault_eyetracking_2019}, we can observe that model-based clustering very high performances on synthetically datasets do not translate to real-world data. On all our datasets, even the less contaminated ones, \emph{MClust} estimates show the largest errors among all the methods tested. \emph{PAM}, which was the best candidate from our previous study, provide good results, with an average error of 1.300 (0:052 visual degrees), offering a satisfying precision of localization of fixations. Nonetheless, most estimates from the BRIL algorithm achieve even higher precision, with average errors below 1.0 for five of them (less than 0:04 visual degrees), and therefore appear as the best choice for estimating the reference calibration coordinates in noisy datasets. We can finally note that \emph{DBSCAN} showed poorer performances (3.634 average error), yet still acceptable depending of the requirements of the application.

\begin{table*}[!bth]

    \caption{Errors measured on three experimental datasets of increasing contamination. The errors are expressed in screen pixels (1 degree of visual angle corresponding to 25 pixels), and averaged across 25 re-samplings of the eye-tracking data for every calibration target in each of the three sessions. Errors correspond to the distance from estimates to the ground-truth coordinates. The standard deviation of the mean is also provided, as well as the sum of squared errors.}
    
  \centering
    \resizebox{\linewidth}{!}{%

       \begin{tabular}{lr@{\hskip 4pt}>{\footnotesize}lrr@{\hskip 4pt}>{\footnotesize}lrr@{\hskip 4pt}>{\footnotesize}lrr@{\hskip 4pt}>{\footnotesize}lr}
           \toprule
          & \multicolumn{3}{c}{\textbf{All}} & \multicolumn{3}{c}{\textbf{Set1}} & \multicolumn{3}{c}{\textbf{Set2}} & \multicolumn{3}{c}{\textbf{Set3}} \\
    \multicolumn{1}{c}{\textbf{Method}} & \multicolumn{2}{c}{\textbf{MeanError (SD)}} & \multicolumn{1}{c}{\textbf{SSE}} & \multicolumn{2}{c}{\textbf{MeanError (SD)}} & \multicolumn{1}{c}{\textbf{SSE}} & \multicolumn{2}{c}{\textbf{MeanError (SD)}} & \multicolumn{1}{c}{\textbf{SSE}} & \multicolumn{2}{c}{\textbf{MeanError (SD)}} & \multicolumn{1}{c}{\textbf{SSE}} \\

    \midrule
    BRIL-Spatial & 0,915 & (0,804) & 556   & 0,927 & (1,064) & 248   & 0,799 & (0,469) & 107   & 1,020 & (0,756) & 201 \\
    BRIL-Tukey & 0,947 & (0,832) & 595   & 0,928 & (1,102) & 258   & 0,851 & (0,498) & 121   & 1,063 & (0,775) & 216 \\
    BRIL-Oja & 0,931 & (0,914) & 637   & 0,893 & (1,020) & 229   & 0,728 & (0,518) & 100   & 1,172 & (1,053) & 309 \\
    BRIL-Projection & 0,938 & (0,935) & 657   & 0,858 & (0,968) & 208   & 0,741 & (0,534) & 104   & 1,215 & (1,136) & 345 \\
    BRIL-L2 & 0,956 & (0,922) & 661   & 0,952 & (1,062) & 253   & 0,768 & (0,543) & 110   & 1,149 & (1,035) & 298 \\
    BRIL-Mahalanobis & 1,052 & (1,024) & 807   & 0,937 & (1,032) & 242   & 1,014 & (0,864) & 221   & 1,205 & (1,145) & 344 \\
    BRIL-MVE & 1,084 & (1,114) & 905   & 1,019 & (1,116) & 284   & 0,894 & (0,847) & 189   & 1,339 & (1,295) & 432 \\
    BRIL-MCD & 1,091 & (1,151) & 942   & 0,808 & (0,891) & 180   & 1,097 & (1,041) & 285   & 1,368 & (1,399) & 477 \\
    PAM   & 1,300 & (1,111) & 1095  & 1,691 & (1,664) & 701   & 0,940 & (0,415) & 132   & 1,269 & (0,703) & 263 \\
    BRIL-Liu & 1,305 & (5,142) & 10528 & 0,938 & (1,120) & 265   & 0,873 & (0,561) & 134   & 2,105 & (8,787) & 10129 \\
    DBSCAN & 3,634 & (9,828) & 41074 & 1,702 & (1,433) & 617   & 4,811 & (2,877) & 3920  & 4,390 & (16,590) & 36538 \\
    MedCW & 8,733 & (22,233) & 213461 & 1,268 & (1,103) & 352   & 2,233 & (0,905) & 725   & 22,696 & (34,547) & 212385 \\
    Mean  & 52,909 & (49,371) & 1961383 & 14,229 & (13,757) & 48776 & 31,721 & (12,493) & 145134 & 112,776 & (37,852) & 1767473 \\
    MClust & 63,861 & (125,099) & 7382355 & 29,795 & (91,250) & 1143470 & 30,507 & (67,868) & 687488 & 131,282 & (165,516) & 5551397 \\
    \bottomrule
    \end{tabular} 
    }
  \label{tab:resultCalibGeneral}
\end{table*}

\section{Conclusion}\label{sec:conclusion}

In this study, we presented an analysis of the use of depth measures in identifying main modes of multivariate distributions. We characterized their robustness and behavior with respect to the sample size, number of clusters, percentage of inliers, and uniform noise. We showed that standard depth medians can provide reliable estimates when contamination levels are relatively low. However, in multimodal distributions with proportions of outliers beyond 50\% (the breaking point of most depth functions analized), these locators are not suited to estimate the center of the main component. Our method, on the other hand, performs well in such critical conditions, withstanding up to $100\times\frac{(K-1)}{K}$ percent of clustered contamination (with $K$ being the total number of clusters). The overall proportion of outliers tolerated by our algorithm can be even higher if additional contamination in a non-clustered form, such as uniform noise, is considered.

Our method consists of the following steps: i) a recursive estimate based on depth measures or convex body minimizers; ii) outlier filtering; and iii) an iterative identification of clusters. The recursive application of depth functions, or convex-body minimizers, aims at converging toward a local density peak, more reliably than the traditional multivariate medians. The purpose of the next filtering steps is to refine this location by discarding as many outliers as possible to guarantee high robustness and precision. We first eliminate the more distant outliers based on their Euclidean distance and a test of unimodality. The remaining samples are, then, filtered using robust distances and a test of multivariate normality. Besides improving the accuracy of the estimate, our filtering procedure allows to identify the samples belonging to a particular component of the multimodal distribution. By removing the cluster found from the overall distribution, and applying the same process recursively, we are able to extract the different components of the mixture. We can finally select the cluster with the highest cardinality as the main mode, thus, avoiding errors that commonly occur when the first depth estimate identified a cluster of outliers.

Our method proved to be successful when applied to a real-world scenario, the calibration of an eye-tracking system in the context of visual experiments with monkeys. Although the datasets were highly contaminated, we still obtained precise estimates for the reference coordinates. By using this new central tendency measure, the user or experimenter are not required to confirm the correct fixations manually. As long as the cumulative durations of the fixations to the actual targets exceeds the time spent on irrelevant positions, our estimate will reliably identify the calibration parameters. For these reasons, we believe that our method could be of high value in eye-tracking studies with inattentive or poorly cooperative subjects, such as infants and animals, where data typically suffer from high contamination. It would also present interest in the case of experimental paradigms which do not include an explicit calibration routine. 
Beyond its use in eye-tracking calibrations, our multivariate mode estimate can find application in various other scenarios requiring a robust measure of central tendency. Moreover, because our algorithm also provides a partition of the data, it may also contribute to the development of new powerful clustering techniques. Compared to current clustering methods, our approach presents several advantages. First, it does not require an explicit number of clusters, and therefore avoids the costly iteration over a range of potential $k$ values in order to select the optimal number (maximizing a criteria as for instance the average silhouette or gap statistics). Our method is also computationally fast, and offers a very high tolerance to noise. Many robust clustering techniques, such as \emph{TClust}~\citep{fritz_tclust:_2012}, require an alpha parameter, which can strongly affect the final partitions and require prior knowledge on the amount of noise, or a careful visual analysis of the data. Our method, on the other, is non-parametric, and will handle any quantity of noise seamlessly. The use and performances of the BRIL algorithm as a clustering technique, named BRIC (Bootstrap and Refine Iterative Clustering), will be investigated in subsequent studies.

\section*{Acknowledgments} \label{sec:acknowledgments}

This paper is a result of research conducted on projects sponsored by CAPES (Coordination for the Improvement of Higher Education Personnel -- Brazilian Federal Government Agency) under grant 88882.306276/2018-01 and CNPq (National Council for Scientific and Technological Development), Brazil, under grants 438850/2018-1 and 432995/2018-8. We also would like to thank FAPESP for supporting the Center of Mathematical Sciences Applied to Industry (CEPID-CeMEAI) under grant 2013/07375-0, which provided all computational resources to perform our experimental analyses.

\bibliographystyle{spbasic}      
\bibliography{ArticleEyetracking}  

\end{document}